\documentclass[aps,prx,superscriptaddress,amsfonts,amsmath,amssymb,showpacs,floatfix,reprint,longbibliography]{revtex4-1}
\usepackage{url}
\usepackage{bm}
\usepackage{graphicx}
\usepackage{amsmath}
\usepackage{amstext}
\usepackage{amssymb}
\usepackage{amsfonts}
\usepackage{amsbsy}
\usepackage{verbatim}
\usepackage{color}
\usepackage[colorlinks=true, urlcolor=blue, linkcolor=blue, citecolor=blue, pdftex]{hyperref}
\usepackage{multirow}
\usepackage{float}
\usepackage{gensymb}
\usepackage{textcomp}
\usepackage{enumitem}
\usepackage[english]{babel}
\usepackage[pdftex]{epsfig}
\usepackage{ragged2e}
\usepackage{times}
\usepackage[table,xcdraw]{xcolor}
\usepackage{epstopdf}
\usepackage{physics}
\usepackage[export]{adjustbox}
\usepackage{dsfont}
\usepackage[utf8]{inputenc}

\definecolor{karlocolor}{rgb}{0.9,0.2,0}
\definecolor{pocokcolor}{rgb}{0.9,0.05,0.05}
\definecolor{addedtextcolor}{rgb}{0.0,0.5,0.5}

\def\gsm{\mathcal{M}_{\textrm{GS}}}

\def\onehalf{\frac{1}{2}}

\begin{document}

\title{
Degenerate manifolds, helimagnets, and multi-{\bf Q} chiral phases in the classical Heisenberg antiferromagnet on the face-centered-cubic lattice}

\author{P\'eter Balla}
\affiliation{Institute for Solid State Physics and Optics, Wigner Research Centre for Physics, H-1525 Budapest, P.O.B. 49, Hungary}
\author{Yasir Iqbal}
\affiliation{Department of Physics, Indian Institute of Technology Madras, Chennai 600036, India}
\author{Karlo Penc}
\affiliation{Institute for Solid State Physics and Optics, Wigner Research Centre for Physics, H-1525 Budapest, P.O.B. 49, Hungary}

\date{\today}

\begin{abstract}
We present a detailed study of the ground state phase diagram of the classical frustrated Heisenberg model on the face-centered-cubic lattice. By considering exchange interactions up till third nearest neighbors, we find commensurate, helimagnetic, as well as noncollinear multi-{\bf Q} orders which include noncoplanar and chiral spin structures. We reveal the presence of subextensively degenerate manifolds that appear at triple points and certain phase boundaries in the phase diagram. Within these manifolds, the spin Hamiltonian can be recast as a complete square of spins on finite motifs, permitting us to identify families of exact ground state spin configurations in real space -- these include  randomly stacked ferro- or antiferromagnetically ordered planes and interacting ferromagnetic chains, among others. Finally, we critically investigate the ramifications of our findings on the example of the Ising model, where we exactly enumerate all the states numerically for finite clusters.
\end{abstract}


\maketitle

\section{Introduction}

A first acquaintance with {\it geometric frustration}, which has played such a pivotal role in modern condensed matter physics, is tacitly made in the context of a classic textbook example of the crystal structure of salt (NaCl), namely, the face centered cubic (fcc) lattice. The elementary motif of any covering of the fcc lattice is a triangle, and this feature renders it impossible for antiferromagnetically (AF) interacting spins to simultaneously satisfy all interactions. Indeed, a system of AF interacting spins treated as $n$-component classical (spin $S\to\infty$) vectors of a fixed length forms an infinitely degenerate one-dimensional ground state manifold at zero temperature~\cite{Sousa-2008}. The investigation of the thermodynamic and critical behavior of the fcc $O(n)$ antiferromagnet has had a long history and been the subject of much debate, especially for the $n=1$ (Ising) model~\cite{Peierls-1936,Easthope-1937,Shockley-1938,Danielian-1961,Danielian-1964} which is now known to undergo a first order transition into a collinear AF state~\cite{Binder-1980,Meirovitch-1984,Kammerer-1996,Ader-2001,Beath-2005,Beath-2006,Beath-2007}, while on the other hand, hardly much is known about the classical $n=2$ ($XY$) model~\cite{Diep-1989,Sousa-2008}. Concerning the physically realizable case of $n=3$ (Heisenberg) spins, after much debate~\cite{Fernandez-1983,Diep-1989,finnish_93,henley_fcc}, there now appears to be a consensus that the model undergoes a first order phase transition into a collinear AF state~\cite{Gvozdikova-2005}. For Heisenberg spins away from the classical limit, i.e., when $1/S\neq0$, only few studies have addressed the role of quantum fluctuations in the large-$S$~\cite{haar_lines} or small-$S$ limits~\cite{Kuzmin-2003}, and nearly seven decades after being first attended to~\cite{Anderson-1950}, the determination of the nature of the ground state of the quantum Heisenberg antiferromagnet on the fcc lattice remains a critically outstanding problem.

Recently, it has been realized that the subextensive degeneracy of the $T=0$ ground state manifold is not unique to first  neighbor ($J_{1}$) antiferromagnetic interactions and that upon inclusion of second neighbor interactions ($J_{2}$) a two-dimensional subextensively degenerate ground state manifold in the form of a spiral surface can be stabilized for $J_{2}/J_{1}=1/2$~\cite{ourspiral}, in addition to three different AF commensurate orders~\cite{Villain-1959}. In this work, we incorporate a third neighbor exchange coupling $J_{3}$ and obtain the $T=0$ global phase diagram of the classical $J_{1}$-$J_{2}$-$J_{3}$ Heisenberg model for all combinations of signs of couplings, which has hitherto not been investigated. Our study reveals a rich phase diagram featuring helimagnetic and noncollinear multiple-{\bf Q} orders which allow for noncoplanar and chiral structures. Our most salient finding is that the phase diagram is host to highly (subextensive) degenerate one- and two-dimensional ground state manifolds in  reciprocal space (henceforth referred to as $\mathbf{q}$ space) occurring at triple points (where several phases meet) and phase boundaries. Remarkably, at these triple points and phase boundaries, we are able to express the Hamiltonian as a positive definite sum of complete squares on finite motifs covering the lattice allowing one to understand the origin of the subextensively degenerate manifold of states. This reformulation also  permits us to explicitly construct large classes of nontrivial, aperiodic ground states in real space, consisting of randomly stacked ordered planes and frustrated ferromagnetically ordered chains in special crystallographic directions. Considering the case of Ising spins, we are able to completely enumerate, on finite clusters, the type of possible configurations in real and Fourier space, and provide indications of an even richer structure for Heisenberg spins. Our work also provides the basis for understanding the origin of a plethora of fcc magnetic structures in a wide  variety of magnetic materials~\cite{Seehra-1988}. In particular, a recent investigation of the half-Heusler compound GdPtBi~\cite{Sukhanov-2020} proposed an antiferromagnetic $J_{1}$-$J_{2}$-$J_{3}$ Heisenberg model, and argued for the indispensability of a $J_{3}$ interaction to match the observed neutron scattering profile while a highly frustrated double perovskite compound Ba$_{2}$CeIrO$_{6}$ is known to develop antiferromagnetic order and has been argued to be located in the vicinity of the $J_{1}$\textendash$J_{2}$ model~\cite{Dominik}.

The remainder of the article is structured as follows:
We introduce the fcc lattice, the Heisenberg model and the Luttinger-Tisza method in Sec.~\ref{sec:model}.
The phase diagram is derived in Sec.~\ref{sec:phase_diagram} and
Sec.~\ref{sec:realfourier_chapter} presents the commensurate phases and the construction of the possible multiple-$\mathbf{Q}$ structures.
In Sec.~\ref{sec:chirality} the different notions of the chirality are discussed.
Sec.~\ref{sec:incommensurate} is devoted to incommensurate phases.
We describe the details of the construction of the spin structures in real and reciprocal space in the ground-state manifolds in Sec.~\ref{sec:manifolds}.
Finally, in Sec.~\ref{sec:conclusions} we conclude with a summary of the results. 
The article ends with the following (mostly technical) Appendices. 
In Sec.~\ref{sec:lattice_conventions} we define our conventions for the lattice and show the Fourier transform of the interactions.
Sec.~\ref{sec:table_of_phase_boundaries} contains a table of the phase boundaries. 
The Ising configurations in the different subextensive manifolds are enumerated in Sec.~\ref{sec:Ising_results} for finite clusters and critically compared to results of Sec.~\ref{sec:manifolds}.

\section{The model and the Luttinger-Tisza method}
\label{sec:model}

\begin{figure}
	\centering
	\includegraphics[width=.9\columnwidth]{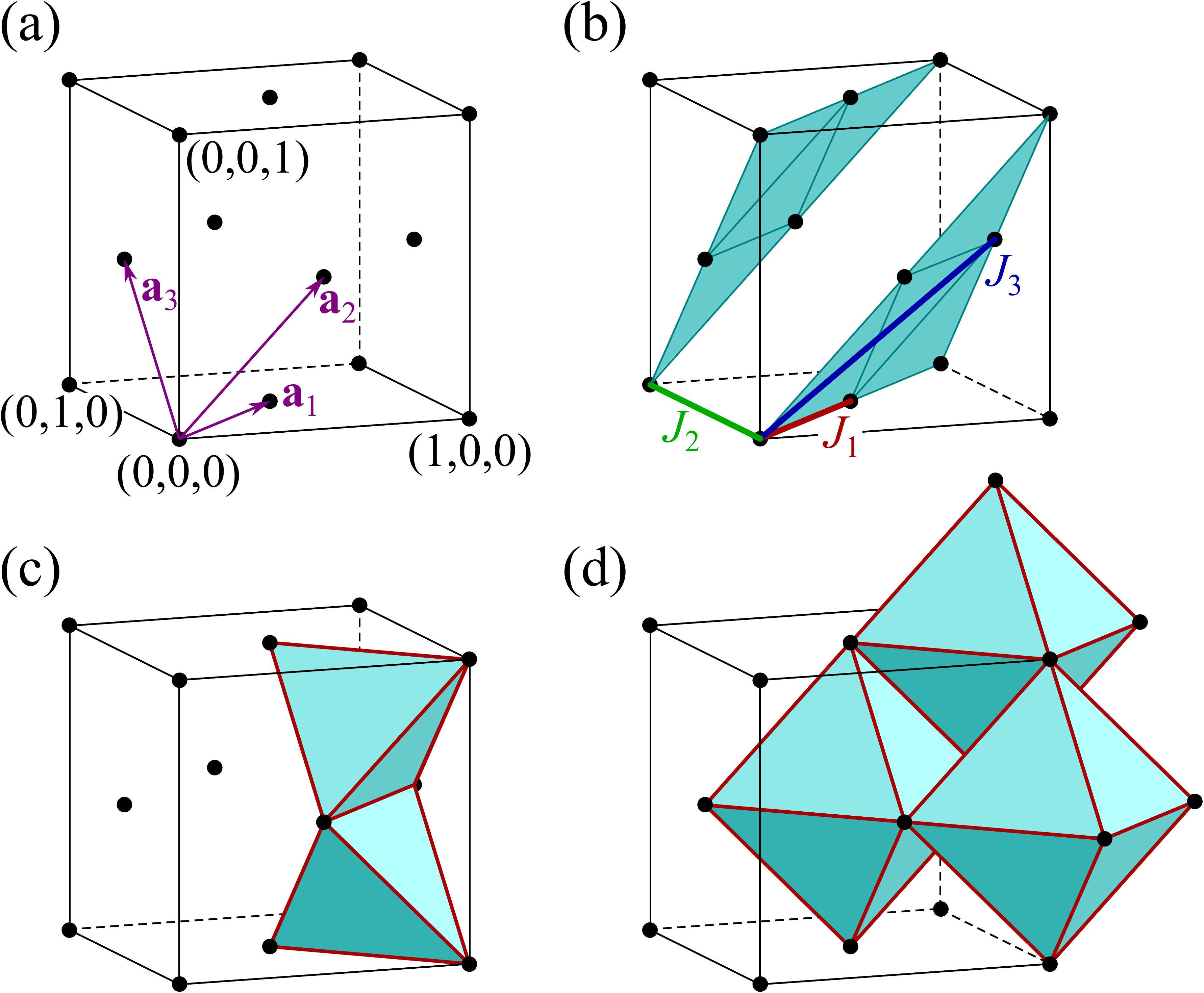}
	\caption{Geometry and exchange interactions of the face-centered-cubic lattice. The enclosing cube is the conventional cell with edge length $a$. (a) Primitive lattice vectors $\mathbf{a}_1=\frac{1}{2}\left(1, 1, 0\right)$, $\mathbf{a}_2=\frac{1}{2}\left(1, 0, 1\right)$, $\mathbf{a}_3=\frac{1}{2}\left(0, 1,1\right)$ connect the first neighbors. (b) The first neighbor $J_1$, second neighbor $J_2$, and third neighbor $J_3$ interactions on the lattice [c.f. the Hamiltonian Eq.~(\ref{Hamiltonian})]. There are 12 first, 6 second and 24 third neighbors, respectively.  $\left(1\overline{1} \overline{1}\right)$ planes are indicated by a light cyan color as a guide to the eye: interactions $J_1$ and $J_3$ are first and second neighbor interactions of the triangular lattices formed by these planes, and $J_2$ connects the planes. The fcc lattice can be covered by edge sharing elementary tetrahedra (c) or  by edge sharing octahedra (d).}
	\label{fig:exchanges}
\end{figure}

The face-centered-cubic (fcc) lattice is an archetypal \textit{frustrated lattice}, it can be built from  $(111)$  triangular  planes, in an $ABCABC$ type stacking style, see Fig.~\ref{fig:exchanges}(a), Appendix \ref{sec:lattice_conventions} and Fig.~\ref{fig:exchanges}(b). A triangular lattice is frequently  called hexagonal to emphasize its sixfold symmetry. This frustration is even more emphasized if we build the lattice from edge sharing tetrahedra, the two differently oriented tetrahedral building blocks are depicted in Fig.~\ref{fig:exchanges}(c). Another way of constructing the lattice is an edge sharing octahedral covering, for a picture of the octahedral building blocks see Fig.~\ref{fig:exchanges}(d).

The Hamiltonian of the classical isotropic Heisenberg model
\begin{equation}
\mathcal{H}=J_1\sum_{\left\langle i,j\right\rangle_1}\mathbf{S}_i\cdot\mathbf{S}_{j}+J_2\sum_{\left\langle i,j\right\rangle_2}\mathbf{S}_i\cdot\mathbf{S}_{j}+J_3\sum_{\left\langle i,j\right\rangle_3}\mathbf{S}_i\cdot\mathbf{S}_{j},\label{Hamiltonian}
\end{equation}
describes three dimensional unit vectors $\left|\mathbf{S}_i\right|=1$  at the sites $\mathbf{R}_i$ of the fcc lattice, interacting with first ($J_1$), second ($J_2$), and third neighbor ($J_3$) interactions. The summation indices $\left\langle i,j\right\rangle_\delta$ with $\delta=1,2,3$  refer to the $\delta$'th neighbor pairs. There are twelve first, six second and twenty-four third neighbor vectors in the fcc structure. One vector of each neighbor set is presented in Fig.~\ref{fig:exchanges}(b).


We wish to find the \textit{ground states} of the model Eq.~(\ref{Hamiltonian}) using the method developed by Luttinger and Tisza~\cite{luttinger_tisza,*luttinger_tisza1}, i.e. by finding the minimum of the exchange interaction in Fourier space $J(\mathbf{q})$~\cite{Bertaut-1961}. We define the  energy in reciprocal space as
\begin{equation}
\mathcal{H} = N\sum_{\mathbf{q}\in\textrm{BZ}}
J(\mathbf{q})\mathbf{S}_{\mathbf{q}} \cdot \mathbf{S}_{-\mathbf{q}}\;,
\label{eq:Hamilton_FT}\\
\end{equation}
where the summation runs over the Brillouin-zone (BZ),  $N$ is the total number of sites of the lattice with periodic boundary conditions,  and 
\begin{equation}
J(\mathbf{q}) = \frac{1}{2}\sum_{\bm{\delta}} J_{\delta} e^{\imath\mathbf{q}\cdot\bm{\delta}}  \;,
\label{eq:jq_def}
\end{equation}
is the Fourier transform of the exchange interactions with lattice separation vectors $\bm{\delta}=\mathbf{R}_i-\mathbf{R}_j$, which is presented in Eq.~(\ref{ft:concrete}).
We used the following convention for the Fourier transform of the spins:
\begin{subequations}
\begin{eqnarray}
\mathbf{S}_{\mathbf{q}}&=&\frac{1}{N}\sum_{i} \mathbf{S}_{i} e^{\imath\mathbf{q}\cdot\mathbf{R}_i},\label{eq:sq_defa}\\ \mathbf{S}_{i}&=&\sum_{\mathbf{q}\in\textrm{BZ}} \mathbf{S}_{\mathbf{q}} e^{-\imath\mathbf{q}\cdot\mathbf{R}_i}.
\label{eq:sq_defb}
\end{eqnarray}
\end{subequations}

In the Luttinger-Tisza method we minimize $J\left(\mathbf{q}\right)$ with respect to  $\mathbf{q}$, i.e. we solve the gradient equation
\begin{equation}
\frac{\partial J\left(\mathbf{q}\right)}{\partial q_\alpha}=0,
\label{eq:jq_gradient}
\end{equation}
for  $\alpha=x, \ y, \ z$. If the equation is satisfied at a point $\mathbf{q}=\mathbf{Q}$ we check the positive semidefiniteness of the Hessian
\begin{equation}
\left.\frac{ \partial^2 J (\mathbf{q})}{\partial q_\alpha \partial q_\beta}\right|_{\mathbf{q}=\mathbf{Q}}\geqslant0.
\label{eq:Hessian}
\end{equation}
The conditions above are necessary, but not sufficient to have a global minimum: to find  a true ground state we have to compare the  different local minima and choose the lowest one.
\begin{table*}
	\caption{Symmetry points and lines with their labels (first column) in the Brillouin zone of the face centered cubic lattice. Second column: number of arms of the star of the point or line (degeneracy). The corresponding energy per site $\varepsilon\left(\mathbf{Q}\right)$ is given in the third column. The fourth column gives the stability region of the commensurate phases, i.e. exchange parameter regions where the Hessian [Eq.~\eqref{eq:Hessian}] is positive definite. The last column gives the conventional names of the commensurate antiferromagnetic phases. Compare this table with the phase diagram given in Fig.~\ref{fig:phase_diagram_descartes}. For pictures of wave-vectors in the Brillouin zone see Fig.~\ref{fig:brillzone}. About notation: we refer to points in the Brillouin zone either by their names and coordinates in units of $2\pi$ or their respective wave-vector, i.e. $ W(1,\onehalf,0)\equiv(2\pi,\pi,0) =\mathbf{Q}_W $.}
	\begin{center}
		\begin{ruledtabular}
			\begin{tabular}{lrccl}
				Label$\left(\mathbf{Q}\right)$ & \# arms & $\varepsilon\left(\mathbf{Q}\right) $ & Local stability & Type \\
				\hline
				$\Gamma\left(0,0,0\right)$ & 1 & $6 J_1 + 3 J_2 + 12 J_3$ & $ J_1 < - J_2 \!-\! 6  J_3 $ & -- \\
				$X\left(1,0,0\right)$ & 3 & $-2 J_1+3 J_2-4 J_3$ & $ J_2 < 4 J_3 < 2 J_1 \!-\! 2 J_2 $ & Type I\\
				$L\left(\frac{1}{2},\frac{1}{2},\frac{1}{2}\right)$ & 4 & $ -3 J_2$ & $ J_1 \!-\! 2 J_2 < 2 J_3  < J_1 \!+\! J_2 $ & Type II \\
				$W\left(1,\onehalf,0\right)$ & 6 & $-2 J_1 +  J_2 + 4 J_3$ & $ 8 J_3 < 2 J_2 < J_1 \!-\! 2  J_3$ & Type III\\
				$\Delta\left(q,0,0\right)$ & 6 & $\frac{2 -2 J_1^2 + 2 J_1 J_2 + J_2^2 - 
					24 J_3^2}{2J_2 + 8 J_3}$ \\
				$\Lambda\left(q,q,q\right)$ & 8 & $-\frac{3 \left(J_1^2+2 J_1 \left(J_2-2 J_3\right)+\left(J_2+2 J_3\right)^2\right)}{8 J_3}$ \\
				$\Sigma\left(q,q,0\right)$ & 12 & $\varepsilon\left(\mathbf{Q}_\Sigma\right)$\footnote{
						The $\varepsilon(\mathbf{Q}_\Sigma=(q_\Sigma,q_\Sigma,0))$ Fourier transform is:
						\begin{eqnarray}
						\varepsilon\left(\mathbf{Q}_\Sigma\right) &=& 
						\frac{J_1^3+6 J_1^2 J_2-66 J_1^2 J_3+12 J_1 J_2^2-120 J_1 J_2 J_3+12 J_1 J_3^2+8 J_2^3+24 J_2^2 J_3-120 J_2 J_3^2+296 J_3^3}{432 J_3^2}
						\nonumber \\ 
						&\phantom{=}& +\frac{\left(-J_1^2-4 J_1 J_2+44 J_1 J_3-4 J_2^2-8 J_2 J_3-100 J_3^2\right)\sqrt{(J_1+2 (J_2+J_3))^2-48 J_3 (J_1-2 J_3)}}{432 J_3^2}.\nonumber
						\end{eqnarray}	
				}
			\end{tabular}
		\end{ruledtabular}
	\end{center}
	\label{tab:sympoints}
\end{table*}
We denote the set  of points $\left\{\mathbf{Q}\right\}$ --the \textit{ordering vectors}-- where $J(\mathbf{q})$ takes its minimal value  by $\mathcal{M}_{\textrm{GS}}$,  and call it as the \textit{ground state manifold}, and choose the Fourier amplitude vectors $\mathbf{S}_{\mathbf{Q}}$-s to satisfy the spin-length constraint $\left|\mathbf{S}_i\right|=1$ for every site. For a given $\mathcal{M}_{\textrm{GS}}$ the amplitudes have to be chosen carefully to satisfy the local constraints (the Luttinger-Tisza method only guaranties the fulfilment of the global constraint $\sum_{i}\left|\mathbf{S}_{i}\right|^2=N$). We give a detailed analysis of the amplitudes and the corresponding orderings in real space in Sec.~\ref{sec:realfourier_chapter}.

When the local constraints are satisfied, the ground state energy per site $\varepsilon (\mathbf{Q})$ equals to the Fourier transform of the exchange constant evaluated  on the $\mathcal{M}_{\textrm{GS}}$. To show this we evaluate Eq.~(\ref{eq:Hamilton_FT}) on the $\mathcal{M}_{\textrm{GS}}$: 
\begin{eqnarray}
	\varepsilon(\mathbf{Q})=\frac{\left\langle\mathcal{H}\right\rangle_0}{N}=\sum_{\mathbf{Q}\in\mathcal{M}_{\textrm{GS}}}J\left(\mathbf{Q}\right)\left|\mathbf{S}_{\mathbf{Q}}^0\right|^2=J\left(\mathbf{Q}\right), 
	\label{ground_state_energy}
\end{eqnarray}
where the $0$ indices refer to the ground state properties, note that $\varepsilon (\mathbf{Q})$ only depends on the ordering vector parametrically. We use the reality of the spin components $\mathbf{S}_{-\mathbf{Q}}^0=\mathbf{S}_{\mathbf{Q}}^{0*}$, $J\left(\mathbf{Q}\right)$ being constant on $\mathcal{M}_{\textrm{GS}}$, and employ the Fourier form of the local spin length constraint $1= \sum_{\mathbf{q}\in \text{BZ}}\left|\mathbf{S}_{\mathbf{q}}\right|^2 = \sum_{\mathbf{Q}\in\mathcal{M}_{\textrm{GS}}}\left|\mathbf{S}_{\mathbf{Q}}\right|^2$ (since $\mathbf{S}_{\mathbf{q}} = 0$ for $\mathbf{q}\notin \mathcal{M}_{\textrm{GS}}$ in the ground state).

The \textit{dimension} of the $\gsm$ is a crucial ingredient in understanding the physics of these systems. Conventional commensurate ferromagnets or antiferromagnets correspond to zero dimensional manifolds, i.e. a handful of points in highly symmetric positions of the BZ. Incommensurate orders (spin spirals, helices, cycloids)~\cite{Dzyaloshinskii-1964a,Dzyaloshinskii-1964b,Khomskii} also have zero dimensional $\gsm$, but now at generic points in the BZ. Magnetic Bragg peaks show up at these points in neutron scattering. But in frustrated systems the possibilities are richer: the $\gsm$ can be a one-dimensional degenerate manifold as for the $J_{1}$-$J_{2}$ model on the square~\cite{ChandraJ1-J2_1988,Ioffe-1988} and honeycomb lattices~\cite{Rastelli-1979,Fouet-2001,ganesh} and the $J_{1}$ only model on the fcc lattice~\cite{haar_lines,alexander1980} where every point on a line is a possible ordering vector, and with carefully chosen amplitudes we can compose ground states with complicated spatial variation. The situation can be even more complex when the dimension of the $\gsm$ is larger: spin spiral surfaces (i.e. two-dimensional $\gsm$-s) were found in $J_{1}$-$J_{2}$ Heisenberg model on the diamond~\cite{bergmann07}, fcc~\cite{finnish_93,ourspiral},  body-centered cubic~\cite{Attig2017}, and hexagonal close packed~\cite{Niggemann_2019} lattices. Furthermore, on the kagome~\cite{Chalker-1992,Zhitomirsky-2008} and pyrochlore~\cite{Reimers-1991,Reimers-1992,Iqbal-2019} lattices the whole BZ is the $\gsm$. These extended manifolds give an opportunity to the system to fluctuate between the degenerate ground states making them candidates for classical spin liquids in some temperature range~\cite{Moessner-1998a,Moessner-1998b,bergmann07,gao16,Iqbal-2018}.

There is another type of degeneracy, even in unfrustrated spin models without extended $\gsm$-s (i.e. simple cubic ferro- or antiferromagnets of first neighbor couplings) where the interactions are isotropic: breaking the global $O(3)$ rotational symmetry of the model results a family of ground states that can be rotated globally to each other in spin space, resulting a degeneracy parametrized by the three dimensional group of rotations (we will refer to this type of degeneracy as trivial, since its presence is independent of frustration).  Besides the trivial degeneracy, ground states having multiple sublattices can still be indeterminate: we can continuously deform them to each other by a set of  local rotations~\cite{villain1980,henley_fcc,henley1989}.  This leftover degeneracy can again be characterized by a (continuous or discrete) set of parameters, and we will give a detailed analysis of this scenario in cases of commensurate orders in our model. In the following we construct the ground state phase diagram of the model in exchange parameter space by minimizing the energy $\varepsilon(\mathbf{q})$ with respect to the wave-vector.

\section{Classical phase diagram and ordering vectors}
\label{sec:phase_diagram}

\begin{figure}
  \centering
  \includegraphics[width=.95\columnwidth]{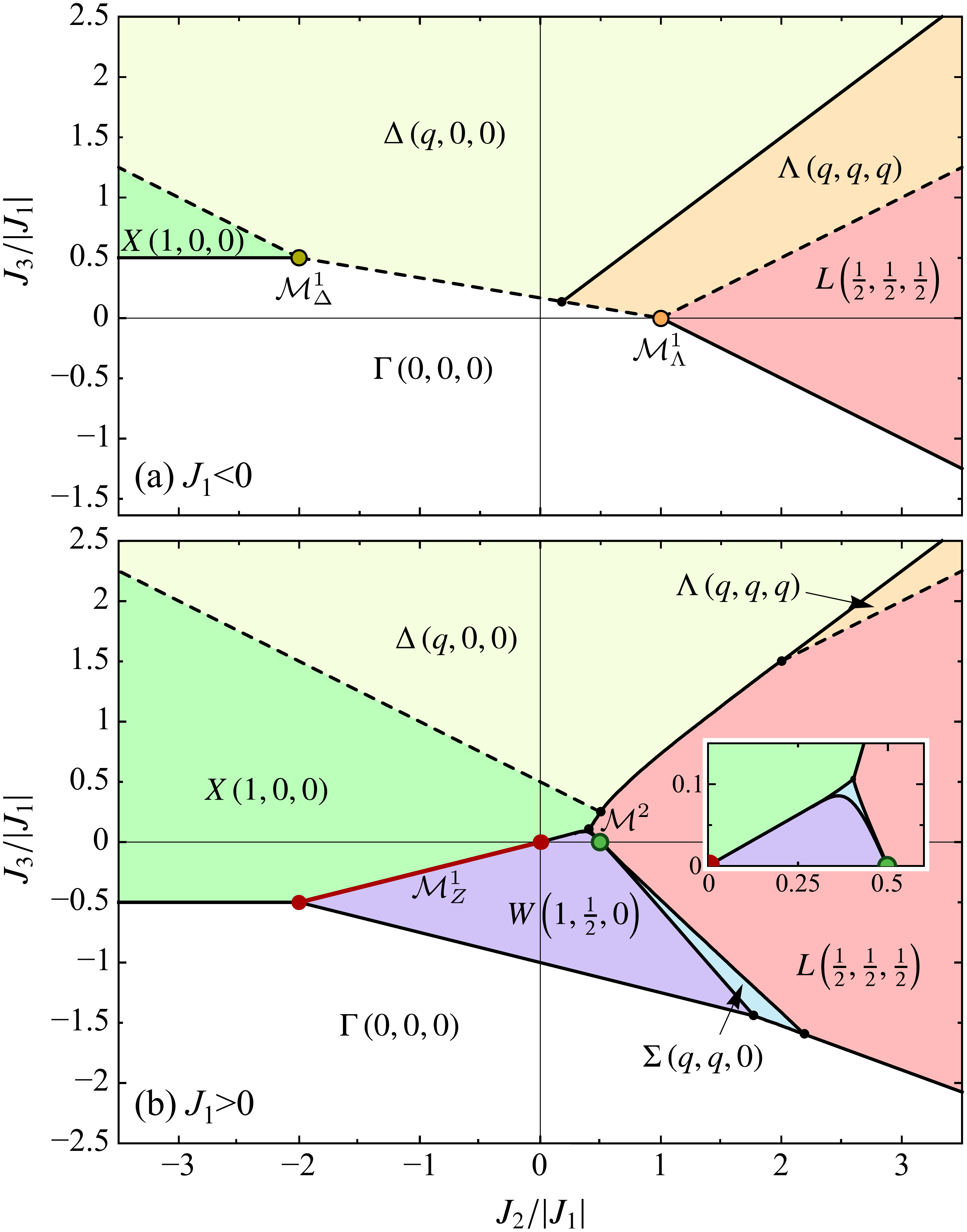}
  \caption{Phase diagram of the classical $J_1-J_2-J_3$ Heisenberg model on the face-centered-cubic lattice for (a) ferromagnetic $J_1<0$ and (b) antiferromagnetic $J_1>0$ first neighbor interactions. Basic information about the phases is collected in Table~\ref{tab:sympoints}. 
Solid  black lines mark first order phase transitions, dashed black lines stand for continuous (second order) transitions, and the equations describing these boundaries are collected in Table~\ref{tab:phase_boundaries}. 
We label the phases by their ordering vectors given in units of $2\pi/a$, as presented in Fig.~\ref{fig:brillzone}.
The four \textit{commensurate} phases  are the ferromagnet $\Gamma(0,0,0)$, and the three types of antiferromagnets: $X(1,0,0)$, $L\left(\frac{1}{2},\frac{1}{2},\frac{1}{2}\right)$ and $W\left(1,\frac{1}{2},0\right)$. 
The commensurate ordering vectors are depicted in Fig.~\ref{fig:brillzone}(a), note that all these phases are already present in the absence of $J_3$~\cite{henley_fcc,yamamoto1972}.  
Introducing a finite $J_3$ introduces the \textit{incommensurate} phases $\Delta (q,0,0)$, $\Lambda (q,q,q)$ and $\Sigma (q,q,0)$, where $q$ has to be optimized according to Eqs.~(\ref{eq:optimizedqs1})-(\ref{eq:optimizedqs3}), and the possible ordering vectors are depicted in Fig.~\ref{fig:brillzone}(d)-(f). 
The phase $\Sigma (q,q,0)$ in (b) has a bow-tie shape (enlarged in the inset) with a neck consisting of a single point $J_2=J_1/2$ and $J_3=0$ (the green dot), and through this point the $L\left(\frac{1}{2},\frac{1}{2},\frac{1}{2}\right)$ and $W\left(1,\frac{1}{2},0\right)$ phases meet. 
The dark red $X-W$ phase boundary emanating from the first-neighbor antiferromagnetic point $J_2=J_3=0$ is degenerate: along this line \textit{any} of the ground states with ordering vectors residing on the one-dimensional manifold $\mathcal{M}_Z^1$ defined by $\mathbf{Q}=(2\pi,q,0)$ with $q\in [-\pi,\pi]$ shown in Fig.~\ref{fig:brillzone}(b) have the same energy. In (a) the two triple points also have one-dimensional ground state manifolds: $\mathcal{M}_\Delta^1$ [Fig.~\ref{fig:brillzone}(d)] and $\mathcal{M}_\Lambda^1$ [Fig.~\ref{fig:brillzone}(e)]. The green dot at $J_3=0$, $J_2=J_1/2>0$ is a phase of even higher degeneracy: the ground states form the two-dimensional manifold shown in Fig.~\ref{fig:brillzone}(c). Basic properties of these manifolds are collected in Table~\ref{tab:triplepoints}.}
  \label{fig:phase_diagram_descartes}
\end{figure}

In this section we construct the classical, zero temperature, ground state phase diagram   of the model Eq.~(\ref{Hamiltonian}) in the $J_1$\textendash$J_2$\textendash$J_3$ parameter space.  We compare the $\varepsilon(\mathbf{Q})$ values for the possible orderings and choose the lowest one for a given set of parameters, these results are collected in Table~\ref{tab:sympoints}, and we present the detailed phase diagram in  Fig.~\ref{fig:phase_diagram_descartes}. 

We analyze the phase boundaries by comparing the ground state energies of the neighboring phases summarized in Table~\ref{tab:phase_boundaries}. The phase boundaries are of second order if the ordering vectors of the two matching phases can be deformed continuously into each other, and of first order if the transition requires a discontinuous jump of the ordering vector. There are special points of the phase diagram: the triple points and the $X(1,0,0)-W(1,\onehalf,0)$ phase boundary that require particular attention: at these points ground state manifolds extend to lines and a surface, signaling a large but subextensive degeneracy of the ground states. A sidenote about notation: we use two sets of notation for the BZ points: we refer to points in the BZ either by their names and  coordinates in units of $2\pi$ or their respective wave-vector, i.e., $\mathbf{Q}_W=(2\pi,\pi,0)=W(1,\onehalf,0)$. Basic information about these phases is collected in Table~\ref{tab:triplepoints}.  In the following we briefly summarize the properties of the phase diagram, but details of the real space picture of the orders are given in Sec.~\ref{sec:realfourier_chapter}.

\begin{table*}
	\caption{Multiple points of the phase diagram, corresponding to degenerate manifolds in $\mathbf{q}$-space. In the first column we give the label of the manifold, see Fig.~\ref{fig:phase_diagram_descartes}. The manifolds themselves are depicted in  Figs.~\ref{fig:brillzone}(b)-(e). In the second column we list phases that meet at the special parameter values given in the third column. In the fourth column the dimension of the manifolds is given, together with their defining equation in $\mathbf{q}$-space, when given in parametric form we  mention only one of the crystallographically equivalent directions. The last column gives the energy per site on the manifolds.}
  \begin{center}
    \begin{ruledtabular}
	  \begin{tabular}{lllcll}
			Label & Touching phases & Constraints on $J$-s  & $\textrm{dim} \ \mathcal{M}_{\textrm{GS}}$ & Definition of $\mathcal{M}_{\textrm{GS}}$ & $\varepsilon(\mathbf{Q})$ \\ \hline
				$\mathcal{M}^1_{Z}\cup \Gamma$ & $\Gamma-X-W$ & $J_2=-2J_1$, $J_3=-\frac{J_1}{2}$, $J_1>0$ & 1 & $\mathbf{Q}=(2\pi,q,0)$ & $-6J_1$ \\
				$\mathcal{M}^1_{Z}$& $X-W$ & $J_3=\frac{J_2}{4}$, $-2J_1<J_2<0$, $J_1>0$ & 1 & $\mathbf{Q}=(2\pi,q,0)$ & $2\left(J_2-J_1\right)$ \\
				$\mathcal{M}^1_{\Delta}$& $\Gamma-\Delta-X$ & $J_2=2J_1$, $J_3=-\frac{J_1}{2}$, $J_1<0$ & 1 & $\mathbf{Q}=(q,0,0)$ & $+6J_1$ \\
				$\mathcal{M}^1_{\Lambda}$& $\Gamma-\Lambda-L$ & $J_2=-J_1$, $J_3=0$, $J_1<0$ & 1 & $\mathbf{Q}=(q,q,q)$ & $+3J_1$\\
				$\mathcal{M}^2$& $L-W-\Sigma$ & $J_2=\frac{J_1}{2}$, $J_3=0$, $J_1>0$ & 2 & $\cos \frac{Q_x}{2}+\cos \frac{Q_y}{2}+\cos\frac{Q_z}{2}=0$ & $-\frac{3}{2}J_1$ 
	  \end{tabular}
	\end{ruledtabular}
  \end{center}
  \label{tab:triplepoints}
\end{table*}

\begin{figure*}
	\centering
	\includegraphics[width=1.75\columnwidth]{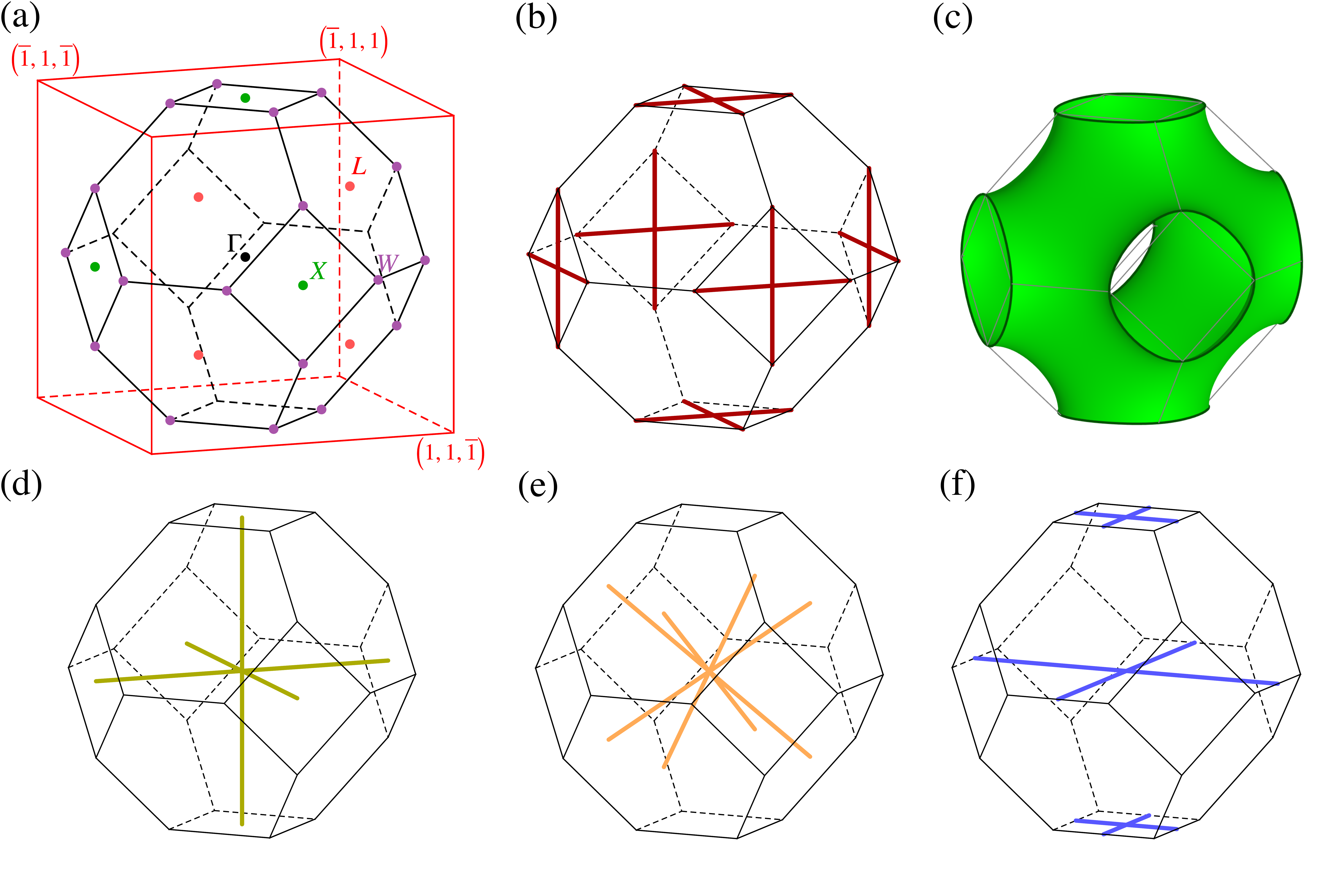}
	\caption{Brillouin zone (truncated octahedron) of the face-centered-cubic lattice together with ordering vectors and ground state manifolds corresponding to the phases in Fig.~\ref{fig:phase_diagram_descartes}. 
	(a) Commensurate ordering vectors, and the enclosing red cube as a guide to the eye. $\Gamma(0,0,0)$ is the ferromagnet (this wavevector's star has only one arm),  $X\left(1,0,0\right)$ is the Type-I antiferromagnet (three-armed star of the wave-vector), $L\left(\onehalf,\onehalf,\onehalf\right)$  is the Type-II antiferromagnet (four-armed star of the wave-vector) and $ W\left(1,\onehalf,0\right)$ is the Type-III antiferromagnet (six-armed star of the wavevector).  (b) Degenerate wave-vectors ($Z$-lines) on the boundary of the $X(1,0,0)-W\left(1,\frac{1}{2},0\right)$ phases forming a one-dimensional manifold $\mathcal{M}_Z^1$ for $J_1>0$ and $J_3=J_2/4$ , c.f. the dark red line in Fig.~\ref{fig:phase_diagram_descartes}(b). \textit{Every} point on the crisscrosses is energetically degenerate. (c) Two-dimensional energetically degenerate manifold $\mathcal{M}^2$ (a Schwarz P surface~\cite{bubble_surfaces, ourspiral}) corresponding to $J_3=0$, $J_2=J_1/2>0$ (green dot in Fig.~\ref{fig:phase_diagram_descartes}(b)). The second row ((d)--(f)) corresponds to incommensurate orderings, in these figures --depending on the exchange parameter values (c.f. Eq.~(\ref{eq:optimizedqs1})-(\ref{eq:optimizedqs3}))-- a single $\pm \mathbf{Q}$ pair of wave-vectors is chosen as the ordering vector of the developing spin spiral. This row also depicts the one dimensional degenerate manifolds of the special points of the phase diagram.   (d) Incommensurate ordering vectors   $\Delta (q,0,0)$ corresponding to spin spirals propagating in the directions of the cubic axes with 6 arms. This is also the manifold $\mathcal{M}_\Delta^1$ of the $\Gamma-\Delta-X$ triple point, see the yellow dot in Fig.~\ref{fig:phase_diagram_descartes}(a). (e) Incommensurate ordering vectors   $\Lambda (q,q,q)$  corresponding to spin spirals propagating in the directions of the body diagonals of the cubic cell with 8 arms. This is also the manifold $\mathcal{M}_\Lambda^1$ of the $\Gamma-\Lambda-L$ triple point, see the orange dot in Fig.~\ref{fig:phase_diagram_descartes}(a). (f) Incommensurate ordering vectors $\Sigma (q,q,0)$ corresponding to spin spirals propagating in the directions of the face diagonals of the cubic cell (only pictured in the horizontal planes for better visibility, this vector has 12 arms). About notation: we refer to points in the Brillouin zone either by their names and coordinates in units of $2\pi$ (fractional coordinates) or their respective wave-vector, e.g., $W(1,\onehalf,0)\equiv \mathbf{Q}_W=(2\pi,\pi,0)$.
}\label{fig:brillzone}
\end{figure*}

Basically, we found three types of phases:
\begin{enumerate}[label=(\roman*)]
\item Four commensurate spin configurations with ordering vectors at the high symmetry points in the BZ~\cite{alexander1980,henley_fcc,yamamoto1972,Villain-1959} ---  the usual ferromagnet, with ordering vector $\Gamma (0,0,0)$, and three kinds of antiferromagnetic orders: with ordering vectors $X(1,0,0)$, $L\left(\onehalf,\onehalf,\onehalf\right)$ and  $W\left(1,\onehalf,0\right)$ [see Fig.~\ref{fig:brillzone}(a)], which are already present in the $J_1-J_2$ models. These are discussed in Sec.~\ref{sec:realfourier_chapter}.
\item Three types of incommensurate spin spirals (helices~\cite{Yoshimori-1959,Dzyaloshinskii-1964a,Dzyaloshinskii-1964b,Khomskii,Uchida-2006}, cycloids) caused by the frustrating effect of $J_3$, with a pitch vector of length fixed by the exchange parameter values and pointing in highly symmetric crystallographic directions, with incommensurate ordering vectors $\Delta(q,0,0)$, $\Lambda\left(q,q,q\right)$  and $\Sigma\left(q,q,0\right)$ [see Fig.~\ref{fig:brillzone}(d)--(f)]. (Sec.~\ref{sec:incommensurate}) 
\item  Four phases with large ground state degeneracy. Three of these phases have one dimensional $\gsm$-s: one is at the $X(1,0,0)-W\left(1,\onehalf,0\right)$ phase boundary, that extends from the first neighbor antiferromagnetic model~\cite{alexander1980,henley_fcc}. The degenerate manifold corresponding to  this phase boundary is called $\mathcal{M}^1_Z$, this manifold is depicted in Fig.~\ref{fig:brillzone}(b): the lines are connecting the $X(1,0,0)$ and $W\left(1,\onehalf,0\right)$ points of the BZ, these lines are sometimes called ``$Z$", hence the name of the manifold, and the upper index "$1$" refers to its dimensionality. The two other one dimensional degenerate manifolds are at the triple points $\Gamma(0,0,0)-\Delta(q,0,0)-X(1,0,0)$ (its $\gsm=\mathcal{M}^1_\Delta$ coincides with the collection of $\Delta(q,0,0)$ depicted in Fig.~\ref{fig:brillzone}(d)) and $\Gamma(0,0,0)-\Lambda(q,q,q)-L\left(\onehalf,\onehalf,\onehalf\right)$ (its $\gsm=\mathcal{M}^1_\Lambda$ coincides with the collection of $\Lambda(q,q,q)$ depicted in Fig.~\ref{fig:brillzone}(e)). There is also a phase with a two dimensional $\gsm=\mathcal{M}^2$~\cite{ourspiral,finnish_93,Dominik} at the triple point $L\left(\onehalf,\onehalf,\onehalf\right)-\Sigma\left(q,q,0\right)-W\left(1,\onehalf,0\right)$: this surface is depicted in Fig.~\ref{fig:brillzone}(c). These degenerate phases can be found at carefully chosen parameter values where other more conventional phases meet (Sec.~\ref{sec:manifolds}).
\end{enumerate}
In the following sections we give a detailed analysis of the possible configurations.

\section{Commensurate orderings and real space description}
\label{sec:realfourier_chapter}
%
In this section we describe and analyze the developing orders in detail. For the four commensurate orderings (c.f. Fig.~\ref{fig:brillzone}(a)) we calculate the Fourier amplitudes, give the constraints on them, count the degeneracies (the number of free parameters describing the order that remain after removing the trivial, global $O\left(3\right)$ of the symmetry breaking).  We describe  the orders in real space and  analyze their symmetry properties. In order to find the number of wave-vectors participating in a given order and to find the constraints on their Fourier amplitudes  we must not distinguish between the equivalent $\mathbf{q}$-vectors, where by equivalence we mean differing only in some reciprocal lattice vector $\mathbf{G}$, i.e., $\mathbf{q}\sim\mathbf{q}'$ if $\mathbf{q}=\mathbf{q}'+\mathbf{G}$. Constraints on the Fourier amplitudes and the number of free parameters can be calculated as follows: we expand the spins $\mathbf{S}_i$ in Fourier space, keeping only the amplitudes of the arms of the star of the respective ordering vector finite. Afterwards we impose the constraints that for every lattice point the spins have to be real unit vectors, and solve the equations~\cite{villain_rules,yamamoto1972,Nussinov2001} for the Fourier amplitudes: 
\begin{subequations}
\begin{eqnarray}
&&\mathbf{S}_{\mathbf{Q}}^0=\mathbf{S}_{-\mathbf{Q}}^{0*},\label{villaina}\\
&&\sum_{\mathbf{Q}}\left|\mathbf{S}_{\mathbf{Q}}^0\right|^2=1,\label{cmplxlength}\label{villainb}\\
&&\sum_{\mathbf{Q}}\mathbf{S}_{\mathbf{Q}}^0\cdot\mathbf{S}_{\mathbf{Q}-\mathbf{q}'}^{0*}=0, \ \forall~\mathbf{q}'\neq 0.
\label{villainc}
\end{eqnarray}
\end{subequations}
In order to make the last equation  useful one has to choose $\mathbf{q}'$ such a way that $\mathbf{Q}-\mathbf{q}'$ lies on the $\gsm$.

Eq.~(\ref{villaina}) is a consequence of the reality of the real space spins $\mathbf{S}_i$. The second set of equations Eqs.~(\ref{villainb}--\ref{villainc}) can be calculated as follows: we substitute the Fourier decomposition Eq.~(\ref{eq:sq_defb})  in the \textit{local} length constraints $|\mathbf{S}_i|^2=1$:
\begin{equation}
\sum_{\mathbf{q},\mathbf{q}''}\mathbf{S}_{\mathbf{q}}\cdot\mathbf{S}_{\mathbf{q}''} \ e^{-\imath (\mathbf{q}+\mathbf{q}'') \cdot \mathbf{R}_i}=|\mathbf{S}_i|^2=1, \ \forall i,
\end{equation}
this is a set of $N$ equations.
We perform a Fourier transform by multiplying the $i$-th equation by $e^{\imath \mathbf{q}' \cdot \mathbf{R}_i}$ and  sum  over $i$:
\begin{equation}
	\sum_{i}\sum_{\mathbf{q},\mathbf{q}''}\mathbf{S}_{\mathbf{q}}\cdot\mathbf{S}_{\mathbf{q}''} \ e^{-\imath (\mathbf{q}+\mathbf{q}''-\mathbf{q}') \cdot \mathbf{R}_i}  =\sum_{i}e^{\imath \mathbf{q}' \cdot \mathbf{R}_i}. 
\end{equation}
Performing the sums yields
\begin{equation}
	\sum_{\mathbf{q}}\mathbf{S}_{\mathbf{q}}\cdot\mathbf{S}_{\mathbf{q}'-\mathbf{q}}=\delta_{ \mathbf{q}',\mathbf{0} },
\end{equation}
which is true for every configuration, even  for the ones on  $\gsm$, resulting Eqs.~(\ref{villainb}--\ref{villainc}).

\subsection{The $\Gamma\left(0,0,0\right)$ ferromagnet}
%
This phase is an ordinary ferromagnet, where all the spins align and only the trivial $O(3)$ degeneracy is present.
%
\subsection{The $X\left(1,0,0\right)$ antiferromagnet, Type-I}
In this phase the nonequivalent $\mathbf{Q}$-vectors form a three-armed star:
\begin{equation}
\mathbf{X}_1=\left(2\pi, 0, 0\right), \ \mathbf{X}_2=\left(0,2\pi,0\right),\ \mathbf{X}_3=\left(0,0,2\pi\right).
\end{equation} 
Since these arms reside on the midpoints of the square-shaped faces  of the BZ~\cite{villain_rules}  we can mix them to construct a triple-$\mathbf{Q}$ order, provided we choose the Fourier amplitudes appropriately. In order to make the spins real unit vectors  we need to consider some constraints on the complex amplitudes $\mathbf{S}_{\mathbf{X}_\alpha}^0$:
\begin{equation}
\mathbf{S}_i=\sum_{\alpha=1}^3\mathbf{S}_{\mathbf{X}_\alpha}^0e^{-\imath \mathbf{X}_\alpha \cdot \mathbf{R}_i}.
\end{equation}
Since $\mathbf{X}_\alpha$ and $-\mathbf{X}_\alpha$ are equivalent ($\mathbf{X}_\alpha\sim -\mathbf{X}_\alpha$) and the phase factors $e^{-\imath \mathbf{X}_\alpha \cdot \mathbf{R}_i}$ are simply $\pm 1$-s, the amplitudes have to be real to ensure the reality of $\mathbf{S}_i$:
\begin{equation}
\mathbf{S}_{\mathbf{X}_\alpha}^{0} \in \mathbb{R}^3, \ \forall \alpha.
\end{equation}
The following four constraints (see Eqs.~(\ref{villainb}--\ref{villainc})) fix the lengths of the spins:
\begin{align}
&\sum_{\alpha=1}^3\left|\mathbf{S}_{\mathbf{X}_\alpha}^{0}\right|^2=1  && \textrm{(1 constraint)},\\
&\mathbf{S}_{\mathbf{X}_\alpha}^{0}\cdot\mathbf{S}_{\mathbf{X}_\beta}^{0}=0, \ \forall \alpha\neq\beta  && \textrm{(3 constraints)},
\label{Xconstraints}
\end{align}
where the last equation follows if we chose $\mathbf{q}'=\mathbf{X}_\alpha-\mathbf{X}_\beta$ in Eq.~(\ref{villainc}).
The three real amplitudes $\mathbf{S}_{\mathbf{X}_\alpha}^0$ mean 9 free parameters. The global $O(3)$ freedom removes 3 of them, and together with the four constraints in Eq.~(\ref{Xconstraints}) we are left with  two free parameters to characterize the degeneracy~\cite{henley_fcc}.

The global $O(3)$ freedom of the symmetry breaking and the mutual orthogonality and normalization of the $\mathbf{S}^0_{\mathbf{X}_\alpha}$-s allows us to parametrize them as:
\begin{eqnarray}
\left(\mathbf{S}^0_{\mathbf{X}_1}|\mathbf{S}^0_{\mathbf{X}_2}|\mathbf{S}^0_{\mathbf{X}_3}\right)=\left(\begin{array}{c|c|c}
  \xi & 0 & 0\\ 
  0 & \eta & 0\\
	0 & 0 & \zeta
 \end{array}\right),
 \label{Xparams}
\end{eqnarray}
where all the parameters are real, and they satisfy the additional constraint: $\xi^2+\eta^2+\zeta^2=1$. The ground state manifold thus can be parametrized by a unit vector $(\xi,\eta,\zeta)$.
In the following we construct and analyze the developing order in real space.

With the parametrization of Eq.~(\ref{Xparams}) the spin sitting on the lattice point 
$\mathbf{R}_i=(x,y,z)$ is (we recall that the coordinates can be integers or half-integers):
\begin{equation}
\mathbf{S}_i=\left(
\begin{array}{c}(-1)^{2x}\xi\\ (-1)^{2y}\eta\\ (-1)^{2z}\zeta
\end{array}\right). \label{Xspins}
\end{equation}
The superlattice vectors of this order form a simple cubic lattice with the same unit cell as the conventional cell of the original fcc lattice,  and the  four sublattices form tetrahedra 
with  spins
\begin{figure}
	\centering
	\includegraphics[width=.95\columnwidth]{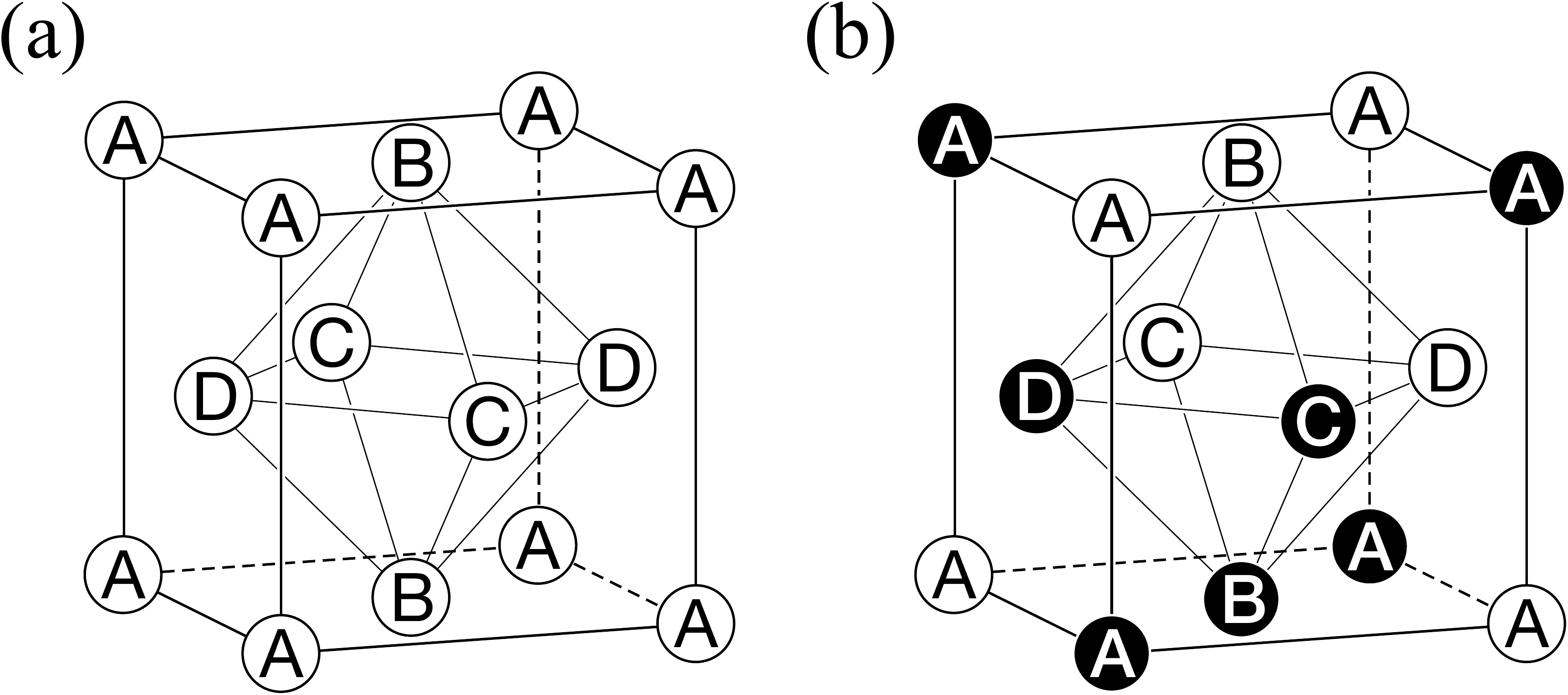}
	\caption{(a) Type-I $X\left(1,0,0\right)$  4-sublattice antiferromagnetic order. The spins on each sublattice are $\mathbf{S}_A$, $\mathbf{S}_B$, $\mathbf{S}_C$ and $\mathbf{S}_D$, repectively. Each sublattice forms a simple cubic lattice with lattice vectors $(1,0,0)$, $(0,1,0)$, and $(0,0,1)$. The  different spins sit on  elementary tetrahedra of the fcc lattice, with the constraint of the total spin of each tetrahedron being $\mathbf{S}_A+\mathbf{S}_B+\mathbf{S}_C+\mathbf{S}_D=\mathbf{0}$. (b) Type-II $L\left(\onehalf,\onehalf,\onehalf\right)$  antiferromagnetic order, with four pairs of antiferromagnetically ordered sublattices: inverted colors correspond to opposite spins, i.e. $\mathbf{S}_{\bar{A}}=-\mathbf{S}_A$ is a white letter on a black disk. 
%
Each sublattice forms an fcc lattice with a doubled lattice constant. 
On each elementary octahedron of the original fcc latice the spins form  antiparallel pairs on opposite vertices of the octahedra.}
	\label{fig:XL_order}
\end{figure}
\begin{eqnarray}
\left(\mathbf{S}_A|\mathbf{S}_B|\mathbf{S}_C|\mathbf{S}_D\right)=\left(\begin{array}{c|c|c|c}
  \xi & -\xi & -\xi & \xi\\ 
  \eta & -\eta & \eta & -\eta\\
	\zeta & \zeta & -\zeta & -\zeta
 \end{array}\right),
 \label{tetraruleX}
\end{eqnarray}
see Fig.~\ref{fig:XL_order}(a). The spins on every elementary tetrahedron  sum up to zero:
\begin{equation}
\mathbf{S}_A+\mathbf{S}_B+\mathbf{S}_C+\mathbf{S}_D=\mathbf{0},
\label{tetrarule_general}
\end{equation}
we refer to this situation as the \textit{tetrahedron rule}. We can check our Fourier space degeneracy counting in real space: the four sublattice spins mean four unit vectors as eight free parameters, and the global $O(3)$ symmetry and also the tetrahedron rule remove three of them, leaving only two free parameters as expected. 

If we use all three arms of the star, creating a triple-$\mathbf{Q}$ order (with $\xi$, $\eta$, $\zeta$ finite) the spins are noncoplanar. For --say-- $\zeta=0$, the configuration is coplanar, and if only $\xi$ remains, it is collinear. Thermal or quantum order by disorder effects select a single arm of the star resulting in a collinear structure~\cite{henley_fcc}.

\subsection{The $ L\left(\onehalf,\onehalf,\onehalf\right)$  antiferromagnet, Type-II}
\label{Lsection}
In this phase the nonequivalent $\mathbf{Q}$-vectors form a four-armed star:
\begin{eqnarray}
&&\mathbf{L}_1=\left(\pi,\pi,\pi\right), \ \mathbf{L}_2=\left(\pi,-\pi,-\pi\right),\nonumber\\
&&\mathbf{L}_3=\left(-\pi,\pi,-\pi\right), \ \mathbf{L}_4=\left(-\pi,-\pi,\pi\right).
\end{eqnarray} 
Since they are on the midpoints of the hexagonal BZ faces we can construct a quadruple-$\mathbf{Q}$ order out of them~\cite{villain_rules}. We expand the spins in Fourier amplitudes:
\begin{equation}
\mathbf{S}_i=\sum_{\alpha=1}^4\mathbf{S}_{\mathbf{L}_\alpha}^0e^{-i \mathbf{L}_\alpha \cdot\mathbf{R}_i}\in \mathbb{R}^3.
\end{equation}
Since $\mathbf{L}_\alpha\sim -\mathbf{L}_\alpha$ and the phase factors are $e^{-i \mathbf{L}_\alpha \cdot\mathbf{R}_i}=\pm 1$, the amplitudes have to be real to ensure the reality of $\mathbf{S}_i$ :
\begin{equation}	
	\mathbf{S}_{\mathbf{L}_\alpha}^{0} \in \mathbb{R}^3, \ \forall \alpha.
\end{equation}
We can express the spins in real space as:
\begin{eqnarray}
\mathbf{S}_i&=&(-1)^{x+y+z}\mathbf{S}_{\mathbf{L}_1}^0+(-1)^{y+z-x}\mathbf{S}_{\mathbf{L}_2}^0
\nonumber\\
&+&(-1)^{x+z-y}\mathbf{S}_{\mathbf{L}_3}^0+(-1)^{x+y-z}\mathbf{S}_{\mathbf{L}_4}^0. 
\label{Lfourq}
\end{eqnarray}
We can substitute the lattice points in the above equation that yield four \textit{independent} sublattices on an elementary tetrahedron. The  tetrahedron rule does not hold, since the $J_1$ interactions cancel. Shifting this tetrahedron by $\bm{\delta}=(1,0,0)$ reverses the directions of the spins, so we have an 8-sublattice antiferromagnet of spin pairs $\mathbf{S}_A$, $\mathbf{S}_B$, $\mathbf{S}_C$ and $\mathbf{S}_D$, and $\mathbf{S}_A=-\mathbf{S}_{\bar{A}}$ (shifted by $(1,0,0)$), and so on:
\begin{subequations}
\begin{eqnarray}
&&\mathbf{S}_A=\mathbf{S}\left(\mathbf{0}\right)=\mathbf{S}_{\mathbf{L}_1}^0+\mathbf{S}_{\mathbf{L}_2}^0+\mathbf{S}_{\mathbf{L}_3}^0+\mathbf{S}_{\mathbf{L}_4}^0,\\
&&\mathbf{S}_B=\mathbf{S}\left(\mathbf{a}_1\right)=-\mathbf{S}_{\mathbf{L}_1}^0+\mathbf{S}_{\mathbf{L}_2}^0+\mathbf{S}_{\mathbf{L}_3}^0-\mathbf{S}_{\mathbf{L}_4}^0,\\
&&\mathbf{S}_C=\mathbf{S}\left(\mathbf{a}_2\right)=-\mathbf{S}_{\mathbf{L}_1}^0+\mathbf{S}_{\mathbf{L}_2}^0-\mathbf{S}_{\mathbf{L}_3}^0+\mathbf{S}_{\mathbf{L}_4}^0,\\
&&\mathbf{S}_D=\mathbf{S}\left(\mathbf{a}_3\right)=-\mathbf{S}_{\mathbf{L}_1}^0-\mathbf{S}_{\mathbf{L}_2}^0+\mathbf{S}_{\mathbf{L}_3}^0+\mathbf{S}_{\mathbf{L}_4}^0.
\end{eqnarray}
\label{L_orders}
\end{subequations}
This type of order is depicted in Fig.~\ref{fig:XL_order}(b). We can expand the Fourier amplitudes as:
\begin{subequations}
	\begin{eqnarray}
	&&\mathbf{S}_{\mathbf{L}_1}^0=\frac{1}{4}\left(\mathbf{S}_A-\mathbf{S}_B-\mathbf{S}_C-\mathbf{S}_D\right),\\
	&&\mathbf{S}_{\mathbf{L}_2}^0=\frac{1}{4}\left(\mathbf{S}_A+\mathbf{S}_B+\mathbf{S}_C-\mathbf{S}_D\right),\\
	&&\mathbf{S}_{\mathbf{L}_3}^0=\frac{1}{4}\left(\mathbf{S}_A+\mathbf{S}_B-\mathbf{S}_C+\mathbf{S}_D\right),\\
	&&\mathbf{S}_{\mathbf{L}_4}^0=\frac{1}{4}\left(\mathbf{S}_A-\mathbf{S}_B+\mathbf{S}_C+\mathbf{S}_D\right).
	\end{eqnarray}
\end{subequations}
The superlattice becomes an fcc lattice doubled in linear size with respect to the original one, with primitive lattice vectors given by:
\begin{equation}
\mathbf{a}_1^{L}=(1,1,0),\ \mathbf{a}_2^{L}=(1,0,1),\ \mathbf{a}_3^{L}=(0,1,1).
\end{equation}

It is easier to calculate the degeneracies in real space: the four independent unit sublattice spins mean 8 real degrees of freedom, the global $O(3)$ removes 3 of them yielding 5 independent real parameters~\cite{henley_fcc}.
 In each of the $(111)$ triangular planes only four spins appear, forming a regular 4-sublattice order~\cite{Ignatenko-2016}.

\subsection{The $W\left(1,\onehalf,0\right)$ antiferromagnet, Type-III}
\label{sec:Wphase}
There are 24 symmetry related vectors (the corners of the BZ) belonging to this type of order, and they fall into 6 equivalency classes  so the star of $\mathbf{W}$ has 6 arms. This can be understood since each corner of the BZ is shared by four truncated octahedra. The arms of the star  form three $\pm$ pairs: $\mathbf{Q}_\alpha = \pm \mathbf{W}_1, \ \pm \mathbf{W}_2, \ \pm \mathbf{W}_3$, the classes are:
\begin{align}
\mathbf{W}_1&  \sim \{(\pi,0,2\pi) , (-\pi,-2\pi,0) , (-\pi,2\pi,0) , (\pi,0,-2\pi) \},\nonumber\\
\mathbf{W}_2&  \sim \{  (2\pi,\pi,0), (-2\pi,\pi,0) , (0,-\pi,-2\pi) , (0,-\pi,2\pi)   \},\nonumber\\
\mathbf{W}_3&  \sim \{ (0,2\pi,\pi),(-2\pi,0,-\pi) , (0,-2\pi,\pi)  , (2\pi,0,-\pi)  \}.
\label{eq:Wvecs}
\end{align}
Since $\mathbf{W}_\alpha\nsim -\mathbf{W}_\alpha$ to ensure reality of the spins in real space we have to combine the $\pm\mathbf{W}_\alpha$ pairs:
\begin{equation}
\mathbf{S}_i=\sum_{\alpha=1}^3 \mathbf{S}_{\mathbf{W}_\alpha}^0 e^{-\imath \mathbf{W}_\alpha \mathbf{R}_i} + \mathbf{S}_{-\mathbf{W}_\alpha}^0e^{+\imath \mathbf{W}_\alpha \mathbf{R}_i},\\
\end{equation}
with
\begin{equation}
\mathbf{S}_{-\mathbf{W}_\alpha}^0=\mathbf{S}_{\mathbf{W}_\alpha}^{0 * }.
\end{equation}
This type of ordering is also called a triple-$\mathbf{Q}$ one. 
To fix the lengths of the spins we have the following  constraints for the complex amplitudes:
\begin{subequations}
\begin{align}
\sum_{\alpha=1}^3\mathbf{S}_{\mathbf{W}_\alpha}^{0*}\cdot\mathbf{S}_{\mathbf{W}_\alpha}^{0} &= \onehalf,
\\
\mathbf{S}_{\mathbf{W}_\alpha}^{0}\cdot\mathbf{S}_{\mathbf{W}_\alpha}^{0} &\in \imath\mathbb{R}, \ \forall \alpha,
\\
\mathbf{S}_{\mathbf{W}_\alpha}^{0}\cdot\mathbf{S}_{\mathbf{W}_\beta}^{0} = \mathbf{S}_{\mathbf{W}_\alpha}^{0*}\cdot\mathbf{S}_{\mathbf{W}_\beta}^{0}&=0, \ \forall~\alpha\neq\beta.
\end{align}
\label{Wcons}
\end{subequations}
Noting that  $\mathbf{S}^{0 *}_{\mathbf{W}_\alpha}=\mathbf{S}^{0}_{-\mathbf{W}_\alpha}$ the first equation follows from Eq.~(\ref{villainb}):
\begin{equation}
	\sum_{\alpha=1}^3\left( \mathbf{S}^{0 *}_{\mathbf{W}_\alpha}\cdot \mathbf{S}^0_{\mathbf{W}_\alpha}+\mathbf{S}^{0 *}_{-\mathbf{W}_\alpha}\cdot \mathbf{S}^{0 }_{-\mathbf{W}_\alpha}\right)=1,
\end{equation} 
and the last two equations follow from Eq.~(\ref{villainc}) by choosing $\mathbf{q}'=2\mathbf{W}_\alpha$ and $\mathbf{q}'=\mathbf{W}_\alpha \pm \mathbf{W}_\beta$.

We can decompose the complex amplitudes into real vectors, 
\begin{equation}
\mathbf{S}_{\pm\mathbf{W}_\alpha}^0 = \mathbf{u}_\alpha \mp \imath\mathbf{v}_\alpha, \quad \mathbf{u}_\alpha, \mathbf{v}_\alpha\in \mathbb{R}^3,
\end{equation} 
and we can express the constraints  Eqs.~(\ref{Wcons}) for the real and imaginary parts of the amplitudes as
\begin{subequations}
\begin{align} 
&\onehalf = \sum_{\alpha=1}^3 \left( \left|\mathbf{u}_\alpha\right|^2+\left|\mathbf{v}_\alpha\right|^2 \right) && \textrm{(1 constraint)},\\
&\left|\mathbf{u}_\alpha\right|=\left|\mathbf{v}_\alpha\right|, \ \forall~\alpha  && \textrm{(3 constraints)},\\
&0  = \mathbf{u}_\alpha \cdot \mathbf{u}_\beta, \ \forall~\alpha\neq\beta && \textrm{(3 constraints)}, \label{uvorth1}\\
&0  = \mathbf{v}_\alpha \cdot \mathbf{v}_\beta, \ \forall~\alpha\neq\beta && \textrm{(3 constraints)}, \label{uvorth2}\\
&0  = \mathbf{u}_\alpha \cdot \mathbf{v}_\beta, \ \forall~\alpha\neq\beta && \textrm{(3 constraints)}.\label{uvorth3}
\end{align}
\end{subequations}
The three pairs of real vectors $\mathbf{u}_\alpha$ and $\mathbf{v}_\alpha$ mean 18 free parameters, the equations above give 13 constraints, and we have the  global $O(3)$ degrees of freedom (3 free parameters), so we are left with 2 free real parameters for the degeneracy degrees of freedom, in perfect analogy with the Type-I phase. 

Using the global $O(3)$ freedom and the orthogonality and normalization of the $\mathbf{u}_\alpha$-s we can parametrize them as:
\begin{eqnarray}
\left(\mathbf{u}_1|\mathbf{u}_2|\mathbf{u}_3\right)=\frac{1}{2}\left(\begin{array}{c|c|c}
  \xi & 0 & 0\\ 
  0 & \eta & 0\\
	0 & 0 & \zeta
 \end{array}\right),
\end{eqnarray}
where all the parameters are real, and they satisfy the additional constraint: $\xi^2+\eta^2+\zeta^2=1$. We use the orthogonality relations between the $\mathbf{u}_\alpha$-s and $\mathbf{v}_\alpha$-s (see Eq.~\eqref{uvorth1}--\eqref{uvorth3}) to calculate the form of the $\mathbf{v}_\alpha$-s (at this point any combination of signs is allowed, resulting  8 possible combinations):
\begin{eqnarray}
\left(\mathbf{v}_1|\mathbf{v}_2|\mathbf{v}_3\right)=\frac{1}{2}\left(\begin{array}{c|c|c}
  \pm \xi & 0 & 0\\ 
  0 &\pm  \eta & 0\\
	0 & 0 &\pm  \zeta
 \end{array}\right).
 \label{plusmins}
\end{eqnarray}
With this parametrization the spins become:
\begin{eqnarray}
\mathbf{S}_i(\xi,\eta,\zeta) &=& \sqrt{2}
  \left(
    \begin{array}{c}
      \xi\cos\left(\mathbf{W}_1 \cdot \mathbf{R}_i \pm \frac{\pi}{4}\right)\\ 
      \eta\cos\left(\mathbf{W}_2 \cdot \mathbf{R}_i \pm \frac{\pi}{4}\right) \\ 
      \zeta\cos\left(\mathbf{W}_3\cdot \mathbf{R}_i \pm \frac{\pi}{4}\right)
    \end{array}
  \right).
  \label{W_spins}
\end{eqnarray} Actually these states are not all physically different, there are only two independent phases that form chiral partners (i.e. they are transformed to each other by space inversion). We are going to show this at the in Appendix \ref{sec:W_algebra}.

\begin{table*}[bt!]
\caption{The effect of the translations by the elementary lattice vectors $\bm{\delta}$ on the triple-$\mathbf{Q}$ states for the $W\left( 1, \onehalf, 0 \right)$ spin configurations, $  t_{\bm{\delta}} \mathbf{S}^{W,\Upsilon}_{\mathbf{R}}(\xi,\eta,\zeta)  
=  \mathbf{S}^{W,\Upsilon}_{\mathbf{R}-\bm{\delta}}(\xi,\eta,\zeta) 
= \mathbf{S}^{W,\Upsilon'}_{\mathbf{R}}(\xi',\eta',\zeta') $ .
\label{tab:Wtrans}
}
\begin{ruledtabular}
\begin{tabular}{ccccccc}
     & \multicolumn{2}{c}{$\bm{\delta}=\left(0,\frac{1}{2},\frac{1}{2}\right)$} & \multicolumn{2}{c}{$\bm{\delta}=\left(\frac{1}{2},0,\frac{1}{2}\right)$} & \multicolumn{2}{c}{$\bm{\delta}=\left(\frac{1}{2},\frac{1}{2},0\right)$} \\
$ \Upsilon $&$ \Upsilon' $&$ (\xi',\eta',\zeta') $&$ \Upsilon' $&$ (\xi',\eta',\zeta') $&$ \Upsilon' $&$ (\xi',\eta',\zeta') $\\
 \hline
$(ppp)	$&$(pmm)$&$(-\xi,\eta,-\zeta)	$&$(mpm)$&$(-\xi,-\eta,\zeta)	$&$(mmp)$&$(\xi,-\eta,-\zeta)	$\\
$(pmm)	$&$(ppp)$&$(-\xi,-\eta,\zeta)	$&$(mmp)$&$(-\xi,-\eta,-\zeta)	$&$(mpm)$&$(\xi,\eta,-\zeta)	$\\
$(mpm)	$&$(mmp)$&$(-\xi,\eta,\zeta)	$&$(ppp)$&$(\xi,-\eta,-\zeta)	$&$(pmm)$&$(-\xi,-\eta,-\zeta)	$\\
$(mmp)	$&$(mpm)$&$(-\xi,-\eta,-\zeta)	$&$(pmm)$&$(\xi,-\eta,\zeta)	$&$(ppp)$&$(-\xi,\eta,-\zeta)	$\\
\\
$(mmm)	$&$(mpp)$&$(-\xi,-\eta,\zeta)	$&$(pmp)$&$(\xi,-\eta,-\zeta)	$&$(ppm)$&$(-\xi,\eta,-\zeta)	$\\
$(mpp)	$&$(mmm)$&$(-\xi,\eta,-\zeta)	$&$(ppm)$&$(\xi,-\eta,\zeta)	$&$(pmp)$&$(-\xi,-\eta,-\zeta)	$\\
$(pmp)	$&$(ppm)$&$(-\xi,-\eta,-\zeta)	$&$(mmm)$&$(-\xi,-\eta,\zeta)	$&$(mpp)$&$(\xi,\eta,-\zeta)	$\\
$(ppm)	$&$(pmp)$&$(-\xi,\eta,\zeta)	$&$(mpp)$&$(-\xi,-\eta,-\zeta)	$&$(mmm)$&$(\xi,-\eta,-\zeta)	$\\
\end{tabular}
\end{ruledtabular}
\end{table*}

In order to understand this real space spin pattern we write the lattice point in Cartesian coordinates [$\mathbf{R}_i=(x,y,z)$, i.e. in our fcc lattice the Cartesian coordinates $x$, $y$, and $z$ can either be integers, or some half integer combinations]. The spins are:
\begin{eqnarray}
\mathbf{S}_i(\xi,\eta,\zeta) &=&\sqrt{2}\left(
    \begin{array}{c}
      \xi\cos\left(\pi(x+2z)\pm\frac{\pi}{4}\right)\\ 
	  \eta\cos\left(\pi(y+2x)\pm\frac{\pi}{4}\right)\\ 
	  \zeta\cos\left(\pi(z+2y)\pm\frac{\pi}{4}\right)
	\end{array}
	\right).
	\label{WCartesian}
\end{eqnarray}
This form is particularly useful for analyzing the properties of the spin structure: we only have  to monitor the phase shifts due to transformations (either in real or spin space) and draw the consequences. From this form it can be seen that this pattern is periodic under a translation of $2$ along the Cartesian directions (we have  to change all $x$, $y$, and $z$ by an even number to achieve a $2\pi$ phase shift in every component), and this is the smallest possible magnetic unit cell, with primitive translations:
\begin{equation}
\mathbf{a}_1^W= \left(2, 0, 0\right),\ \mathbf{a}_2^W= \left(0,2,0\right),\   \mathbf{a}_3^W= \left(0,0,2\right).
\end{equation}	
The resulting superlattice is simple cubic, and the magnetic unit cell contains 32 points of the original fcc lattice. There are eight spin directions, that come in four $\pm$ pairs (like in the Type-II phase), but they form a complicated 32-sublattice order (we do not even  try to visualize this spin pattern here). The four spin directions are not independent: there is still a tetrahedron rule in action (like in Eq.~\eqref{tetrarule_general}): spins on every elementary tetrahedron sum up to zero. So out of 8 parameters describing the four sublattice spins the tetrahedron rule removes three parameters, and the global $O(3)$ removes another three yielding the correct number of two parameters of the unit vector $(\xi,\eta,\zeta)$. We are left with the task to decide how many physically different phases the discrete $\pm$ parameters in  Eq.~\eqref{WCartesian} yield.

Let us  denote a state in Eq.~\eqref{WCartesian} by their respective sign triplet, i.e., $(mmm)$ and $(ppp)$ have signs $(---)$ and $(+++)$. We want to study the effect of space inversion ($I$) and translations $t_{\bm{\delta}}$ by a lattice vector $\bm{\delta}=\left(\delta x, \delta y,\delta z\right)$ on the signs in Eq.~\eqref{WCartesian}. We are going to prove that there are only \textit{two} nonequivalent phases, transformed into each other by space inversion centered on a lattice point, and those phases cannot be transformed to each other by any lattice translation, i.e. they are really physically different.

The action of a translation by the elementary lattice vectors on a spin configuration is defined as
\begin{equation}
  (t_{\bm{\delta}} \mathbf{S})_{\mathbf{R}} =  \mathbf{S}_{\mathbf{R}-\bm{\delta}}\;.
    \label{eq:transS}
\end{equation}

As an example, let us consider the $\bm{\delta} = \left(0,\frac{1}{2},\frac{1}{2}\right)$ case.
\begin{align}
\mathbf{S}^{W,(ppp)}_{\mathbf{R}_i-\bm{\delta}}(\xi,\eta,\zeta)
&= \sqrt{2}
  \left(
    \begin{array}{c}
      \xi\cos\left[\pi(x+2z-1)+\frac{\pi}{4}\right]\\ 
	  \eta\cos\left[\pi(y+2x-\frac{1}{2})+\frac{\pi}{4}\right] \\
	  \zeta\cos\left[\pi(z+2y-\frac{3}{2})+\frac{\pi}{4}\right]
	\end{array} 
  \right)
  \nonumber\\
&= \sqrt{2}
  \left(
    \begin{array}{c}
     -\xi\cos\left[\pi(x+2z)+\frac{\pi}{4}\right]\\ 
	  \eta\cos\left[\pi(y+2x)-\frac{\pi}{4}\right] \\
	 -\zeta\cos\left[\pi(z+2y)-\frac{\pi}{4}\right]
	\end{array} 
  \right)
  \nonumber\\
  &=
 \mathbf{S}^{W,(pmm)}_{\mathbf{R}_i}(-\xi,\eta,-\zeta),
\end{align}
so we can write
\begin{equation}
  t_{\left(0,\frac{1}{2},\frac{1}{2}\right)} \mathbf{S}^{W,(ppp)}_{\mathbf{R}_i}(\xi,\eta,\zeta)=  \mathbf{S}^{W,(pmm)}_{\mathbf{R}_i}(-\xi,\eta,-\zeta).
\end{equation}
This and other cases are collected in Tab.~\ref{tab:Wtrans}. 
Of particular importance in our analysis is the effect of translations on the sign structure of the $\pi/4$ phase factors entering the parametrization of the spin configurations defined in Eq.~\eqref{WCartesian}.
Scrutinizing Tab.~\ref{tab:Wtrans}, we find that configurations that have an odd number of $m$-s (i.e. $(mmm)$, $(mpp)$, $(pmp)$ and $(ppm)$) can be translated to each other, so can the ones with an even number of $m$-s (i.e. $(ppp)$, $(pmm)$, $(mpm)$ and $(mmp)$). Since the translations preserve the  parity of the number of $m$s, the configurations fall into two disjoint sets (characterized by even or odd number of $m$s).

 This two sets, however, are connected by the inversion $I$. Namely,
the $\mathbf{R}_i \rightarrow -\mathbf{R}_i$ operation of the inversion  can be counteracted by reversing the sign of the $\pi/4$ phase in every component of Eq.~\eqref{WCartesian}
, i.e. it  changes each $m$ to $p$ and vice versa, and therefore the number of $m$s is changed by an odd number (for example $(mmm)\rightarrow(ppp)$, $(mmp)\rightarrow(ppm)$, and so on).  So the states of different parity of number of $m$s that form the two disjoint sets can only be transformed into each other  by inversion. The importance of this distinction will become clear in Sec.~\ref{sec:chirality}. A more abstract way to describe the effect of the space-group transformations is given in Appendix \ref{sec:W_algebra}.

The fact that the $W\left( 1, \onehalf, 0 \right)$ type order is defined within a 32 site unit cell, naturally raises the question as to how it can be obtained from a spin configuration [obeying the tetrahedron rule of Eq.~\eqref{tetrarule_general}] defined on an elementary tetrahedron. To this end, we consider the translations by $\bm{\delta} = (1,0,0)$, $(0,1,0)$ and $(0,0,1)$:
\begin{subequations} 
\begin{align}
  t_{(1,0,0)} \mathbf{S}^{W}_i(\xi,\eta,\zeta) &=  \mathbf{S}^{W}_i(-\xi,\eta,\zeta) = \Theta C_2(x)\cdot\mathbf{S}^{W}_i(\xi,\eta,\zeta), \\
  t_{(0,1,0)} \mathbf{S}^{W}_i(\xi,\eta,\zeta) &=  \mathbf{S}^{W}_i(\xi,-\eta,\zeta) =\Theta C_2(y)\cdot\mathbf{S}^{W}_i(\xi,\eta,\zeta),\\
  t_{(0,0,1)} \mathbf{S}^{W}_i(\xi,\eta,\zeta) &=  \mathbf{S}^{W}_i(\xi,\eta,-\zeta) =\Theta C_2(z)\cdot\mathbf{S}^{W}_i(\xi,\eta,\zeta).
  	  \label{eq:t001S}
\end{align}
\end{subequations}
In this cases the signs of the $\pi/4$ phase do not change, and we can omit the $(ppp)$, and so on. Under the effect of translations some of the spin components change sign. These changes in the sign can be captured by $C_2$ rotation in the spin space augmented by time reversal.
 Hence we obtain an important property of the triple-$\mathbf{Q}$ spin configuration, namely that they are invariant under the above mentioned operations.
This provides a prescription of generating the spin configuration defined within the 32 site magnetic unit cell starting from four spins on an tetrahedron. We just need to be careful about the choice of the three axes we do the $C_2$ rotations about: they are the three $C_2$ symmetry axes of the initial tetrahedron rule obeying configuration.

Finally, combining all of the three translations
above, we get that the translation by the $\bm{\delta} = (1,1,1)$ reverses the spins, 
 \begin{equation}
  t_{\left(1,1,1\right)} \mathbf{S}^{W}_i(\xi,\eta,\zeta)=  \mathbf{S}^{W}_i(-\xi,-\eta,-\zeta) = \Theta \mathbf{S}^{W}_i(\xi,\eta,\zeta),
\end{equation}
where $\Theta$ denotes the time reversal operation. 

Let us note, that  for double-$\mathbf{Q}$ state the unit cell reduces to 16 sites, e.g., setting $\zeta=0$, Eq.~(\ref{eq:t001S}) tells us that the spin configuration is invariant with respect to translation by $\bm{\delta} = (0,0,1)$, halving the lattice vectors defining the unit cell in this directions. The single-$\mathbf{Q}$ state has a 4-site unit cell, e.g., for $\eta=\zeta=0$ the lattice vectors of the unit cell are   $(1,\frac{1}{2},\frac{1}{2})$, $(0,1,0)$, and $(0,0,1)$.

\section{Non-coplanar states and chirality}
\label{sec:chirality}

In this section we discuss the  non-coplanar but commensurate spin configurations.  We have seen in Sec. \ref{sec:realfourier_chapter} that choosing multiple arms of the stars of a commensurate ordering vector we get non-collinear or even non-coplanar states on this Bravais-lattice: such multiple-$\mathbf{Q}$ structures can be created e.g. by choosing $\xi$, $\eta$ and $\zeta$ in Eq.~(\ref{Xspins}) finite (and analogous constructions work for  Eq.~(\ref{Lfourq}) or  Eq.~(\ref{W_spins})). 
In general, non-coplanar states are rarely observed in isotropic spin systems, since order by disorder (either quantum or thermal) mechanisms tend to select the collinear (or coplanar) configurations~\cite{henley_fcc}.
 Nevertheless, counterexamples exist in models including ring-exchanges on Bravais lattices, e.g. the tetrahedral phase on the triangular lattice \cite{korsh,Momoi_PRL,Kubo_Momoi}, and longer range exchanges on non-Bravais lattices such as cuboc orders on the kagome lattice \cite{Domenge2005,Janson_2009}.
   Disorder mechanisms may also favor noncollinearity~\cite{henley1989}, so the fate of these states depends on further details.
A non-coplanar state can be \textit{chiral}. In the following  we will review some notions of chirality and its relationship to  non-coplanar orders found here.


\subsection{Scalar chirality}

If a magnetic order is non-coplanar, it has a finite \textit{scalar chirality} \cite{villain_rules,WWZ1989,Wiegmann1988,Baskaran} defined on an oriented triangular plaquette  of vertices $ABC$ as 
\begin{equation}
  \chi_{ABC}=\mathbf{S}_A\cdot(\mathbf{S}_B \cross \mathbf{S}_C) ,
\end{equation}
i.e. as the \textit{signed} volume of the parallelepiped spanned by the three spins. 
This also shows that a finite value of the scalar chirality on a triangle is equivalent to a non-coplanar spin configuration. 
On the fcc lattice care has to be taken of how to define the orientation of the triangular plaquettes. 
Since no triangle is shared between two octahedra, one can use the right hand rule for outward pointing normals on the faces of the octahedra.
All the commensurate phases can be chiral in the sense that one can construct non-coplanar spin configurations bu using multiple arms of the star of the ordering vector, these are the \textit{mutiple-$\mathbf{Q}$} phases.

\subsubsection{The $X(1,0,0)$ phase}
 
 The Type-I X order, given by Eq.~(\ref{tetraruleX})  and shown in Fig.~\ref{fig:XL_order}(a), is non-coplanar when $\xi$, $\eta$, and $\zeta$ are all nonzero. The scalar chirality $\chi$ is then finite with $|\chi| = | 4\xi\eta\zeta |$ on \textit{all} faces of \textit{every} tetrahedron, the sign alternating on the two types of tetrahedra.
 
\subsubsection{The $L\left( \onehalf, \onehalf, \onehalf \right)$ phase}

  This phase is composed of pairs of opposite pointing spins located on antipodes of the  octahedra, as shown in Fig.~\ref{fig:XL_order}(b). As a consequence, the scalar chirality is equal on all six faces of an octahedron. There are four types of octahedra, so there can be four different values of scalar chiralities in the lattice for non-coplanar spins.

\subsubsection{The $W\left( 1, \onehalf, 0 \right)$ phase}

\begin{figure}
  \centering
  \includegraphics[width=0.6\columnwidth]{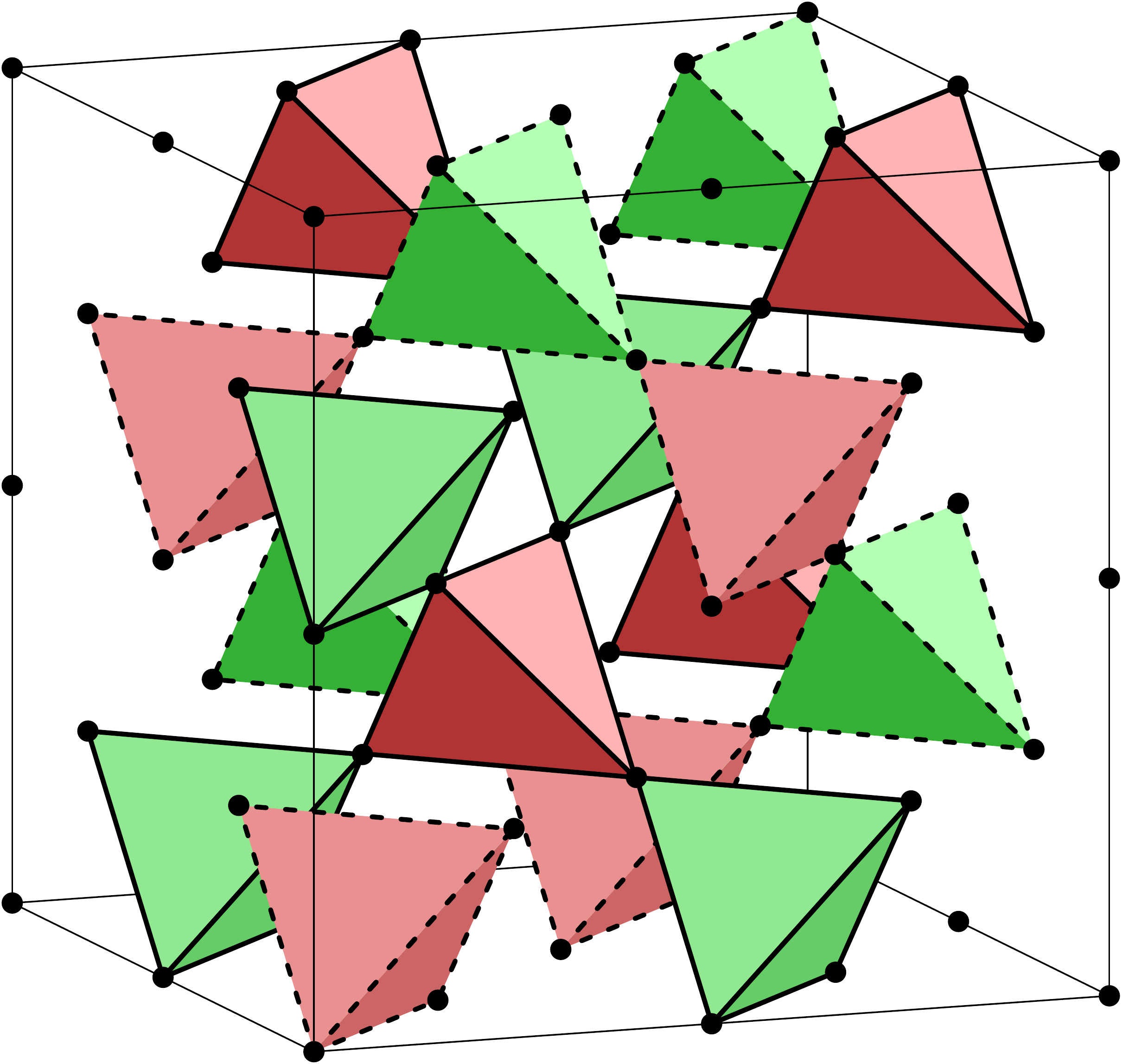}
  \caption{
	Tetrahedra with triangular faces having nonzero scalar chirality $\chi$  in the triple-$\mathbf{Q}$ $W(1,\frac{1}{2},0)$ state, here shown within a 32 site cubic unit cell defined by the lattice vectors $(2,0,0)$, $(0,2,0)$, and $(0,0,2)$. 
		The red and green color depict triangles with equal $|\chi|$, but opposite signs of $\chi$.
	The network  of the corner-sharing alternating red and green tetrahedra builds two interpenetrating pyrochlore lattices, distinguished by solid and dashed lines.
}
  \label{fig:Wchirality}
\end{figure}

The direct evaluation of the scalar chirality $\chi$ results in a  pattern displayed in Fig.~\ref{fig:Wchirality}. The $\chi$s on the faces of a tetrahedron are all identical, with three types of tetrahedra:
on 1/4 of all tetrahedra the chiralities are alternating between the  $\chi=4 \xi \eta \zeta$ and $\chi=-4 \xi \eta \zeta$, while on the remaining tetrahedra the $\chi$ vanishes. 
The chirality pattern changes sign when translated by $(\pm 1,0,0)$, $(0,\pm 1,0)$, and $(0,0,\pm 1)$ lattice vectors, and is invariant under the translations  by $(1,1,0)$ and equivalent vectors.

\subsection{Chiral enantiomers}
\label{sec:enantiomers}

Another concept of chirality more closely related to handedness is the following \cite{villain_rules}: for a given spin configuration apply a mirror plane on the configuration that is a symmetry of the underlying lattice. If the resulting spin pattern cannot be transformed back to the original configuration by any \textit{proper} space group operation to the original one, we call the configuration \textit{chiral}  [A sidenote about nomenclature: in three-dimensional space a rotation is called proper (an element of $SO(3)$) if it is orientation-preserving. Point group operations that change the the orientation of a basis are either called improper rotations or rotoreflections, this includes the inversion $I: \mathbf{R}_i \rightarrow -\mathbf{R}_i$, the orthogonal group is the direct product $O(3)=SO(3) \times \{E, I\}$, with $E$ being the identity.]. This notion of chirality is the straightforward generalization of the concept of chirality defined for molecules: a molecule is chiral if it cannot be rotated to its mirror image. Such pairs of reflection-related partners are called enantiomers, or enantiomorphic/chiral partners. In the following we will expand this concept and analyze the commensurate orders according to it.

A space group operation (proper or improper) acting on a lattice point $\mathbf{R}_i$ is denoted by $g_{\bm{\delta}} \equiv \{\mathbf{G}| \bm{\delta}\}$:
\begin{equation}
	\mathbf{R}_i'=g_{\bm{\delta}}\mathbf{R}_i \equiv \{\mathbf{G}| \bm{\delta}\}\mathbf{R}_i  = \mathbf{G}\cdot  \mathbf{R}_i+\bm{\delta},
\end{equation}
where $\mathbf{G}$ is the $O(3)$ matrix of a point group element followed by the lattice translation $\bm{\delta}$. On a spin --being an axial vector-- the point group element acts as $\mathbf{S}'=\mathbf{G}^A \cdot \mathbf{S}$, where $\mathbf{G}^A=(\det \mathbf{G}) \mathbf{G}$ is the axial-vector representative of the group element $g$ (it agrees with $\mathbf{G}$ for proper rotations and it is $-\mathbf{G}$ for improper rotations). The inverse of a space group operation is $g_{\bm{\delta}}^{-1}= \{\mathbf{G}^{-1}| -\mathbf{G}^{-1}\cdot \bm{\delta}\}$. The transformation rule for a spin pattern (sometimes called as an active view of a transformation: we grab the spin pattern  together with the lattice points and transform them as a rigid body) reads:
\begin{equation}
\mathbf{S}'_{\mathbf{R}_i}=\{\mathbf{G}| \bm{\delta}\}\mathbf{S}_{\mathbf{R}_i}=\mathbf{G}^A \cdot \mathbf{S} _{g_{\bm{\delta}}^{-1}\mathbf{R}_i}.
\end{equation}
Any improper rotation is the product of a proper rotation and inversion, e.g. a mirror plane with normal $\hat{\mathbf{n}}$ is a composition of a twofold rotation about $\hat{\mathbf{n}}$ composed with inversion, and inversion acts in spin space as the identity, $I: \mathbf{S} \rightarrow \mathbf{S}$.  Every Bravais-lattice is inversion-symmetric, hence instead of mirror planes we can use space inversion to define chirality, and this definition has a practical advantage: spins remain intact under inversion.  Now we can formulate chirality for spin patterns defined on Bravais-lattices: let us apply inversion to a spin pattern $\mathbf{S}_{\mathbf{R}_i}$:
\begin{equation}
	\mathbf{S}'_{\mathbf{R}_i} = \{I| \mathbf{0}\} \mathbf{S}_{\mathbf{R}_i} =  \mathbf{S}_{-\mathbf{R}_i}.\label{spinchange}
\end{equation}
If there is no \textit{proper} space group element  
$\{\mathbf{G}| \bm{\delta}\}$, with $\mathbf{G} \in SO(3)$, that can compensate for the change in spin pattern in Eq.~(\ref{spinchange}),
\begin{equation}
	\{\mathbf{G}| \bm{\delta}\}\mathbf{S}'_{\mathbf{R}_i} \neq  \mathbf{S}_{\mathbf{R}_i}, \label{spinchangeback}
\end{equation}
then the pattern is chiral. Here we have considered the case when the mirror plane contains a lattice point, i.e. the inversion used instead of the mirror plane is centered at that point, but for other cases --when for example the mirror plane is a perpendicular bisector plane of  a bond-- the argumentation --\textit{mutatis mutandis}-- still applies.

Let us now turn to the question if the non-coplanar phases show handedness or not. The spin configuration in the $X(1,0,0)$ phase and in the $L(\frac{1}{2},\frac{1}{2},\frac{1}{2})$ phases is unchanged if inverted about a lattice site. Therefore neither of these states is chiral in the sense defined above.
  
 The situation is different for $W\left( 1, \onehalf, 0 \right)$ phase. In Sec.~\ref{sec:Wphase} we have examined the effect of translations and inversions on the spin configurations and found that there are two disjoint sets, which can be transformed into each other by inversion. Having introduced the concept of the handedness, we can now identify the spin configurations in the two sets as being of opposite handedness, i.e.  they are enantiomers.

\subsection{Time-reversal and chirality}

 Finally, Ref.~\cite{Domenge2005,messio_colored} considers yet another definition for the chirality: whether the operation of time reversal (called "spin inversion" in Ref.~\cite{Domenge2005,messio_colored}) on a spin configuration can be undone using an SO(3) rotation acting on the spins. Using this definition, any non-coplanar $X(1,0,0)$, $L(\frac{1}{2},\frac{1}{2},\frac{1}{2})$, and $W\left( 1, \onehalf, 0 \right)$ spin configuration is chiral. However, if we allow for translations, the $L(\frac{1}{2},\frac{1}{2},\frac{1}{2})$ and $W\left( 1, \onehalf, 0 \right)$ are not chiral in this sense, since they are invariant under time reversal combined with a translation.The cuboc orders considered in Refs.~\cite{Domenge2005,Janson_2009} are chiral by means of the time-reversal symmetry considered here \cite{messio_colored}, and are also chiral in the sense defined in Sec.~\ref{sec:enantiomers}, since time reversal and spatial inversion about the center of a hexagon is equivalent in this case: both of them flip every spin and this cannot be undone by any proper space group operation.

\section{Incommensurate Phases}
\label{sec:incommensurate}

All the commensurate phases were found in the $J_1-J_2$ models~\cite{Villain-1959,henley_fcc,yamamoto1972}, see the $J_3=0$ lines in Fig.~$\ref{fig:phase_diagram_descartes}$(a) and (b). A novel feature of the $J_3\neq 0$ model is the appearence of incommensurate orderings with propagation vectors along special, highly symmetric directions~\cite{kamiya_2015,Kaplan-1959,Iqbal-2019bcc}. For the three incommensurate orderings (c.f. Fig.~\ref{fig:brillzone}(d)--(e)) we give the dependence of the wave-vectors on the exchange parameters and give their accessible ranges.

In all these phases --since the ordering vectors $\mathbf{Q}$ are incommensurate-- we have to include both $\pm\mathbf{Q}$ to ensure reality of the spin components, the resulting coplanar spin pattern becomes:
\begin{equation}
\mathbf{S}_i=\mathbf{s}_1\cos\left( \mathbf{Q}\cdot\mathbf{R}_i+\varphi\right) \pm \mathbf{s}_2\sin\left( \mathbf{Q}\cdot\mathbf{R}_i+\varphi\right),\label{incomm_spiral}
\end{equation}
where $\mathbf{s}_1$ and $\mathbf{s}_2$ are arbitrary orthogonal unit vectors spanning the plane of spin rotations, and $\varphi$ is an arbitrary phase, and $\pm$ accounts for the two possible chiral enantiomers (in the sense explained in Sec.~\ref{sec:enantiomers}). The model being isotropic, there is nothing to fix the plane of rotation to the wave-vectors or to the crystallographic axes (another manifestation of the spontaneous breaking of the global $O(3)$ symmetry). Since the wave-vectors are incommensurate we cannot build any multiple-$\mathbf{Q}$ ground states of their stars~\cite{villain_rules}. 

The three possible incommensurate ordering directions are $\Delta\left(q,0,0\right)$ with its  6-armed star (see Fig.~\ref{fig:brillzone}(d)), $\Lambda\left(q,q,q\right)$ with its 8-armed star (see Fig.~\ref{fig:brillzone}(e)) and $\Sigma\left(q,q,0\right)$ with its 12-armed star (see Fig.~\ref{fig:brillzone}(f)). Note in Fig.~\ref{fig:phase_diagram_descartes}(a) that  $\Delta\left(q,0,0\right)$ smoothly interpolates between the phases $\Gamma(0,0,0)$ and $X(1,0,0)$ so the possible $q$-values exhaust the whole $\Delta\left(q,0,0\right)$-star. Similarly the phase $\Lambda\left(q,q,q\right)$ smoothly connects the phases $\Gamma(0,0,0)$ and $\Lambda\left(\onehalf,\onehalf,\onehalf\right)$ and  the possible $q$-values exhaust the whole $\Lambda\left(q,q,q\right)$-star. The situation with the  $\Sigma\left(q,q,0\right)$ star is quite different, the possible $q$-values are confined to $ 1.28\pi \lesssim q \leq 2\pi$, and the transition between $\Gamma(0,0,0)$ and $\Sigma\left(q,q,0\right)$ is of first order. 

The  vector $\mathbf{Q}=\left(\pi,\pi,0\right)$, although not a special point in the BZ is compatible with a N\'eel-type commensurate antiferromegnetic ordering, called the Type-IV phase fcc antiferromagnet, realized in CoN~\cite{con_experiment,haar_lines}. Unfortunately the possible $q$-values of the $\Sigma\left(q,q,0\right)$ phase are far from $\pi$, and we could not even stabilize this type of N\'eel-order by introducing quantum fluctuations (in the spirit of~\cite{mila_wavestab}).  

The optimized $q$-values of the incommensurate ordering vectors  are given by:
\begin{subequations}	
  \begin{align}
	\cos \frac{q_\Delta}{2} &=\frac{-J_1-2 J_3}{J_2+4 J_3},\label{eq:optimizedqs1}\\
	\cos q_\Lambda &=-\frac{J_1+J_2+2 J_3}{4J_3},\\
	\cos \frac{q_\Sigma}{2} &=
  	\frac{\sqrt{(J_1+2 (J_2+J_3))^2-48 J_3 (J_1-2 J_3)}}{24 J_3}
	 \nonumber\\
	 &\phantom{=} -\frac{J_1+2 (J_2+J_3)}{24 J_3}.
  \label{eq:optimizedqs3}
  \end{align}
\end{subequations}

\section{Ground states of the extended  manifolds}
\label{sec:manifolds}

\begin{figure*}
	\centering
	\includegraphics[width=1.8\columnwidth]{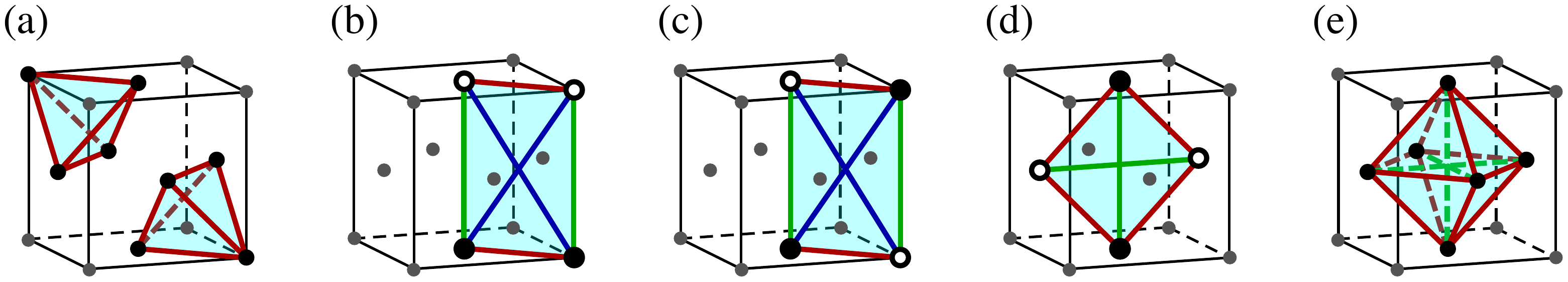}
	\caption{Finite motifs used to cover the fcc lattice in the exchange parameter regions with high ground state degeneracies. Red, green and blue lines denote first, second, and third neighbor bonds of a motif, respectively. By building up the crystal from these motifs we cover bonds  and sites multiple times,  this overcounting  is summarized for every motif in Table~\ref{tab:motifs}. (a) Elementary tetrahedra (index "tetra" in formulas). From each lattice point we  draw two differently oriented tetrahedra to cover every first neighbor bond twice. (b) A signed rectangle (index "rect$_1$" in formulas): with 6 differently oriented rectangles put on every site we can cover every first, second and third neighbor bond. "Signed" here means that when writing the complete squares of the spin sums in the Hamiltonian we have to assign a minus sign to the spins sitting in the vertices denoted by  white dots, black dots get a plus sign, see Eq.~(\ref{eq:rect1}). Together with the tetrahedra we use this motif to construct the ground states of the phase corresponding to the manifold $\mathcal{M}_Z^1$, see Fig. \ref{fig:brillzone}(b). (c) A signed rectangle (index "rect$_2$" in formulas): very similar to the former one, but the signs are distributed differently, see Eq.~(\ref{eq:rect2}). We cover the lattice with this single motif for the phase with $\gsm=\mathcal{M}_\Delta^1$. (d) A signed square: (index "square" in formulas), by 3 differently oriented squares per site we cover the lattice  for the phase with $\gsm=\mathcal{M}_\Lambda^1$, see Eq.~(\ref{eq:square}). (e) Elementary octahedron ("octa" in formulas): we cover the lattice with one edge-sharing octahedron per site for the phase with the two-dimensional $\gsm=\mathcal{M}^2$, see Eq.~(\ref{eq:octa}).}
	\label{fig:motifs}
\end{figure*}

For the pure  fcc model ($J_1>0$, $J_2=J_3=0$) the degenerate  manifold $\mathcal{M}_Z^1$ has already been  found~\cite{Danielian-1961, alexander1980}, and a class of ground states were constructed from   $(100)$-directed,  noninteracting AFM planes. In this section we describe the phases with extended $\gsm$-s of energetically degenerate ordering vectors (see Fig.~\ref{fig:brillzone}(b)--(e)) that correspond to large ground state degeneracies at special points in the phase diagram in Fig.~\ref{fig:phase_diagram_descartes}. We explain these degeneracies by a real space construction of covering the lattice with finite motifs (see Fig.~\ref{fig:motifs}), and write the Hamiltonian as a positive definite sum over these motifs. Minimizing the Hamiltonian imposes local constraints on the spins on these motifs: \textit{any} state that satisfies these constraints is an allowed ground state. We extend the construction presented in~\cite{alexander1980} for the other degenerate manifolds and construct ground states from noninteracting planes, and also find ground states consisting of ferromagnetic chains (though the chains are now interacting).

We have solved the models for the extended manifolds for Ising spins $\mathbf{S}_i \in \left\{ 1,-1\right\}$, for finite, symmetric clusters. Details of these results are presented in Appendix \ref{sec:Ising_results}. We also have performed numerical simulations for planar ($O(2)$ or $XY$) spins $\mathbf{S}_i=(S_x,S_y)_i$, $S_x^2+S_y^2=1$ to guide our intuition about the possible ground states.

\begin{table}
	\caption{\label{tab:motifs} Motifs used to cover the lattice (see Fig.~\ref{fig:motifs}) together with their symbols used in formulas, and their overcounting of the sites and of the  $J_1$ first, $J_2$ second, and  $J_3$ third  neighbor bonds. E.g. in a tetrahedral covering we put two differently oriented tetrahedra on each site, and as a consequence each $J_1$  bond is shared by two tetrahedra, and no longer bonds are covered. In the last column we give the reference as a subfigure for the picture of the motif   in Fig.~\ref{fig:motifs}.}
	\begin{center}
		\begin{ruledtabular}
			\begin{tabular}{llcccccc}
				Motif & Symbol & Site & $J_1$ & $J_2$ & $J_3$ & Subfigure \\ \hline
				Tetrahedron & tetra & 2 & 2 & -- & -- & (a) \\
				Signed rectangle & rect$_{1}$ & 6 & 2 & 4 & 1 & (b) \\
				Signed rectangle & rect$_{2}$ & 6 & 2 & 4 & 1 & (c) \\
				Square & square & 3 & 2 & 2 & -- & (d) \\ 
				Octahedron & octa & 1 & 2 & 1 & -- & (e)	
			\end{tabular}
		\end{ruledtabular}
	\end{center}
\end{table}
\subsection{The $J_1=2J_2>0$, $J_3=0$ point: the  two dimensional $\mathcal{M}^2$ ground state manifold}
\label{subsec:M2}
At the point $J_1=2J_2>0$, $J_3=0$ (the green dot in Fig.~\ref{fig:phase_diagram_descartes}(b)) the ordering vectors of the possible ground states form the  two dimensional $\mathcal{M}^2$ manifold, defined by
\begin{equation}
\cos \frac{Q_x}{2}+\cos \frac{Q_y}{2}+\cos\frac{Q_z}{2}=0,\label{surf_eq}
\end{equation}
as depicted in Fig.~\ref{fig:brillzone}(c). This is the only point of the phase diagram with such a large degeneracy~\cite{finnish_93,Dominik}, though extending the parameters to include the fourth neighbor coupling $J_4$ will allow an additional two-dimensional manifold to appear \cite{ourspiral}. The $W\left(1,\onehalf,0\right)$ and $L\left(\onehalf,\onehalf,\onehalf\right)$ points are parts of this manifold, and this is the point of the phase diagram where the $W\left(1,\onehalf,0\right)$ and $L\left(\onehalf,\onehalf,\onehalf\right)$ phases meet through the  neck of the $\Sigma(q,q,0)$ phase~\cite{yamamoto1972}. At this point the Hamiltonian reads: 
\begin{align}
\mathcal{H}=\frac{J_1}{4}\left(\sum_{\left\langle i,j\right\rangle_1}4 \, \mathbf{S}_i\cdot\mathbf{S}_{j}+\sum_{\left\langle i,j\right\rangle_2}2 \, \mathbf{S}_i\cdot\mathbf{S}_{j}\right). \label{surf1}
\end{align}
We can express this Hamiltonian as a sum of complete squares of spins forming edge-sharing octahedra covering the lattice (see Fig.~\ref{fig:motifs}(e)):
\begin{equation}
  \mathcal{H}=\frac{J_1}{4}\sum_{\textrm{octa}}\left(\mathbf{S}_1+\mathbf{S}_2+\mathbf{S}_3+\mathbf{S}_4+\mathbf{S}_5+\mathbf{S}_6\right)^2-\frac{3}{2} J_1 N,
  \label{eq:octa}
\end{equation}
where $\mathbf{S}_1, \ \dots , \ \mathbf{S}_6$ refer to the six  spins on the sites of an  octahedron.
Since every first neighbor bond is covered twice, and every second neighbor bond  once (see Table~\ref{tab:motifs}), Eq.~(\ref{eq:octa}) exactly reproduces Eq.~(\ref{surf1}), and this is why we have chosen the octahedral covering for these particular values of exchange parameters. Since $J_1>0$  Eq.~(\ref{eq:octa}) is minimized if the spins sum up to zero, 
\begin{equation}
 \mathbf{S}_1+\mathbf{S}_2+\mathbf{S}_3+\mathbf{S}_4+\mathbf{S}_5+\mathbf{S}_6 = \mathbf{0},
   \label{eq:octahedron_rule}
\end{equation}
on \textit{every} octahedron -- we refer to this rule as the \textit{octahedron rule}.
 Every such configuration is a ground state, and every ground state has this property. The additional constant $-\frac{3}{2} J_1N$ gives the ground state energy. 

The following ordered phases automatically satisfy the octahedron rule:
 the  $L\left(\onehalf,\onehalf,\onehalf\right)$ and $W\left(1,\frac{1}{2},0\right)$-type ground states, see Fig.~\ref{fig:XL_order}(b) and Eq.~(\ref{WCartesian}). A general spiral with $\mathbf{Q}\in \mathcal{M}^2$ also satisfies the octahedron rule: this can be checked by putting an arbitrary $\mathbf{Q}$ in Eq.~(\ref{incomm_spiral}), and summing up the spins  on octahedra: the sum vanishes if and only  if  $\mathbf{Q}$ satisfies the defining equation (\ref{surf_eq})  of $\mathcal{M}^2$.
Order by disorder effects (either thermal or quantum) at the harmonic level select the $L\left(\onehalf,\onehalf,\onehalf\right)$ points on the $\mathcal{M}^2$ surface~\cite{ourspiral}. 

\subsection{The $\Gamma(0,0,0)-\Lambda(q,q,q)-L(\onehalf,\onehalf,\onehalf)$ triple point: the one dimensional $\mathcal{M}^1_\Lambda$ ground state manifold}
\label{Lambda_subsection}

\begin{figure}
	\centering
	\includegraphics[width=.95\columnwidth]{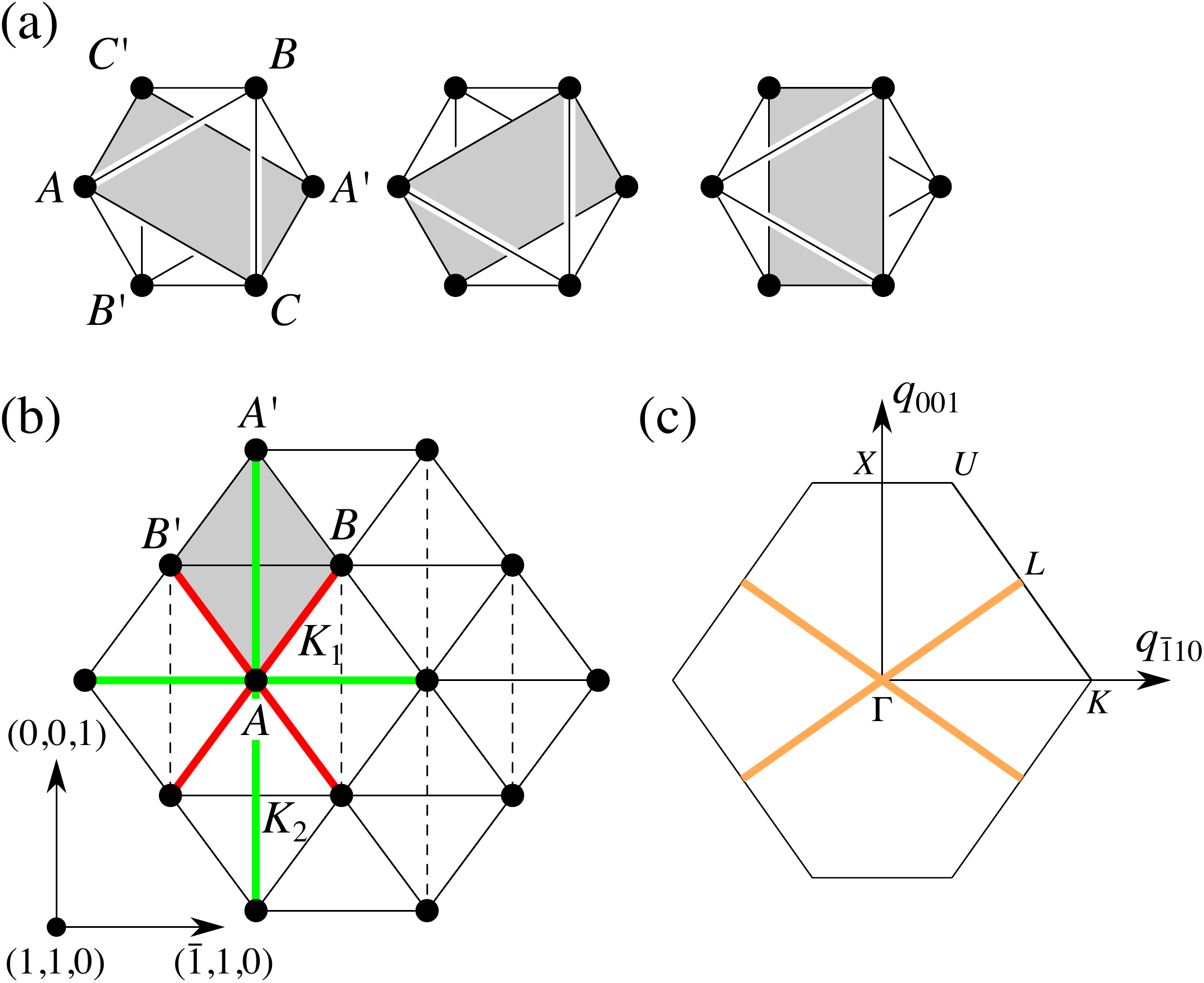}
	\caption{(a) Covering octahedra of the face-centered-cubic lattice (also depicted in Fig.~\ref{fig:exchanges}(d) and Fig.~\ref{fig:motifs}(e)) showing the three possible orientations of the signed squares (see Fig.~\ref{fig:motifs}(d))  inscribed. (b) The face-centered-cubic lattice viewed from the $(110)$ direction. This is a two dimensional lattice  the $(110)$ ferromagnetic chains form in a class of solutions of the model in the $\Gamma-\Lambda-L$ point of the phase diagram. The bond strengths of the effective two dimensional  Hamiltonian Eq.~(\ref{triang_2D}) for the chains are denoted by  $K_1$ for  the first neighbor red bonds, and by $K_2$ for the second neighbor green bonds. The gray rhombus is the projection of one of the covering signed squares also depicted in Fig.~\ref{fig:motifs}(d), minus signs have to be associated to one pair of opposite vertices, say to $A$ and $A'$.  (c) Brillouin zone of the lattice depicted in Fig.~\ref{fig:111_octahedra}(b), together with the ground state manifold (orange cross) of the Hamiltonian Eq.~(\ref{triang_2D}), this manifold is  the section of $\mathcal{M}^1_\Lambda$ (see Fig.~\ref{fig:brillzone}(e)) with the $(110)$ $\mathbf{q}$-plane passing through the origin, this BZ is not a perfect hexagon. Symmetry points of the original three dimensional BZ (see Fig.~\ref{fig:brillzone}(a)) are shown, together with some less commonly known points $K\left(\bar{\frac{3}{4}},\frac{3}{4},0\right)$ and $U\left(\bar{\frac{1}{4}},\frac{1}{4},1\right)$.}
	\label{fig:111_octahedra}
\end{figure}

\begin{figure*}
	\centering
	\includegraphics[width=1.8\columnwidth]{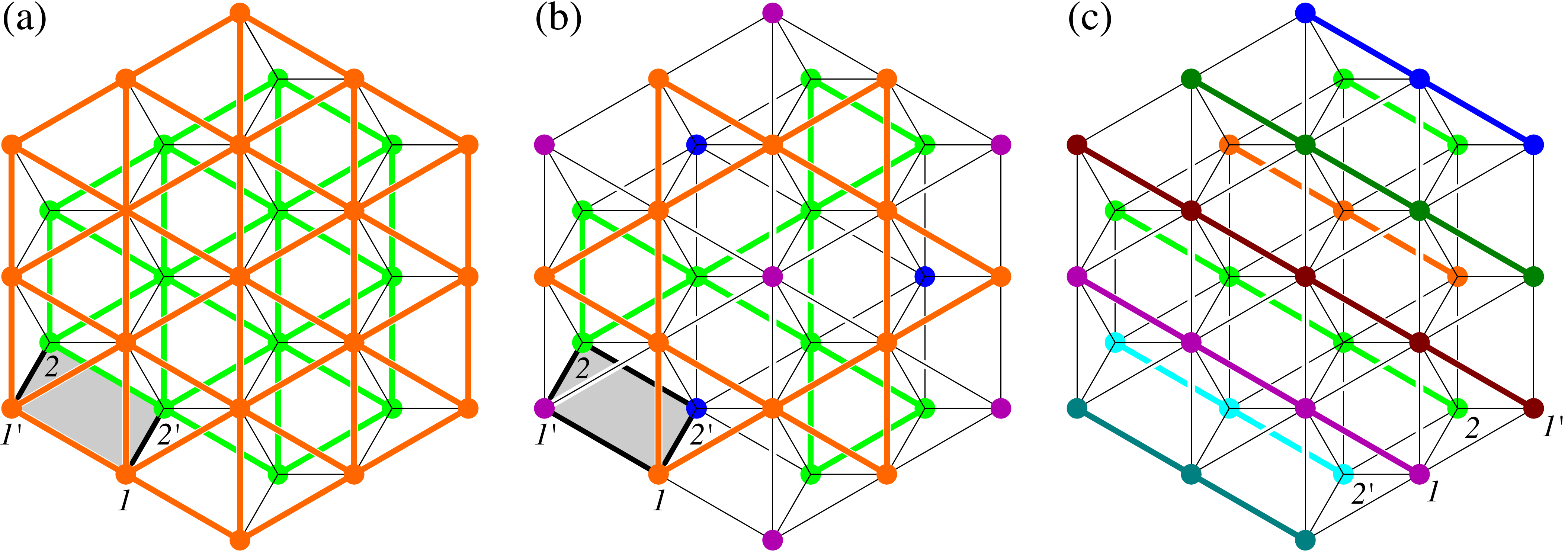}
	\caption{The face-centered-cubic lattice viewed from the $(111)$ direction. (a) Consecutive $(111)$ planes are highlighted in orange and green, the octahedra  connecting the planes are denoted by thin, black hexagons. One can recognize the 6 first neighbor ($J_1$) orange in-plane bonds emanating from the central site, and the 3 black $ J_1$ lines connecting it to the green plane (for the enumeration of all the interplane bonds, see Table~\ref{tab:111stacking}). The gray parallelogram shows the projection of a covering  square also depicted in Fig.~\ref{fig:motifs}(d). (b) Kagome sublattices of majority ferromagnetically ordered spins ($\mathbf{S}_{1}$ orange, and green $\mathbf{S}_{2}$) of the triangular $(111)$ planes. Minority spins are denoted by purple $\mathbf{S}_{1'}$ and blue $\mathbf{S}_{2'}$ dots. (c) Highlighted $(\bar{1}10)$ lines on the $(111)$ planes. Ferromagnetic order develops along these chains  in a class of solutions of the model in the $\Gamma-\Lambda-L$ point of the phase diagram, the effective interaction between these chains is given in Eq.~(\ref{triang_2D}).}
	\label{fig:kagome_stacking}
\end{figure*}

At the triple point $J_2=-J_1>0$, $J_3=0$ (the orange dot in in the phase diagram Fig.~\ref{fig:phase_diagram_descartes}(a))  the $\gsm$ is $\mathcal{M}^1_\Lambda$.  The possible ordering vectors $\Lambda(q,q,q)$ smoothly interpolate between $\Gamma(0,0,0)$  and $L\left(\onehalf,\onehalf,\onehalf\right)$, hence the shape of the manifold, see Fig.~\ref{fig:brillzone}(e). The Hamiltonian reads
\begin{equation}
\mathcal{H}=J_1\sum_{\left\langle i,j\right\rangle_1}\mathbf{S}_i\cdot\mathbf{S}_{j}-J_1\sum_{\left\langle i,j\right\rangle_2}\mathbf{S}_i\cdot\mathbf{S}_{j}.
\end{equation}
We can cover the lattice by signed squares, with signs  distributed according to Fig.~\ref{fig:motifs}(d):
\begin{equation}\mathcal{H}=-\frac{J_1}{4}\sum_{\textrm{square}}\left(\mathbf{S}_1-\mathbf{S}_2+\mathbf{S}_3-\mathbf{S}_4\right)^2+3J_1 N,\label{eq:square}
\end{equation} 
to every site we associate 3 squares, lying in each of the $\left\{ 100 \right\} $ planes. This way  every first and second neighbor bond is covered twice, see Table~\ref{tab:motifs}. This Hamiltonian is minimized if and only if the   
\begin{equation}
 \mathbf{S}_1-\mathbf{S}_2+\mathbf{S}_3-\mathbf{S}_4=\mathbf{0}
 \label{eq:squarecondition}
\end{equation}  
sums vanish on every square, and the  ground state energy per site is given by the additional constant $\varepsilon=+3J_1$.

The three equations on the signed squares are not independent, and  instead of them we can use the octahedra containing these squares to cover  the lattice (see Fig.~\ref{fig:exchanges}(d) and Fig.~\ref{fig:111_octahedra}(a)). Out of the 3 square equations  on orthogonal squares only two are independent per octahedron. Using the notations of Fig.~\ref{fig:111_octahedra}(a) for the sites of the octahedron, the ground state spin configuration shall satisfy the equations:
\begin{equation}
\mathbf{S}_{A} +  \mathbf{S}_{A'} =  \mathbf{S}_{B} +  \mathbf{S}_{B'} = \mathbf{S}_{C} +  \mathbf{S}_{C'} = 2 \mathbf{m} \,, \label{eq:SA+SA'}
\end{equation}
where $\mathbf{m}$ is proportional to the magnetization of an octahedron.
We can solve them introducing the $\mathbf{a}$, $\mathbf{b}$, and $\mathbf{c}$  vectors
\begin{subequations}
	\label{eq:eabc} 
	\begin{align}
	\mathbf{S}_{A} & =  \mathbf{m} + \mathbf{a}\,, &  \mathbf{S}_{A'} & =  \mathbf{m} - \mathbf{a}\,,\\
	\mathbf{S}_{B} & =  \mathbf{m} + \mathbf{b}\,, &  \mathbf{S}_{B'} & =  \mathbf{m} - \mathbf{b}\,, \\
	\mathbf{S}_{C} & =  \mathbf{m} + \mathbf{c}\,, &  \mathbf{S}_{C'} & =  \mathbf{m} - \mathbf{c}\,. 
	\end{align}
\end{subequations}
The length constraint $|\mathbf{S}_{A}|^2 = |\mathbf{S}_{A'}|^2 = 1$ becomes $\left(\mathbf{m} \pm \mathbf{a}\right)\cdot\left(\mathbf{m} \pm \mathbf{a}\right) = 1$, and similar equations for $\mathbf{b}$ and $\mathbf{c}$ hold. Adding and subtracting these equations, we get
\begin{subequations}
	\label{eq:abce_constraint}
	\begin{align}
	|\mathbf{m}|^2 + |\mathbf{a}|^2 &= 1\,,& 
	\mathbf{m}\cdot\mathbf{a} &= 0\,, \label{eq:eda}\\
	|\mathbf{m}|^2 + |\mathbf{b}|^2 &= 1\,,& 
	\mathbf{m}\cdot\mathbf{b} &= 0\,, \\
	|\mathbf{m}|^2 + |\mathbf{c}|^2 &= 1\,,& 
	\mathbf{m}\cdot\mathbf{c} &= 0\,. 
	\end{align}
\end{subequations}
The   vectors above can differ on every octahedron provided they satisfy some compatibility conditions: sharing an edge creates a dependence among them, but we omit the octahedron index for clarity. For $N_s$ component spins the $\mathbf{a}, \mathbf{b}, \mathbf{c}$, and $\mathbf{m}$ counts $4N_s$ degrees of freedom, and there are $6$ constraints in Eq.~(\ref{eq:abce_constraint}), so we expect $4N_s-6$ free continuous parameters to describe  the ground state of an octahedron.
 
 \begin{table}[b]
	\caption{Number of bonds connecting a single point to its neighbors on the nearby $(111)$ planes, see Fig.~\ref{fig:exchanges}(b) and especially Fig.~\ref{fig:kagome_stacking}(a). The first column gives the separation of consecutive planes: "0" means the $(111)$ plane containing the chosen point, "1" means the two first neighbor $(111)$ planes, "2" means the two second neighbor $(111)$ planes. The second column gives the number of first neighbor bonds connecting the chosen point to the planes of the given separation, and so on.}
	\begin{center}
		\begin{ruledtabular}
			\begin{tabular}{cccc}
				Separation & $J_1$ & $J_2$ & $J_3$ \\ \hline
				0 & 6 & 0 & 6 \\
				1 & 6 & 6 & 12 \\
				2 & 0 & 0 & 6
			\end{tabular}
		\end{ruledtabular}
	\end{center}
	\label{tab:111stacking}
\end{table}
 
 A ferromagnetic order  trivially satisfies the rule given by Eq.~(\ref{eq:SA+SA'}), and this is not surprising: the $\Gamma(0,0,0)$ point is part of this manifold. In the ferromagnet $\mathbf{a}=\mathbf{b}=\mathbf{c}=\mathbf{0}$. If $\mathbf{m}=\mathbf{0}$ we get an $L\left(\onehalf,\onehalf,\onehalf\right)$ order: all the Type II states described in  Sec.~\ref{Lsection} and shown in Fig.~\ref{fig:XL_order}(b) can be constructed this way. Among others one can  choose the single ordering vector $(\pi,\pi,\pi)$ and get  a set of alternating $(111)$ ferromagnetic planes: see Eq.~(\ref{Lfourq}) with only the amplitude $\mathbf{S}^0_{\mathbf{L}_1}$ nonvanishing, and  Fig.~\ref{fig:XL_order}(b) with $\mathbf{S}_B=-\mathbf{S}_C=\mathbf{S}_D=\mathbf{S}_A$. This suggests other possible candidate ground states:  we can try to construct a family of ground states by stacking ferromagnetic $(111)$ planes, these planes form triangular lattices and they are depicted in Fig.~\ref{fig:kagome_stacking}(a). 

Assuming a  state consisting of ferromagnetically ordered $(111)$ planes, and representing  a plane by a single effective spin $\mathbf{s}_i$ of unit length, where now the index ``$i$" enumerates the consecutive planes one can derive an effective one-dimensional model:
\begin{eqnarray}
\mathcal{H}^{(111)}_{\Lambda} &=&
\frac{1}{4}\left(6J_1+6J_2+12J_3\right)\sum_{i=1}^{L^{(111)}} \mathbf{s}_{i }\cdot \mathbf{s}_{i +1}
\nonumber\\ 
&+&
\frac{1}{2}\left(6J_1+6J_3\right)L^{(111)},
\label{lambda_1d}
\end{eqnarray}
where the effective exchanges can be inferred either from Table~\ref{tab:111stacking} or from Fig.~\ref{fig:kagome_stacking}(a), $L^{(111)}$ is the number of $(111)$ planes in the crystal, and the additional constant derives from the in-plane couplings. Substituting the actual values $J_2=-J_1$ and $J_3=0$, we see that the first term  disappears, so the planes disentangle, and the second term gives $+3J_1$ for the correct ground state energy per site of the original model. The resulting ground state is of the form $F_1 F_2 F_3 F_4\dots$, where $F_i$ denotes the  independent ferromagnetic planes. This independence of planes can be further rationalized by noting that only the first neighbor  planes are connected by the covering squares (see Fig.~\ref{fig:kagome_stacking}(a)).  
Using the notation of Fig.~\ref{fig:kagome_stacking}(a) one can see that $\mathbf{S}_{1'}=\mathbf{S}_{1}$ and $\mathbf{S}_{2'}=\mathbf{S}_{2}$, since these pairs lie on FM planes. Therefore Eq.~(\ref{eq:SA+SA'}) is automatically satisfied since $\mathbf{S}_{1}+\mathbf{S}_{2}=\mathbf{S}_{1'}+\mathbf{S}_{2'}$.

Such a state can be cooked up by choosing  ordering vectors solely from the  $(q,q,q)$ line of the $\mathcal{M}^1_\Lambda$ manifold:
\begin{equation}
\mathbf{S}_{\left(x,y,z\right)}=\sum_{q \in [-\pi,\pi)}\mathbf{S}^0_{\left(q,q,q\right)}e^{-\imath q \left( x+y+z\right)},
\end{equation}
and depending on the complexity of the real space pattern, any symmetric set of points on the $\Lambda(q,q,q)$ line can be present in the expansion, as long as we care about the choice of the Fourier amplitudes to produce a real space pattern of unit length spins. Of course we could have chosen any of the symmetry related $ \left\langle 111\right\rangle $ directions. 

In the finite cluster Ising solution (see Appendix \ref{sec:Ising_results} for details) we have found the $\left\{ 111 \right\} $ stacking of independent FM planes: these involve only one line of ordering vectors of $\mathcal{M}_\Lambda^1$. We have found another type of solution where  up and down spins form two interpenetrating pyrochlore lattices: the unit cell consists of 8 sites, see Fig.~\ref{fig:XL_order}(b) with  $\mathbf{S}_A=\mathbf{S}_B=\mathbf{S}_C=\mathbf{S}_D=1$ and $\mathbf{S}_{\bar A}=\mathbf{S}_{\bar B}=\mathbf{S}_{\bar C}=\mathbf{S}_{\bar D}=-1$.

In the numerical simulations on planar spins we found another interesting class of ground states: a  $3/4$ majority fraction of the spins on $(111)$ planes ordered ferromagnetically on a \textit{kagome} sublattice of the triangular layer (see the orange and green dots in Fig.~\ref{fig:kagome_stacking}(b)), and the minority spins (purple and blue dots)  seemed to be independent of the majority spins, and  a similar structure was formed on every $(111)$ plane. In the following we use the notations of Fig.~\ref{fig:kagome_stacking}(b). We can exploit the \textit{octahedral constraint} of Eq.~(\ref{eq:abce_constraint}): we assume the kagome-style ordering described above on consecutive planes indexed by 1, 2, 3,\dots, 
and monitor how the consequences of the constraints propagate as we step  down on octahedra between the planes. We fix the majority spins $\mathbf{S}_1$ and minority spins $\mathbf{S}_{1'}$ on the first plane. For Ising spins fixing the spins on layer ``1" determines the spins on all of the consecutive planes, and the resulting pattern is the quadruple-$\mathbf{Q}$ order described in the previous paragraph.  For $XY$ spins, if $\mathbf{S}_{1'} \neq -\mathbf{S}_{1}$ we have four possible choices for $\left\{\mathbf{S}_{2'},  \mathbf{S}_{2}\right\}$, resulting in a $\mathbb{Z}_4$ degree of freedom. If $\mathbf{S}_{1'} = - \mathbf{S}_{1}$ we are free to choose any   $\mathbf{S}_{2'}= -\mathbf{S}_{2}$, resulting an $O(2)$ degree of freedom. For $O(3)$ spins, if $\mathbf{S}_{1'} \neq -\mathbf{S}_{1}$ we have an $O(2)\times \mathbb{Z}_2$   freedom of choice for $\left\{\mathbf{S}_{2'},  \mathbf{S}_{2}\right\}$. If $\mathbf{S}_{1'} = - \mathbf{S}_{1}$ we are free to choose any   $\mathbf{S}_{2'}= -\mathbf{S}_{2}$ parametrized by the unit sphere $S^2$.

In the numerical study on the planar spins we have found ground states formed by seemingly independent ferromagnetic chains~\cite{Anderson-1956,McClarty-2015} lying in the $\left\langle 111 \right\rangle$ planes, pointing in one of the $\left\langle 110 \right\rangle$ directions, a set of such lines are depicted in Fig.~\ref{fig:kagome_stacking}(c). This numerical finding suggests the following strategy: we \textit{assume} a FM ordering along the $(110)$ chains (the bonds along the chain are $J_1<0$ ferromagnetic), and derive an effective two dimensional model where we substitute the chains  by a single effective spin $\mathbf{s}_i $ of unit length, where the index ``$i$" refers to points of the  lattice  formed by the chains, for a picture of the  lattice see Fig.~\ref{fig:111_octahedra}(b). The effective interactions $K_{\delta}$ ($\bm{\delta}$ points to the neighboring chains) are in general very complicated (each point is connected to 16 others, usually by multiple bonds) but the actual exchange parameters ($J_2=-J_1$ and $J_3=0$) come to help us and result in a remarkably simple set of nonvanishing effective exchanges:
\begin{subequations}
\begin{eqnarray}
&&K_1=2J_1+2J_3=2J_1<0,\\
&&K_2=J_1+2J_2=-J_1>0,
\end{eqnarray} 
\end{subequations}
 where the indices refer to the bonds depicted in Fig.~\ref{fig:111_octahedra}(b). These values can be inferred from the gray rhombus in Fig.~\ref{fig:111_octahedra}(b) depicting the projection of one of the covering squares, also shown in Fig.~\ref{fig:motifs}(d) (in order to have the correct effective exchanges one needs to take into account all the three differently oriented squares). This lattice is topologically equivalent to a first and second neighbor FM-AFM model with bond strengths  $K_{2}=-K_1/2>0$ on the square lattice. The effective two dimensional model reads:
\begin{equation}
\mathcal{H}_{2D}^{(111)}=\frac{1}{2}\sum_{i,\bm{\delta}} K_{\delta}  \mathbf{s}_i \cdot\mathbf{s}_{i +\bm{\delta}}+J_1 N^{(111)},
\label{triang_2D}
\end{equation}
 where the additional constant derives from the couplings within a chain. This model is strongly frustrated having a codimension-one  $\gsm$, depicted in Fig.~\ref{fig:111_octahedra}(c): this manifold is nothing but the section of $\mathcal{M}^1_\Lambda$ with the $(110)$ $\mathbf{q}$-plane passing through the origin (the $q_{110}=0$ plane with notation of Fig.~\ref{fig:111_octahedra}(b) and (c)). The large ground state degeneracy can be further rationalized by noting that this Hamiltonian can be written as a sum of complete squares on signed rhombi (see the gray rhombus in Fig.~\ref{fig:111_octahedra}(b)): the resulting  rhombus rule $\mathbf{S}_{A}+\mathbf{S}_{A'}-\mathbf{S}_{B}-\mathbf{S}_{B'}=\mathbf{0}$ is just the signed square rule inherited from the  three dimensional problem.  Any state obeying the rhombus rule is a ground state for the $(110)$ chains, and this is consistent with the numerical finding of seemingly random chains in the $XY$-model that actually obey the rhombus rule.

To summarize we propose the following candidate ground states for the $\Gamma(0,0,0)-\Lambda(q,q,q)-L(\onehalf,\onehalf,\onehalf)$ triple point:
\begin{itemize}
\item Stacking of independent ferromagnetic $\left\{ 111 \right\} $ planes in the style $F_1 F_2 F_3 F_4\dots$. This type of ordering is realized in all the Ising, $XY$ and $O(3)$ models.
\item Almost independent ferromagnetic kagome sublattices in the  $\left\{ 111 \right\} $  planes. This type of ordering is realized in the $XY$ and $O(3)$ models. For Ising spins this order reduces to the commensurate quadruple-$\mathbf{L}$ structure of intercalating pyrochlore lattices. 
\item Interacting ferromagnetic chains in the $\left\langle 110 \right\rangle $ directions, these are absent in the Ising models.	
\end{itemize}

\subsection{The $\Gamma(0,0,0)-\Delta(q,0,0)-X(1,0,0)$ triple point: the one dimensional $\mathcal{M}^1_\Delta$ ground state manifold}
\label{gxdeltasubsection}

\begin{figure}
	\centering
	\includegraphics[width=.92\columnwidth]{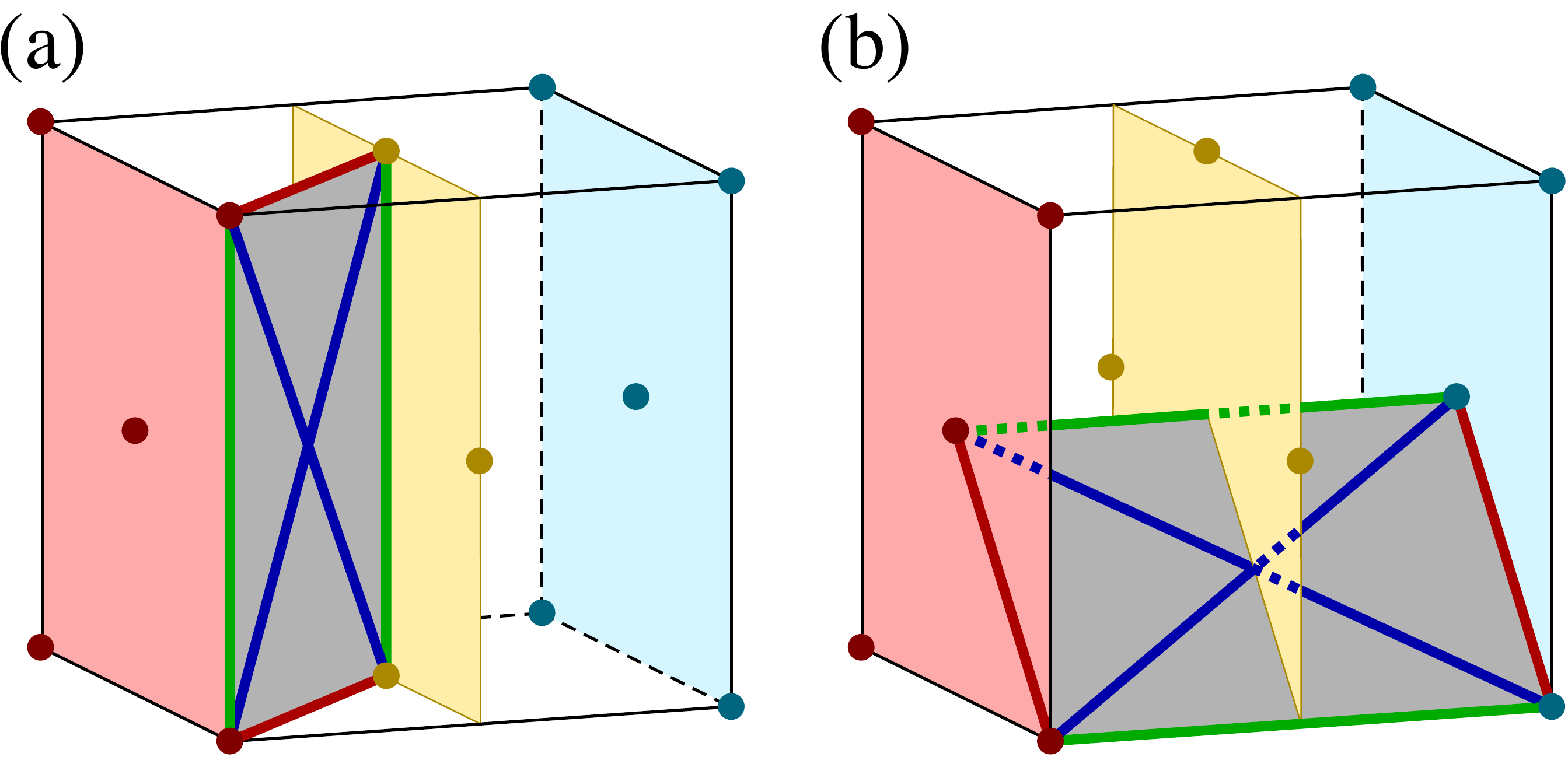}
	\caption{Consecutive $(100)$ planes of the face-centered-cubic lattice. (a) First neighbor planes are connected with signed rectangle motifs rect$_1$ and rect$_2$ (see Fig.~\ref{fig:motifs}(b) and (c)), first (red, $J_1$), second (green, $J_2$) and third (blue, $J_3$) bonds connecting the consecutive planes, only one of the possible 4 orientations of the connecting rectangles is shown.  Only one in-plane $J_2$ bond is shown in the planes. (b) Second neighbor planes are connected with signed rectangle motifs rect$_1$ and rect$_2$ (see Fig.~\ref{fig:motifs}(b) and (c)), first (red, $J_1$), second (green, $J_2$) and third (blue, $J_3$) bonds connecting the second neighbor planes, only one of the possible 4 orientations of the connecting rectangles is shown in the planes. Only one in-plane $J_1$ bond is shown.}
	\label{fig:100_planes}
\end{figure}

\begin{figure}
	\centering
	\includegraphics[width=.95\columnwidth]{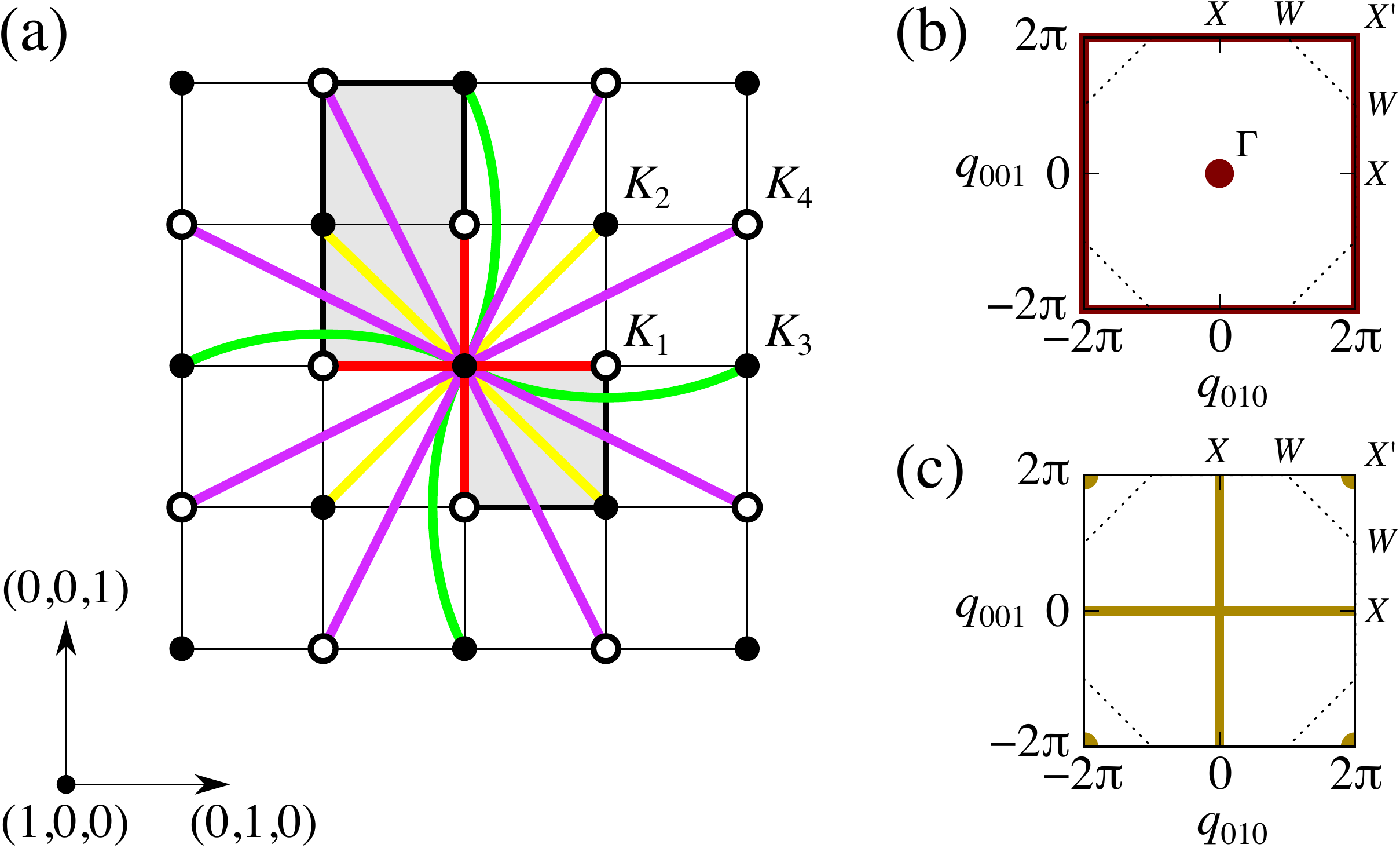}
	\caption{ Assuming $(100)$-directed, ferromagnetic chains we get a two dimensional model Eq.~(\ref{square_2D}) on the square lattice for the effective spins $\mathbf{s}_i $ representing the magnetizations of the chains. (a) View of the face-centered-cubic lattice from the $(100)$ direction. The black and white dots represent the $(100)$ chains, lattice points on the differently colored chains are shifted by a vector $(1/2,0,0)$, but the points are equivalent in the two dimensional effective model. Primitive vectors of the square lattice are $(1/2,0)$ and $(0,1/2)$. Effective interactions $K_\delta$ in Eq.~(\ref{square_2D}) are represented by colored bonds. The gray rectangle shows the projection of one covering rectangle motif of the original model, see Fig.~\ref{fig:motifs}(b) and (c). Note that black and white dots here have nothing to do to the sign distribution on rectangles. The gray square shows the projection of the tetrahedron in the original model, see Fig.~\ref{fig:motifs}(a). (b) Brillouin zone of the lattice depicted in Fig.~\ref{fig:100_chains}(a), together with the ground state manifold (red square) of the Hamiltonian Eq.~(\ref{square_2D}), on the $X-W$ phase boundary of the original model. At the $\Gamma-X-W$ triple point this manifold extends with red $\Gamma(0,0)$ point. This manifold is  the section of $\mathcal{M}^1_Z$ (see Fig.~\ref{fig:brillzone}(b)) with the $(100)$ $\mathbf{q}$-plane passing through the origin. Symmetry points of the original three dimensional BZ (see Fig.~\ref{fig:brillzone}(a)) are shown. (c) Brillouin zone of the lattice depicted in Fig.~\ref{fig:100_chains}(a), together with the ground state manifold (dark yellow cross) of the Hamiltonian Eq.~(\ref{square_2D}), at the $\Gamma-\Delta-X$ triple point. This manifold is  the section of $\mathcal{M}^1_\Delta$ (see Fig.~\ref{fig:brillzone}(d)) with the $(100)$ $\mathbf{q}$-plane passing through the origin. Symmetry points of the original three dimensional BZ (see Fig.~\ref{fig:brillzone}(a)) are shown.}
	\label{fig:100_chains}
\end{figure}
The triple point $J_2=2J_1$, $J_3=-J_1/2$, $J_1<0$ is denoted by a yellow dot in the phase diagram Fig.~\ref{fig:phase_diagram_descartes}(a).  The possible ordering vectors $\Delta(q,0,0)$ smoothly interpolate between $\Gamma(0,0,0)$  and $X(1,0,0)$, hence the shape of the manifold $\mathcal{M}^1_\Delta$, see Fig.~\ref{fig:brillzone}(d). The Hamiltonian reads
\begin{equation}
\mathcal{H}=J_1\sum_{\left\langle i,j\right\rangle_1}\mathbf{S}_i\cdot\mathbf{S}_{j}+2J_1\sum_{\left\langle i,j\right\rangle_2}\mathbf{S}_i\cdot\mathbf{S}_{j}-\frac{J_1}{2}\sum_{\left\langle i,j\right\rangle_3}\mathbf{S}_i\cdot\mathbf{S}_{j}.
\end{equation}
We can cover the lattice by signed rectangles where the signs are distributed according to Fig.~\ref{fig:motifs}(c):
\begin{equation}
\mathcal{H}=-\frac{J_1}{4}\sum_{\textrm{rect}_2}\left(\mathbf{S}_1-\mathbf{S}_2+\mathbf{S}_3-\mathbf{S}_4\right)^2+6J_1 N. \label{eq:rect2}
\end{equation}
Since $-J_1/4>0$ this Hamiltonian is minimized if and only if $\mathbf{S}_1-\mathbf{S}_2+\mathbf{S}_3-\mathbf{S}_4=\mathbf{0}$ on every rectangle, 
and the ground state energy per site is $+6J_1$. A ferromagnetic order trivially satisfies the above rectangle rule (the $\Gamma(0,0,0)$ point is part of the manifold), as does any $X(1,0,0)$ (Type I) order. If we  choose a single arm of the $X$-star, i.e. only $\xi \neq 0$ in Eq.~(\ref{Xspins}) we get state of alternating ferromagnetic $(100)$ planes, see  Eq.~(\ref{tetraruleX}) with $\eta=\zeta=0$ and Fig.~\ref{fig:XL_order}(a) with $-\mathbf{S}_B=-\mathbf{S}_C=\mathbf{S}_D=\mathbf{S}_A$. This suggests the possibility to build a state of $(100)$ ferromagnetic planes (the sites on these planes form square lattices). Although one cannot \textit{a priori} exclude antiferromagnetism on the planes: choosing $\xi=0$ but $\eta \neq 0$ and $\zeta \neq 0$ in Eq.~(\ref{tetraruleX}) results  in an antiferromagnetic pattern on the  $(100)$ planes, with $\mathbf{S}_B=-\mathbf{S}_C$ and $\mathbf{S}_D=-\mathbf{S}_A$ in Fig.~\ref{fig:XL_order}(a).

\begin{table}
	\caption{Number of bonds connecting a single point to its neighbors on the nearby $(100)$ planes, see Fig.~\ref{fig:exchanges}(b) and especially Fig.~\ref{fig:100_planes}. The first column gives the separation of consecutive planes: "0" means the $(100)$ plane containing the chosen point, "1" means the two first neighbor $(100)$ planes (see Fig.~\ref{fig:100_planes}(a)), "2" means the two second neighbor $(100)$ planes (see Fig.~\ref{fig:100_planes}(b)). The second column gives the number of first neighbor bonds ($J_1$) connecting the chosen point to the points on the neighboring planes of the given separation, an so on.}
	\begin{center}
		\begin{ruledtabular}
			\begin{tabular}{cccc}
				Separation & $J_1$ & $J_2$ & $J_3$ \\ \hline
				0 & 4 & 4 & 0 \\
				1 & 8 & 0 & 16 \\
				2 & 0 & 2 & 8 
			\end{tabular}
		\end{ruledtabular}
	\end{center}
	\label{tab:100stacking}
\end{table}

Along the same line of reasoning presented in Subsection \ref{Lambda_subsection}. we can construct a family of states of ferromagnetic  $(100)$ planes, see Fig.~{\ref{fig:100_planes}}. Representing  a plane by a single effective spin $\mathbf{s}_i $ of unit length, where now the index ``$i$" enumerates the consecutive planes one can derive an effective one-dimensional model:
\begin{eqnarray}
\mathcal{H}^{(100)}_{X} &=&
\frac{1}{4}\left(8J_1+16J_3\right)\sum_{i=1}^{L^{(100)}} \mathbf{s}_{i }\cdot \mathbf{s}_{i +1}
\nonumber\\ 
&+&
\frac{1}{4}\left(2J_2+8J_3\right)\sum_{i=1}^{L^{(100)}} \mathbf{s}_{i }\cdot \mathbf{s}_{i +2}\nonumber\\
&+&
\frac{1}{2}\left(4J_1+4J_2\right)L^{(100)},
\label{X_1d}
\end{eqnarray}
where the effective exchange can be inferred either from Table~\ref{tab:100stacking} or Fig.~\ref{fig:100_planes}, and $L^{(100)}$ is the number of $(100)$ planes in the crystal. Substituting the actual values $J_2=2J_1$ and $J_3=-J_1/2$, we see that the first two terms  disappear, so the planes disentangle, and the last term gives $+6J_1$ for the correct ground state energy per site of the original model. The resulting ground state is of the form $F_1 F_2 F_3 F_4\dots$, where $F_i$ denotes the  independent ferromagnetic planes. This independence of the planes can be further rationalized by noting that both the first and second neighbor  planes are connected by the covering rectangles,  and  the rectangle rule is satisfied \textit{bondwise} on every ferromagnetic plane: see the rectangles in Fig.~\ref{fig:100_planes}, and remember that the signs are distributed according to Fig.~\ref{fig:motifs}(c) and the planes are ferromagnetic.

Such a state of ferromagnetically aligned independent $(100)$ planes can be Fourier decomposed as
\begin{equation}
\mathbf{S}_{\left(x,y,z\right)}=\sum_{q \in [-2\pi,2\pi)}\mathbf{S}_{\left(q,0,0\right)}e^{-\imath q x},
\end{equation}
and depending on the complexity of the real space pattern, any symmetric set of points on the $\Delta(q,0,0)$ line can be present in the expansion. Of course we could have chosen any of the symmetry related $ \left\langle 100\right\rangle $ directions.

Stacking antiferromagnetic planes is more restrictive:  a rect$_2$ can connect neighboring planes by $J_1$ bonds, in this case the $J_2$ bonds lie in-plane (and connect parallel spins, automatically satisfying the $J_2$ in-plane bonds), see Fig.~\ref{fig:100_planes}(a). Another possibility for  a rect$_2$ to connect second neighbor planes by $J_2$ bonds, and the $J_1$ bonds lie in-plane (and connect antiparallel spins), see Fig.~\ref{fig:100_planes}(b). Second neighbor AFM planes are \textit{locked}: they have to have the same AFM pattern to satisfy the rectangle rule. This leaves us with the two possibilities of stacking: an alternating set of two independent AFM planes $A_1 A_2 A_1 A_2\dots$, or we can put independent FM planes between the  AFM ones: $A_1 F_1 A_1 F_2 A_1 F_3\dots$. 

In the finite cluster Ising solutions (see Appendix \ref{sec:Ising_results} for details) we have found the $\left\{ 100 \right\} $ stacking of independent FM planes: $ F_1 F_2 F_3 F_4\dots$ and the FM stacking with intercalating AFM planes: $A F_1 A F_2 A\dots$. The alternating  AFM stacking is missing here: for Ising spins it is an alternating FM stacking  $F_1 F_2 F_1 F_2\dots$ viewed from a perpendicular direction. 

We have performed numerical simulations for planar spins ($XY$-model): besides the aforementioned planar structures we found (seemingly disordered) ferromagnetic \textit{chains} along the $ \left\langle 100\right\rangle $ directions, corresponding to a Fourier pattern of points on \textit{two} perpendicular lines of  $\mathcal{M}_\Delta^1$ in $\mathbf{q}$-space. Thus we try to construct a family of states consisting of ferromagnetic  $(100)$ directed chains. These chains sit on a square lattice  of primitive vectors $(0,1/2,0)$ and $(0,0,1/2)$, see Fig.~\ref{fig:100_chains}(a). We represent  a chain by a single effective spin $\mathbf{s}_i $ of unit length, where now the indices $i$ refer to points of the square lattice and $\bm{\delta}$ to the neighbors of the lattice, and we map the system to the effective two dimensional model:
\begin{eqnarray}
&&\mathcal{H}_{2D}^{(100)}=\frac{1}{2}\sum_{i,\bm{\delta}} K_{\delta}  \mathbf{s}_i \cdot\mathbf{s}_{i +\bm{\delta}}+J_2N^{(100)},\label{square_2D}\\
&&K_1=2J_1, \ K_2=J_1+2J_3, \ K_3=J_2, \ K_4=2J_3.
\end{eqnarray}
For the $\mathcal{M}^1_{\Delta}$ manifold  the effective interactions are: $K_1=2J_1$, $K_2=0$, $K_3=2J_1$, and  $K_4=-J_1$, see Fig.~\ref{fig:100_chains}(a) for a picture of the generated interactions. Strong, ferromagnetic $J_2=2J_1<0$ bonds connect along the chains and $N^{(100)}$ is the number of $(100)$ chains in the crystal. This model has a codimension-one $\gsm$: in the BZ of the square lattice  the minima reside on the cross connecting the BZ center to the midpoints of the zone boundary  together with the zone corner, see Fig.~\ref{fig:100_chains}(c), note that this manifold is nothing but the intersection of $\mathcal{M}_\Delta^1$ with the $(100)$ $\mathbf{q}$-plane passing through the origin. The energy per site is $6J_1=-6\left| J_1 \right|$ ($4J_1$ comes from the interactions and $J_2=2J_1$ from the additional constant). This Hamiltonian can also be written as a sum of squares on signed rectangles inherited from the rect$_2$-s projected to the $(100)$ plane, see Fig.~\ref{fig:100_chains}(a). This is consistent with the numerical findings of the planar spins: all the configurations found obeyed this projected rectangle rule, but seemed otherwise disordered. 

To summarize we propose the following candidate ground states for the $\Gamma(0,0,0)-\Delta(q,0,0)-X(1,0,0)$ triple point:
\begin{itemize}
	\item Stacking of independent ferromagnetic $\left\{ 100\right\} $ planes in the style $F_1 F_2 F_3 F_4\dots$. This type of ordering is realized in all the Ising, $XY$ and $O(3)$ models.
	\item Stacking of independent ferromagnetic layers separated by the same antiferromagnetic layers on the $\left\{ 100\right\} $ planes in  an $A F_1 A  F_2 A  F_3 \dots$ style. This type of ordering is realized in all the Ising, $XY$ and $O(3)$ models.
	\item  Stacking of an alternating set of two independent $(100)$ AFM planes $A_1 A_2 A_1 A_2\dots$, reaalized in the $XY$ and $O(3)$ models. 
	\item Interacting ferromagnetic chains in the $\left\langle 100 \right\rangle $ directions, these are absent in the Ising models.	
\end{itemize}

\subsection{The $X(1,0,0)-W(1,\frac{1}{2},0)$ phase boundary (with endpoints): the one dimensional $\mathcal{M}^1_Z$ ground state manifold}
\label{XWsubsection}
On the phase boundary line separating  the $W\left(1,\onehalf,0\right)$ and $X(1,0,0)$ phases (the red line in Fig.~\ref{fig:phase_diagram_descartes}(b)) the ordering vectors of the possible ground states form the  one dimensional $\mathcal{M}^1_Z$ manifold, see Fig.~\ref{fig:brillzone}(b). Note that this manifold connects the points $X(1,0,0)$ and $W\left(1,\onehalf,0\right)$ in the BZ. On this phase boundary given by  $J_1>0$, $J_3=J_2/4$, $-2\leq J_2\leq 0$ the Hamiltonian reads:
\begin{equation}
\mathcal{H}=J_1\sum_{\left\langle i,j\right\rangle_1}\mathbf{S}_i\cdot\mathbf{S}_{j}+J_2\sum_{\left\langle i,j\right\rangle_2}\mathbf{S}_i\cdot\mathbf{S}_{j}+\frac{J_2}{4}\sum_{\left\langle i,j\right\rangle_3}\mathbf{S}_i\cdot\mathbf{S}_{j}.\label{eq:mzcover1}
\end{equation}
Since we have two free parameters ($J_1$ and $J_2$) expressing this Hamiltonian as the sum of complete squares on  finite motifs is a little bit tricky. Here we use the elementary edge sharing tetrahedra of the fcc lattice and signed rectangles: Fig.~\ref{fig:motifs}(a) and (b). Two tetrahedra and two rectangles share a nearest neighbor bond, and four rectangles share a second neighbor bond, and each third neighbor bond is covered once by a rectangle, see Table~\ref{tab:motifs}. "Signed" means that in the complete squares on these  rectangles we associate a minus sign to the spins sitting on the sites denoted by white dots in Fig.~\ref{fig:motifs}(b), and plus signs to the black dots. The Hamiltonian becomes:
\begin{align}
\mathcal{H}&=\left(\frac{J_1}{4}+\frac{J_2}{8}\right)\sum_{\textrm{tetra}}\left(\mathbf{S}_1+\mathbf{S}_2+\mathbf{S}_3+\mathbf{S}_4\right)^2
\nonumber\\&\phantom{=}
-\frac{J_2}{8}\sum_{\textrm{rect}_1}\left(\mathbf{S}_1'+\mathbf{S}_2'-\mathbf{S}_3'-\mathbf{S}_4'\right)^2
\nonumber\\&\phantom{=}
+2(J_2-J_1)N. 
\label{eq:rect1}
\end{align}
Since the prefactors are all positive the Hamiltonian is minimized if and only if the spins sum up to zero on every tetrahedron and on every rectangle (with the appropriate signs), and the additional constant $2(J_2-J_1)N$
gives the ground state energy. The spin sum on rect$_1$ can be built by subtracting the spin sums of two edge-sharing tetrahedra, so every configuration that satisfies the tetrahedron rule automatically satisfies rectangle rule. 

At $J_2=J_3=0$ we do not need the rectangles, and only the tetrahedron rule survives~\cite{henley_fcc} (this is the point where the $\Sigma$-phase touches the $X-W$ line). As an example~\cite{alexander1980}, we can make ground states of $(100)$ \textit{independent} antiferromagnetically ordered planes in this endpoint: the spins form a checkerboard pattern on the planes of an $A_1 A_2 A_3\dots$  stacking  style, where $A_i$ refers to the $i$th antiferromagnetic plane. This construction extends without modification to the whole $X(1,0,0)-W\left(1,\onehalf,0\right)$ boundary.  This configuration is indeed a ground state, since both the tetrahedron and the rectangle rules are satisfied \textit{bondwise} (for the motifs and sign distribution see Fig.~\ref{fig:motifs}(a) and (b), for the planes connected by the rectangles see Fig.~\ref{fig:100_planes}(a) and (b)): spins on first neighbor bonds in a $(100)$ plane are antiparallel and on second neighbor bonds are parallel. Just like in Section \ref{gxdeltasubsection} the planes disentangle, and the in-plane contribution of interactions gives the correct ground state energy per spin as $2(J_2-J_1)$.

A configuration of this stacking of AFM $(100)$ planes can be Fourier expanded by combining ordering vectors from the $(100)$ directed lines of the $\mathcal{M}_Z^1$ manifold (see Fig.~\ref{fig:brillzone}(b)) and this nicely explains the shape of $\mathcal{M}^1_Z$. Of course we also could have chosen the stacking direction of planes as $(010)$ or $(001)$. These states appear in the Ising solution, of course there are only two choices of AFM configurations in each plane. 

In the  simulations of planar spins we found $(100)$-directed  FM chains, in a seemingly disordered distribution. Applying the effective two dimensional model for the chains forming a $(100)$ square lattice one gets Eq.~(\ref{square_2D}) with effective interactions $K_1=2J_1$,  $K_2=J_1+J_2/2$, $K_3=J_2$, and  $K_4=J_2/2$ with  $J_1>0$ and $-2 < J_2\leq 0$ (here we exclude the $J_2=-2$ $\Gamma-X-W$ triple point, and discuss it in Subsection \ref{gxwsubsection}), see Fig.~\ref{fig:100_chains}(a) for a picture of the generated interactions. This model has a codimension-one $\gsm$: in the BZ of the square lattice  the minima reside on the BZ boundary, see Fig.~\ref{fig:100_chains}(c), but be careful: the zone center $\Gamma(0,0)$ is \textit{not} part of the manifold. The Fourier transform of the effective exchange has a local but not global minimum at the BZ center, which gets lower and lower as we move along the $X-W$ line towards the $\Gamma-X-W$ point, and this minimum becomes degenerate with the $\gsm$ on the BZ boundary as we finally reach $J_2=-2J_1$. Note that this manifold is nothing but the intersection of $\mathcal{M}_Z^1$ with the $(100)$ $\mathbf{q}$-plane passing through the origin. The ground state energy per site is $2(J_2-J_1)$ (where $J_2-2J_1$ comes from the interactions and $J_2$ from the additional constant in Eq.~(\ref{square_2D})). This Hamiltonian can also be written as a sum of squares on signed rectangles inherited from the tetrahedra and rect$_1$-s projected to the $(100)$ plane, see Fig.~\ref{fig:100_chains}(a) (the projected tetrahedron rule prohibits $\Gamma(0,0)$ being a global minimum). This is consistent with the numerical findings of the planar spins: all the configurations found obeyed this projected rectangle and tetrahedron rule, but seemed otherwise disordered.

To summarize we have found the following candidate ground states for the $ X(1,0,0)- W\left( 1,\onehalf,0 \right)$ line:
\begin{itemize}
	\item Stacking of independent antiferromagnetic $\left\{ 100\right\} $ planes in the style $A_1 A_2 A_3 A_4\dots$. This type of ordering is realized in all the Ising, $XY$ and $O(3)$ models. 
	\item Interacting ferromagnetic chains in the $\left\langle 100 \right\rangle $ directions, these are absent in the Ising models.	
\end{itemize}

At the   endpoint $J_2=-2J_1$ and $J_3=-J_1/2$ the tetrahedron rule vanishes in Eq.~(\ref{eq:rect1}), and the only constraint is that spins on rectangles have to satisfy the equation $\mathbf{S}_1'+\mathbf{S}_2'-\mathbf{S}_3'-\mathbf{S}_4'=\mathbf{0}$, this less restrictive condition offers other possibilities (e.g. the appearance of a net magnetization),  we devote the next subsection to its analysis.

\subsection{The $\Gamma(0,0,0)-X(1,0,0)-W\left(1,\frac{1}{2},0\right)$ triple point: the one dimensional $\mathcal{M}^1_Z\cup \Gamma$ ground state manifold} 
\label{gxwsubsection}
 
The $\Gamma(0,0,0)-X(1,0,0)-W\left(1,\frac{1}{2},0\right)$ triple point bears striking resemblance to the triple point $\Gamma(0,0,0)-\Delta(q,0,0)-X(1,0,0)$ presented in Subsection \ref{gxdeltasubsection}, and has a much richer structure than the rest of the $X(1,0,0)-W\left(1,\onehalf,0\right)$ line. Here the parameters are $J_2=-2J_1$, $J_3=-J_1/2$, $J_1>0$, and the $\gsm$ is $\mathcal{M}^1_Z\cup \Gamma$,  see Table~\ref{tab:triplepoints}. See Fig.~\ref{fig:brillzone}(b) for the degenerate manifold, and Fig.~\ref{fig:phase_diagram_descartes}(b) for the point in the phase diagram: the red dot where the $X-W$ boundary line hits the $\Gamma(0,0,0)$ phase.
We can cover the lattice by signed rectangles (here the tetrahedron rule does not apply), where  the signs are distributed according to Fig.~\ref{fig:motifs}(b) now:
\begin{equation}
\mathcal{H}=\frac{J_1}{4}\sum_{\textrm{rect}_1}\left(\mathbf{S}_1+\mathbf{S}_2-\mathbf{S}_3-\mathbf{S}_4\right)^2-6J_1 N ,\label{eq:rect2}
\end{equation}
since $J_1/4>0$ this Hamiltonian is minimized when $\mathbf{S}_1+\mathbf{S}_2-\mathbf{S}_3-\mathbf{S}_4=\mathbf{0}$ on every rectangle, and the ground state energy per site is $-6J_1$. This rule is compatible with ferromagnetism. 

We can map the   models $\Gamma(0,0,0)-X(1,0,0)-W\left(1,\frac{1}{2},0\right)$ and $\Gamma(0,0,0)-\Delta(q,0,0)-X(1,0,0)$ to each other by changing the sign of $J_1$ but keeping the other two interactions intact. 
In the following we exploit the relationship between the two models, and for the details we refer to \ref{gxdeltasubsection}.  

Stacking of FM and AFM $(100)$ planes works in complete analogy with Subsection \ref{gxdeltasubsection}, we only need to interchange the words antiferromagnetic and ferromagnetic, and instead of rect$_2$ we have to use rect$_1$. The possible orderings constructed by stacking FM/AFM $(100)$ planes (confirmed by the solution of the Ising model and numerical results on the $XY$-model) are 
of the form of an alternate stacking of two independent FM planes: $F_1 F_2 F_1 F_2\dots$, of independent AFM layers: $A_1 A_2 A_3 A_4\dots$, and of independent AFM layers separated by FM planes of fixed magnetization direction: $F A_1 F A_2 F A_3\dots$. 
 
In the numerical solution of the planar model we find $(100)$ chains again, and we can apply the effective two dimensional model of Eq.~(\ref{square_2D}) on the square lattice, but now with parameters $K_1=2J_1$,  $K_2=0$, $K_3=-2J_1$, and  $K_4=-J_1$. Note, that strong ferromagnetic $J_2=-2J_1<0$ bonds connect along the chains again. This model also has a codimension one $\gsm$: in the BZ of the square lattice  the minima reside on the BZ boundary together with the $\Gamma(0,0)$ point, see Fig.~\ref{fig:100_chains}(b).  Note that this manifold is nothing but the intersection of $\mathcal{M}_Z^1$ with the $(100)$ $\mathbf{q}$-plane passing through the origin extended with the $\Gamma(0,0,0)$ point. The energy per site is $-6J_1$. This Hamiltonian can also be written as a sum of squares on signed rectangles inherited from the rect$_1$-s, projected to the $(100)$ plane, see Fig.~\ref{fig:100_chains}(a). This is consistent with the numerical findings of the planar spins: all the configurations found obeyed this projected rectangle rule, but seemed otherwise disordered. 

All the ground states found above can be mapped to the  ground states of the $\Gamma(0,0,0)-\Delta(q,0,0)-X(1,0,0)$ model by  choosing chains along one of the $\left\langle 100\right\rangle$ directions and changing the signs of all the spins on every second chain in a checkerboard pattern (i.e. we flip the spins on all the white $(100)$ chains in Fig.~\ref{fig:100_chains}(a)).

To summarize we propose the following candidate ground states for the $\Gamma(0,0,0)-X(1,0,0)-W\left(1,\frac{1}{2},0\right)$ triple point:
\begin{itemize}
	\item Stacking of independent antiferromagnetic $\left\{ 100\right\} $ planes in the style $A_1 A_2 A_3 A_4\dots$. This type of ordering is realized in all the Ising, $XY$ and $O(3)$ models.
	\item Stacking of independent antiferromagnetic layers separated by the same ferromagnetic layers on the $\left\{ 100\right\} $ planes in  an $A_1 F A_2  F A_3  F \dots$ style. This type of ordering is realized in all the Ising, $XY$ and $O(3)$ models.
	\item  Stacking of an alternating set of two independent $(100)$ ferromagnetic planes $F_1 F_2 F_1 F_2\dots$, reaalized in the $XY$ and $O(3)$ models. 
	\item Interacting ferromagnetic chains in the $\left\langle 100 \right\rangle $ directions, these are absent in the Ising models.	
\end{itemize}

\section{Conclusions}
\label{sec:conclusions}

We presented a detailed study of the ground state phase diagram of the classical isotropic $J_1$-$J_2$-$J_3$ Heisenberg model on the face-centered-cubic lattice. We found and analyzed in detail --in real and Fourier space-- the commensurate Type I, II and III structures, where the multiple-$\mathbf{Q}$ orderings allow for noncoplanar and even chiral structures. Besides the commensurate orders, the introduction of the third neighbor coupling resulted in incommensurate spin spiral orders with propagation vectors along high symmetry axes of the crystal.

We also found ground state manifolds in $\mathbf{q}$-space of dimension one and two with subextensive degeneracies at phase boundaries. In all cases, we could express the Hamiltonian as a positive definite sum of complete squares on finite motifs covering the lattice. This reformulation provided us with hints to explicitly construct large classes of nontrivial, aperiodic ground states in real space, consisting of randomly stacked ordered planes and frustrated ferromagnetically ordered chains in special crystallographic directions. We described relations of the real space patterns to the $ \mathbf{q}$-space picture.

We thoroughly analyzed the model for Ising spins on finite clusters in the phases with extended manifolds, and determined the number and type of possible configurations in real and Fourier space. We performed numerical simulations on $XY$ models and confirmed the validity of our analytical results. Numerical studies on $O(3)$ spins revealed even richer structures than considered here, these need further investigations.  

It is interesting to compare the commensurate orders found here for the fcc lattice to the construction of \textit{regular magnetic orders} in Ref.~\cite{messio_colored}. Besides the trivial ferromagnetic order only the Type I antiferromagnet can be regular, and the latter only if we choose $|\xi|=|\eta|=|\zeta|=1/\sqrt{3}$ in Eq.~(\ref{Xspins}), i.e. for equally weighted Bragg peaks: this is the three dimensional analogue of the tetrahedral state presented in Refs.~\cite{korsh,Momoi_PRL,Kubo_Momoi,messio_colored}.

Our work sets the stage for future studies aimed at investigating the finite-temperature classical phase diagram of the $J_{1}$-$J_{2}$-$J_{3}$ model including an investigation of its critical phenomenon which has, till date, largely focused only on the nearest neighbor model. In particular, the triple points and phase boundaries which are host to a subextensively degenerate manifold of ground states would provide for a promising route towards potentially realizing classical as well as $S=1/2$ quantum spin liquids on the fcc lattice~\cite{Sun-2018,Dominik,Sonnenschein-2020}, in the scenario when order-by-disorder fails to lift the degeneracy as is known to occur for the pyrochlore~\cite{Moessner-1998a,Moessner-1998b,Iqbal-2019} and hyper-hyperkagome lattices~\cite{Chillal-2020}. The triple points occurring in the $J_{1}$-$J_{2}$-$J_{3}$ Heisenberg model on the simple cubic and body-centered-cubic lattices are known to give way to a quantum paramagnetic phase for $S=1/2$~\cite{Laubach-2015,Iqbal-2016,Oitmaa-2017,Iqbal-2019bcc}. Given the fact that three of the degenerate manifolds involve a ferromagnetic phase implies that in the scenario that long-range dipolar magnetic orders are absent, multipole orders such as quadrupolar~\cite{Andreev-1984,Shannon-2006,Sindzingre-2007,Sindzingre-2010,Shindou-2009,Shindou-2011,Richter-2010,Shannon-2004,Penc-2010,Seabra-2016,Iqbal-2016b}, and octupolar~\cite{Zhitomirsky-2008} orders could be stabilized in both classical and quantum models. The role of disorder in stabilizing noncollinear phases will also be an interesting endeavor for future studies~\cite{henley1989,Smirnov-2017}.

\acknowledgements

We acknowledge financial support by the Hungarian Nemzeti Kutatási Fejlesztési és Innovációs Hivatal (NKFIH) Grant No.~K124176, by the BME-Nanonotechnology and Materials Science FIKP grant of Emberi Erőforrások Minisztériuma (EMMI) (BME FIKP-NAT), and by the Science and Engineering Research Board (SERB), DST, India for support through Startup Research Grant No.~SRG/2019/000056 and MATRICS Grant No.~MTR/2019/001042. This research was supported in part by the National Science Foundation under Grant No.~NSF~PHY-1748958, the Abdus Salam International Centre for Theoretical Physics (ICTP) through the Simons Associateship scheme, and the International Centre for Theoretical Sciences (ICTS) during a visit for participating in the program “Novel phases of quantum matter” (Code: ICTS/topmatter2019/12) and “The 2nd Asia Pacific Workshop on Quantum Magnetism” (Code: ICTS/apfm2018/11). We thank F.~Becca, H.~O.~Jeschke, F.~Mila, T.~Müller, J.~Reuther, J.~Richter, R.~Thomale, and M. Zhitomirsky for helpful discussions and collaboration on related topics. 

\appendix
\section{Convention for the lattice and Fourier transform of the exchange interactions}
\label{sec:lattice_conventions}

We choose the following $\mathbf{a}_i$ as primitive lattice translations of the face-centered-cubic (fcc) lattice:
\begin{equation}
\mathbf{a}_1=\frac{1}{2}\left(1, 1, 0\right), \mathbf{a}_2=\frac{1}{2}\left(1, 0, 1\right),   \mathbf{a}_3=\frac{1}{2}\left(0, 1,1\right),
\end{equation}
depicted in Fig.~\ref{fig:exchanges}(a). We will refer to the lattice points by their Cartesian components in units of the lattice constant: $\mathbf{R}_i=(x,y,z)$, note that either all the Cartesian coordinates are integers, or two of them are half-integers, so that $x+y+z$ is always an integer. The corresponding reciprocal lattice vectors are:
\begin{equation}
\mathbf{b}_1=2\pi\left(1, 1, -1\right),  \mathbf{b}_2=2\pi\left(1, -1, 1\right), \mathbf{b}_3=2\pi\left(-1, 1,1\right),
\end{equation}
we will refer to any point in reciprocal space by its $q$-triplet, e.g. $\mathbf{q}=\left(q_x, q_y, q_z\right)=(2\pi,2\pi,-2\pi)=\mathbf{b}_1$.  Special points and lines in the Brillouin zone (BZ) have more or less commonly used labels, we will refer to them either by their labels, or the labels with their Cartesian coordinates in parenthesis in units of $2\pi$, e.g.  one of the BZ corners of the fcc lattice can be referred to as $W$, $W\left(1,\frac{1}{2},0\right)$, or $(2\pi,\pi,0)$.

The Fourier transform of the exchange interaction  for the fcc lattice with first, second and third neighbor interactions presented in Eq.~(\ref{Hamiltonian}) is defined by
\begin{equation}
J(\mathbf{q}) = \frac{1}{2}\sum_{\bm{\delta}} J_{\delta} e^{\imath\mathbf{q}\cdot\bm{\delta}}  \;,
\end{equation}
and it reads:
\begin{widetext}
\begin{align}
J(\mathbf{q}) &  = 
2J_1 \left( \cos \frac{q_x}{2} \cos \frac{q_y}{2} + \cos \frac{q_x}{2} \cos \frac{q_z}{2} + \cos \frac{q_y}{2} \cos \frac{q_z}{2} \right)+ J_2 \left( \cos q_x + \cos q_y + \cos q_z \right)
\nonumber\\&\phantom{ = }
+ 4 J_3 \left( 
\cos q_x \cos \frac{q_y}{2} \cos \frac{q_z}{2}
+ \cos \frac{q_x}{2} \cos q_y \cos \frac{q_z}{2}
+ \cos \frac{q_x}{2} \cos \frac{q_y}{2} \cos q_z 
\right).\label{ft:concrete}
\end{align}
\end{widetext}

\section{Table of phase boundaries}
\label{sec:table_of_phase_boundaries}
The analytical expressions for the boundaries between the different phases shown in Fig. \ref{fig:phase_diagram_descartes} are collected in Table \ref{tab:phase_boundaries}. 
\begin{table*}
	\caption{The boundaries between  phases having ordering vectors $\mathbf{Q}_A$ and $\mathbf{Q}_B$. The optimized values of the incommensurate ordering vectors are given by $\cos \left( q_\Delta/2\right)=\frac{-J_1-2 J_3}{J_2+4 J_3}$,
		$\cos\left( q_\Lambda\right)=-\frac{J_1+J_2+2 J_3}{4J_3}$ and $\cos\left( q_\Sigma/2\right)=\frac{\sqrt{(J_1+2 (J_2+J_3))^2-48 J_3 (J_1-2 J_3)}-J_1-2 (J_2+J_3)}{24 J_3}$. 
		The third column gives the equations of the phase boundaries, and the last column gives the order of the transition.
		Compare this table with the phase diagram given in Fig.~\ref{fig:phase_diagram_descartes}. For pictures of wave-vectors in the Brillouin zone see Fig.~\ref{fig:brillzone}.}
	\begin{center}
		\begin{ruledtabular}
			\begin{tabular}{cccc}
				$\mathbf{Q}_A$&$\mathbf{Q}_B$&$\varepsilon\left(\mathbf{Q}_A\right)=\varepsilon\left(\mathbf{Q}_B\right)$& Type\\ \hline
				$\Gamma\left(0,0,0\right)$ & $\Delta\left(q_\Delta,0,0\right)$ &  $J_1+J_2+6 J_3=0$  & 2nd\\
				$\Gamma\left(0,0,0\right)$ & $\Lambda\left(q_\Lambda,q_\Lambda,q_\Lambda\right)$& $J_1+J_2+6 J_3=0$  & 2nd\\
				$X\left(1,0,0\right)$ & $\Delta\left(q_\Delta,0,0\right)$ & $J_1-J_2-2 J_3=0$  & 2nd\\
				$L\left(\onehalf,\onehalf,\onehalf\right)$ & $\Lambda\left(q_\Lambda,q_\Lambda,q_\Lambda\right)$ & $J_1+J_2-2 J_3=0$  & 2nd\\ 
				$\Gamma\left(0,0,0\right)$ & $X\left(1,0,0\right)$& $J_1+2 J_3=0$  & 1st\\
				$\Gamma\left(0,0,0\right)$ & $W\left(1,\onehalf,0\right)$&$4 J_1+J_2+4 J_3=0$  & 1st\\
				$\Gamma\left(0,0,0\right)$ & $L\left(\onehalf,\onehalf,\onehalf\right)$ & $J_1+J_2+2 J_3=0$  & 1st\\
				$X\left(1,0,0\right)$ & $L\left(\onehalf,\onehalf,\onehalf\right)$& $J_1-3 J_2+2 J_3=0$  & 1st\\
				$\Delta\left(q_\Delta,0,0\right)$ & $L\left(\onehalf,\onehalf,\onehalf\right)$& $J_1^2-J_1 J_2-2 J_2^2-6 J_2 J_3+12 J_3^2 =0$  & 1st\\
				$\Delta\left(q_\Delta,0,0\right)$ & $\Lambda\left(q_\Lambda,q_\Lambda,q_\Lambda\right)$& $3 J_2-4 J_3=0$& 1st\\
				$\Gamma\left(0,0,0\right)$ & $\Sigma\left(q_\Sigma,q_\Sigma,0\right)$ & $19 J_1+6 J_2+46 J_3+8 \sqrt{6 J_1^2+5 J_1 J_2+2 J_2^2}=0$ & 1st\\
				$X\left(1,0,0\right)$ & $\Sigma\left(q_\Sigma,q_\Sigma,0\right)$ & $-11 J_1+10 J_2-14J_3+8 \sqrt{2 J_1^2-3 J_1 J_2+2 J_2^2}=0$ & 1st\\
				$L\left(\onehalf,\onehalf,\onehalf\right)$ & $\Sigma\left(q_\Sigma,q_\Sigma,0\right)$ & $\varepsilon\left(\mathbf{Q}_L\right)=\varepsilon\left(\mathbf{Q}_\Sigma\right)$\footnote{The equation for the phase boundary $\varepsilon\left(\mathbf{Q}_L\right)=\varepsilon\left(\mathbf{Q}_\Sigma\right)$ is:
					\begin{equation}
						J_1^4 + 3 J_1^3 J_2 - 2 J_1^2 J_2^2 - 12 J_1 J_2^3 - 8 J_2^4 - 64 J_1^3 J_3 +
						58 J_1^2 J_2 J_3 + 104 J_1 J_2^2 J_3- 24 J_2^3 J_3 + 376 J_1^2 J_3^2 -
						28 J_1 J_2 J_3^2 - 728 J_2^2 J_3^2 - 768 J_1 J_3^3 - 264 J_2 J_3^3  +  528 J_3^4 = 0.	\nonumber	
				\end{equation}} & 1st\\
				$W\left(1,\onehalf,0\right)$ & $\Sigma\left(q_\Sigma,q_\Sigma,0\right)$ & $\varepsilon\left(\mathbf{Q}_W\right)=\varepsilon\left(\mathbf{Q}_\Sigma\right)$\footnote{The equation for the phase boundary $\varepsilon\left(\mathbf{Q}_W\right)=\varepsilon\left(\mathbf{Q}_\Sigma\right)$ is:
					\begin{equation}
					(J_2 - 4 J_3) (J_1^3 + 2 J_1^2 J_2 - 4 J_1 J_2^2 - 8 J_2^3 -
					50 J_1^2 J_3 + 120 J_1 J_2 J_3 - 72 J_2^2 J_3 + 172 J_1 J_3^2 -
					232 J_2 J_3^2 - 152 J_3^3) = 0.	\nonumber
					\end{equation}
				}  & 1st
			\end{tabular}
		\end{ruledtabular}
	\end{center}
	\label{tab:phase_boundaries}
\end{table*}

\section{The algebraic way to classify the triple-$\mathbf{Q}$ $W\left(1,\onehalf,0\right)$ states}
\label{sec:W_algebra}

We can rewrite Eq.~\eqref{WCartesian} as 
\begin{equation}
  \mathbf{S}^{\mathbf{n}}_i = \sqrt{2}
  \left(
    \begin{array}{c}
      \xi\cos\left(\mathbf{W}_1 \cdot \mathbf{R}_i + (2 n_x +1) \frac{\pi}{4} \right)\\ 
	  \eta\cos\left(\mathbf{W}_2 \cdot \mathbf{R}_i + (2 n_y +1) \frac{\pi}{4} \right)\\ 
	  \zeta\cos\left(\mathbf{W}_3 \cdot \mathbf{R}_i + (2 n_z +1) \frac{\pi}{4} \right)
    \end{array}
   \right),
  \label{eq:WCartesian_n}
\end{equation}
where $\mathbf{n} = (n_x,n_y,n_z)$ is a vector of integers  defined mod 4, and $\xi$, $\eta$, and $\zeta$ are non-negative real numbers. Any combination of signs in front of the $\pi/4$ phase and of the signs of $\xi$, $\eta$, and $\zeta$ in Eq.~\eqref{WCartesian} is now encoded in the vectors $\mathbf{n}$.

The action of the space group elements on the spin configurations can be represented in the way the $\mathbf{n}$ change. The action of a translation $\bm{\delta}$, described by Eq.~(\ref{eq:transS}), becomes
\begin{equation}
  (t_{\bm{\delta}} \mathbf{S}^{\mathbf{n}})_{\mathbf{R}} =  \mathbf{S}^{\mathbf{n}}_{\mathbf{R}-\bm{\delta}} =  \mathbf{S}^{t_{\bm{\delta}} \mathbf{n}}_{\mathbf{R}} 
    \label{eq:transSn}
\end{equation}
where 
\begin{equation}
  t_{\bm{\delta}} \mathbf{n} = \mathbf{n} \oplus \mathbf{n}_{\bm{\delta}} = \mathbf{n} + \mathbf{n}_{\bm{\delta}} \mod 4.
  \label{eq:ndelta}
\end{equation}
We introduced the shorthand notation $\oplus$ as addition modulo 4.
Comparing Eqs.~(\ref{eq:WCartesian_n}), (\ref{eq:transSn}), and (\ref{eq:ndelta}), we get 
\begin{equation}
  \mathbf{n}_{\bm{\delta}} = -\frac{2}{\pi}\left(\mathbf{W}_1\cdot\bm{\delta}, \mathbf{W}_2\cdot\bm{\delta},\mathbf{W}_3\cdot\bm{\delta}\right) \mod 4
\end{equation}  
 and the actual values of the $\mathbf{n}$ vector for the elementary translations are
\begin{subequations} 
\begin{align}
  \mathbf{n}_{(0,1/2,1/2)} &= (2,3,1), \\
  \mathbf{n}_{(1/2,0,1/2)} &= (1,2,3), \\
  \mathbf{n}_{(1/2,1/2,0)} &= (3,1,2).
\end{align}
\end{subequations}
For the inversion we get
\begin{equation}
  (I \mathbf{S}^{\mathbf{n}})_{\mathbf{R}} =  \mathbf{S}^{\mathbf{n}}_{-\mathbf{R}} =  \mathbf{S}^{-\mathbf{n}\oplus \mathbf{n}_I}_{\mathbf{R}} 
    \label{eq:ISn}
\end{equation}
with $\mathbf{n}_I = (3,3,3)$, while for the time reversal 
\begin{equation}
  (\Theta \mathbf{S}^{\mathbf{n}})_{\mathbf{R}} =  -\mathbf{S}^{\mathbf{n}}_{\mathbf{R}} =  \mathbf{S}^{\mathbf{n}\oplus \mathbf{n}_\Theta}_{\mathbf{R}} ,
    \label{eq:ThetaSn}
\end{equation}
with $\mathbf{n}_\Theta = (2,2,2)$.

These operations define a  group acting on $Z_4 \otimes Z_4 \otimes Z_4 = Z_4^{\otimes 3}$ configuration space spanned by the $\mathbf{n}$ vectors. Starting from a configuration labeled by $\mathbf{n}=(0,0,0)$, the translations generate an orbit consisting of 32 configurations  constrained by the $n^x+n^y+n^z=\text{even}$ condition --- this gives half of the all possible   $4^3=64$  configurations in the $Z_4^{\otimes 3}$. Since the number of sites in the unit cell is also 32, we may conclude that all the translated configurations are different. The other half of the configurations (the other orbit) can be generated by translations starting from $\mathbf{n}=(1,1,1)$, changing the condition to $n^x+n^y+n^z = \text{odd}$, and inversion provides a one-to-one map between the two orbits. Thus the possible 64 configurations separate into two disjoint partitions of $Z_4^{\otimes 3}$, showing the two different handednesses.

\section{The degeneracy of the manifolds for Ising or collinear spins on finite clusters}
\label{sec:Ising_results}

\begingroup
\squeezetable
\begin{table}[tb!]
	\caption{Number of planes parallel to one of the $\{100\}$ or $\{111\}$ directions in finite clusters given by the $\mathbf{g}_1$, $\mathbf{g}_2$, and $\mathbf{g}_3$ vectors. $L$ is the linear size of the cluster. All clusters respect the full point group symmetry $O_h$ of the fcc lattice.
	\label{tab:clusters}}
	\begin{ruledtabular}
		\begin{tabular}{cccccccc}
			\multicolumn{3}{c}{Cluster geometry} & No. & \multicolumn{2}{c}{Parallel planes} \\
			$\mathbf{g}_1$ & $\mathbf{g}_2$ & $\mathbf{g}_3$ & sites & $L^{(100)}$ & $L^{(111)}$ \\
			\hline
			$ (0,L,L)$ & $(L,0,L)$ & $(L,L,0)$ & $L^3$ & $L$ & $L$ &\\
			$(2L,0,0)$ & $(0,2L,0)$ & $(0,0,2L)$ & $4 L^3$ & $2L$ & $L$ & \\
			$( - 2L,2L,2L)$ & $(2L, - 2L,2L)$ & $(2L,2L, - 2L)$ & $16 L^3$ & $2L$ & $L$ & \\
		\end{tabular}
	\end{ruledtabular}
\end{table}
\endgroup

To see the way the degeneracy in $\mathbf{q}$-space of the manifolds manifests itself in real space, we have considered Ising spins on finite clusters. The degeneracy of the Ising spins is also the degeneracy for collinear $O(3)$ spin configurations, which is just half of the degeneracy of the Ising manifold if we factor out the trivial $O(3)$ global rotation. 

First, we generate the linear set of equations defining the manifold (e.g. the tetrahedron rule) on a finite cluster with periodic boundary conditions. The finite clusters are defined by the superlattice vectors 
$\mathbf{g}_1$, $\mathbf{g}_2$, and $\mathbf{g}_3$, such that $\mathbf{S}_{i+ \mathbf{g}_1} = \mathbf{S}_{i}$ and so on. The clusters of different shape are listed in Table~\ref{tab:clusters}, together with the number of planes in different directions.  Next, since the set of linear equation is homogeneous, we search for the  null space (or kernel) they define. The dimension of the null space $D_{\text{NS}}$ depends on the type of the manifold and on the shape of the cluster, and the $N-D_{\text{NS}}$ spins in the cluster can be expressed as linear combinations of $D_{\text{NS}}$ linearly independent spins. This would suggest that the number of Ising configurations is $2^{D_{\text{NS}}}$ -- however not all of the solutions satisfy the spin length constraint. In order not to miss a configuration, we generate by computer all the $2^{D_{\text{NS}}}$ linear combinations, and keep only those that give Ising spins on every site. We have collected the numerical results in Table~\ref{tab:Isingnumerics} and discuss the different manifolds below. The findings for the 1D manifolds are summarized in Table~\ref{tab:discrete_degs}.
 
\subsection{The  $\mathcal{M}^2$ 2D manifold with the octahedron rule}

We discussed $\mathcal{M}^2$ in Sec.~\ref{subsec:M2}. The spins shall obey the octahedron rule (Eq.~(\ref{eq:octahedron_rule}). The numerical results are summarized in the last two columns of Table~\ref{tab:Isingnumerics}. Seemingly, the dimension of the null space depends randomly on the size of the cluster. Connecting the real space picture to the reciprocal space reveals that the dimension of the null space is equal to the number of discrete $\mathbf{q}$ points which satisfy Eq.~(\ref{surf_eq}), i.e. which lie on  the two-dimensional manifold shown in Fig.~\ref{fig:brillzone}(c). 
The number of Ising spin configurations is typically much larger then in 1D manifolds.

\subsection{The $\Gamma(0,0,0)-\Lambda(q,q,q)-L(\onehalf,\onehalf,\onehalf)$ $\mathcal{M}^1_\Lambda$ manifold }

The "signed square" constraint (\ref{eq:squarecondition}) provides 3 equations per site, but the number of linearly independent equations grows linearly with the system size, more precisely linearly with the number of the $\{111\}$ planes, as seen in Table~\ref{tab:Isingnumerics} and summarized in Table~\ref{tab:discrete_degs}. 

This is in perfect agreement with the results of Sec.~\ref{Lambda_subsection}, that this manifold consists of independent up- or down-pointing ferromagnetic triangular $\{111\}$ planes. 
Since there are four $\langle 111\rangle$ directions, this leads to $4\times 2^L - 6$ configurations in this manifold, the constant 6 is compensating for the multiple counting of the 8 periodic single-$\mathbf{Q}$ $L\left(\onehalf,\onehalf,\onehalf\right)$ states.
For an even number of planes 8 additional states appear which do not have the layer structure. They are quadruple-$\mathbf{Q}$ $L\left(\onehalf,\onehalf,\onehalf\right)$ states, with 4 amplitudes equal in absolute value. The up and down spins form two interpenetrating pyrochlore lattices, the unit cell consists of 8 sites (e.g. the  $A=B=C=D=1$ and $\bar A=\bar B = \bar C = \bar D = - 1$ in Fig.~\ref{fig:XL_order}(b) represents one of the 8 states). 
	
Let us now see what do Eqs.~(\ref{eq:abce_constraint}) tell us for Ising spins. In this case $a, b, c, m \in \mathbb{R}$ and Eq.~(\ref{eq:eda}) becomes $e^2+a^2=1$ and $e a =0$. These two equations are satisfied by either 
$e^2=1$ and $a=0$, or $e=0$ and $a^2= 1$.
The first solution implies $b=0$ and $c=0$, and the spins in the octahedron are all identical to $e = \pm1$, resulting in a ferromagnetic configuration. 
When $e=0$, there are eight solutions, $a=\pm 1$, $b=\pm 1$, and $c=\pm 1$ corresponding to the choice of the three signs, and the solutions describe structures where on the opposite $\{111\}$ faces of the octahedra we have opposite spins. We can use these octahedral building blocks to tile the fcc lattice. The possible configurations are the uncoupled ferromagnetic $\{111\}$ fcc planes, including the fully polarised ferromagnetic phase as a special case, and the  quadruple-$\mathbf{Q}$ $L\left(\onehalf,\onehalf,\onehalf\right)$ phases, in full agreement with the numerical findings presented in the preceding paragraph.

\subsection{The $\Gamma(0,0,0)-\Delta(q,0,0)-X(1,0,0)$  $\mathcal{M}^1_\Delta$ manifold}

This manifold has been discussed in Sec.~\ref{gxdeltasubsection}. Here there are 6 equations per site, as there are six rectangles per site, see Eq.~(\ref{eq:rect2}) and Table~\ref{tab:motifs}. The dimension of the null space grow linearly with the number of $(100)$ planes, $D_{\text{NS}} = 3 L^{(100)}- 2$, where the factor three comes from the equivalent $(100)$, $(010)$,  and $(001)$ planes. The Ising configurations in the manifold consists of:
	\begin{enumerate}
		\item $F_1F_2F_3F_4\dots F_L$ configurations. The FM layers can have arbitrary directions, and their number is $3\times 2^L - 4$. The periodic states are the 6 single-$\mathbf{X}$ states and the two fully polarized $\Gamma(0,0,0)$ states.
		\item $AF_1 A F_2 A\dots F_{L/2}$ like configuration with alternating ferro- and antiferromagnetic layers, where ferromagnetic layers are independent, but the antiferromagnetic layers are locked with respect to each other. The number of states is $3\times4\times 2^{L/2} - 16$. Among these configurations there are the 8 quadruple-$\mathbf{Q}$ states made from the three $\mathbf{X}_\alpha$-s and the  $\mathbf{Q}=(0,0,0)$ vector.
	\end{enumerate}
	Altogether there are $ 3 \times 4^L + 12\times 2^L - 20$ configurations. For cluster that are not compatible with the 4-site  cubic unit cell, only the $F_1F_2F_3F_4\dots F_L$ states are allowed, so the degeneracy is $3\times 2^{L} - 4$.

\subsection{The $\mathcal{M}^1_Z$ manifold with the tetrahedron rule}

 The states of the  Heisenberg model with only nearest neighbor interactions belong to this manifold. The ground states satisfy the tetrahedron rule -- the sum of the spins on every elementary tetrahedron is zero. This constraint gives two linear equations per site, but these equations are not linearly independent. Since the tetrahedra are edge sharing, the number of linearly independent equations is greatly reduced, and scales with the linear size of the cluster, more precisely with the number of $\left\{100\right\}$ parallel planes. The dimension of the null space (shown in Table~\ref{tab:discrete_degs}) is $3L^{(100)} - 3$ when there are $L^{(100)}$ parallel planes, the factor of 3 comes from the 3 equivalent directions of the parallel planes in clusters respecting the full cubic symmetry. 
	
	We may now count the degeneracy of the manifold assuming Ising spins. Choosing a direction, say $[100]$ 	and the corresponding set of parallel planes, each of $(100)$ planes is antiferromagnetic with a $Z_2$ degree of freedom (we can exchange spins on the two sublattices), the total number of Ising configurations is $2^L$. Since we have three possible directions, the total number of Ising configurations is $3 \times 2^{L^{(100)}} - 6$,  the 6  compensates for the multiple counting of the periodic configurations consisting of antiferromagnetic planes in two directions and ferromagnetic planes in the third direction -- these are the single-$\mathbf{X}$ states.
	
	We can extend the covering with tetrahedra  by signed rectangles, as discussed in Sec.~\ref{XWsubsection}, to allow for second and third neighbor exchanges. Even though the number of linear equations increases by 6 per site, they do not lower the dimension of the null space, neither do they change the number of Ising configurations.

\subsection{The $\mathcal{M}^1_Z \cup \Gamma$ manifold}

This is the endpoint of the line $\mathcal{M}^1_{Z}$ in the $J_1-J_2-J_3$ parameter space, where  the tetrahedron rule is lost  and only the rectangles remain, see Sec.~\ref{gxwsubsection} for details. The $\Gamma$ point appears as an allowed $\mathbf{Q}$ vector. The number of equations is 6 per site and the dimension of the null space has increased by one compared to the pure $\mathcal{M}^1_{Z}$ manifold. The allowed Ising configurations are:
	\begin{enumerate}
		\item $A_1A_2A_3\dots A_L$ like configurations: these are inherited from the $\mathcal{M}^1_{Z}$ manifold and their number is $3\times 2^L - 6$. As we noted, the 6 periodic single-$\mathbf{X}$ states belong to this class.
		\item $FA_1FA_2F\dots F A_{L/2}$ configurations where ferromagnetic and antiferromagnetic planes alternate along a $\langle100\rangle$ direction when the number of planes is even. While the rectangles lock the spins in the ferromagnetic layers, the $L/2$ antiferromagnetic layers retain their $Z_2$ degree of freedom, and the tetrahedron rule is violated. The number of these states is $12 \times 2^{L/2} - 16$. The factor  12 comes from the 3 choices of the $\langle100\rangle$ directions, the polarisation of ferromagnetic planes (a factor of 2), and the choice of the first plane to be ferromagnetic or antiferromagnetic (another factor of 2). The periodic states are the 4-sublattice one-half magnetization plateau configurations (8 of them), that consist of the $\Gamma(0,0,0)$ and $\mathbf{X}$ quadruple-$\mathbf{Q}$ structure, with coefficients equal in absolute value.
		\item Pure ferromagnets: 2 Ising degeneracy. 
	\end{enumerate}

Altogether there are $3 \times 4^L + 12\times 2^L - 20$ configurations, just like for $\mathcal{M}^1_{\Delta}$.
For clusters that are not compatible with the four site cubic unit cell, the frustration only allows the 2 FM configurations.

\begin{table*}[b!]
	\caption{Discrete degeneracy of the one dimensional manifolds for Ising spins in finite size clusters respecting the full cubic symmetry of the fcc lattice. The degeneracy depends on the number of the $(100)$ planes  (or $(111)$ planes in the case of the $\mathcal{M}^1_{\Lambda}$ manifold). Depending on the even/odd number of the planes, the frustration reduces the degeneracy.
		\label{tab:discrete_degs}}
	\begin{ruledtabular}
		\begin{tabular}{cccccc}
			Manifold & $\text{Eq}/N$ & \multicolumn{2}{c}{Planes} & Dim. of & Number of Ising \\
			type & & No. & Parity & null space & configurations \\
			\hline
			\multirow{2}{*}{$\mathcal{M}^1_{Z}$} & \multirow{2}{*}{8} & \multirow{2}{*}{$L^{(100)}$} 
			& even & $3L^{(100)} \!-\! 3$ & $3\!\times\! 2^{L^{(100)}} \!-\! 6$ \\
			&& & odd & 0 & 0 \\
			\\
			\multirow{2}{*}{$\mathcal{M}^1_{Z}\cup \Gamma$} & \multirow{2}{*}{6} & \multirow{2}{*}{$L^{(100)}$} 
			& even & $3L^{(100)} \!-\! 2$ & $ 3 \!\times\! 2^{L^{(100)}} \!+\! 12\!\times\! 2^{L^{(100)}/2} \!-\! 20$ \\
			&& & odd & $1$ & $ 2 $ \\
			\\
			\multirow{2}{*}{$\mathcal{M}^1_{\Delta}$} & \multirow{2}{*}{6} & \multirow{2}{*}{$L^{(100)}$}
			& even & $3L^{(100)} \!-\! 2$ & $ 3 \!\times\! 2^{L^{(100)}} + 12 \!\times\! 2^{L^{(100)}/2} \!-\! 20$ \\
			&& & odd & $3L^{(100)} \!-\! 2$ & $3\!\times\! 2^{L^{(100)}} \!-\! 4$ \\
			\\
			\multirow{2}{*}{$\mathcal{M}^1_{\Lambda}$} & \multirow{2}{*}{3} & \multirow{2}{*}{$L^{(111)}$} 
			& even & $4L^{(111) }\!-\! 3$ & $4\!\times\! 2^{L^{(111)}} \!+\! 2$ \\
			&&  & odd & $4L^{(111)} \!-\! 3$ & $4\!\times\! 2^{L^{(111)}} \!-\! 6$ \\
		\end{tabular}
	\end{ruledtabular}
\end{table*}

\begin{table*}
\caption{Numerical enumeration of the Ising configurations. The first two columns are the number of $(100)$ and $(111)$ planes in the cluster, the next three columns show the geometry of the clusters with periodic boundary conditions, the number of sites in the cluster is shown in sixth row. The following rows list the dimension  of the null space $D_{\text{NS}}$ (i.e. the number of linearly independent equations) and the number of Ising configurations.
\label{tab:Isingnumerics}}
\begin{ruledtabular}
\begin{tabular}{cccccrrrrrrrrrrr}
  \multicolumn{2}{c}{No. planes}  & \multicolumn{3}{c}{Cluster geometry} 	& No. 	& \multicolumn{2}{c}{$\mathcal{M}^1_{Z}$} & \multicolumn{2}{c}{$\mathcal{M}^1_{Z}\cup \Gamma$} & \multicolumn{2}{c}{$\mathcal{M}^1_{\Delta}$} & \multicolumn{2}{c}{$\mathcal{M}^1_{\Lambda}$} & \multicolumn{2}{c}{$\mathcal{M}^2$} \\
 $L^{(100)}$ & $L^{(111)}$ & $\mathbf{g}_1$ 	& $\mathbf{g}_2$ 	& $\mathbf{g}_3$ 	& sites 	& $D_{\text{NS}}$ 	& Ising& $D_{\text{NS}}$ 	& Ising& $D_{\text{NS}}$ 	& Ising& $D_{\text{NS}}$ 	& Ising& $D_{\text{NS}}$ 	& Ising \\
\hline
 2 	& 2 	& $(2,2,0)$ 	& $(0,2,2)$		& $(2,0,2)$ 	& 8 	& 3 	& 6 	& 4 	& 16 	& 4 	& 16 	& 5 	& 18 	& 4 	& 16 	\\
 3 	& 3 	& $(3,3,0)$ 	& $(0,3,3)$ 	& $(3,0,3)$ 	& 27 	& 0 	& 0 	& 1 	& 2 	& 7 	& 20 	& 9 	& 26 	& 12 	& 0 	\\
 4 	& 4 	& $(4,4,0)$ 	& $(0,4,4)$ 	& $(4,0,4)$ 	& 64 	& 9 	& 42 	& 10 	& 76 	& 10 	& 76 	& 13 	& 66 	& 10 	& 140 	\\
 5 	& 5 	& $(5,5,0)$ 	& $(0,5,5)$ 	& $(5,0,5)$ 	& 125 	& 0 	& 0 	& 1 	& 2 	& 13 	& 92 	& 17 	& 122 	& 0 	& 0  	\\
 6 	& 6 	& $(6,6,0)$ 	& $(0,6,6)$ 	& $(6,0,6)$ 	& 216 	& 15 	& 186 	& 16 	& 268 	& 16 	& 268 	& 21 	& 258 	& 40 	& 12688 \\
 7 	& 7 	& $(7,7,0)$ 	& $(0,7,7)$ 	& $(7,0,7)$ 	& 343 	& 0 	& 0 	& 1 	& 2 	& 19 	& 380 	& 25  	& 506  	& 0 	& 0  	\\
 8 	& 8 	& $(8,8,0)$ 	& $(0,8,8)$ 	& $(8,0,8)$ 	& 512 	& 21 	& 762 	& 22 	& 940 	& 22 	& 940 	& 29  	& 1026 	& 34 	& 2637	\\
\\
 4 	& 2 	& $(4,0,0)$ 	& $(0,4,0)$ 	& $(0,0,4)$ 	& 32 	& 9 	& 42 	& 10 	& 76 	& 10 	& 76 	& 5 	& 18 	& 10 	& 140 	\\
 6 	& 3 	& $(6,0,0)$ 	& $(0,6,0)$ 	& $(0,0,6)$ 	& 108 	& 15 	& 186 	& 16 	& 268 	& 16 	& 268 	& 9 	& 26 	& 12 	& 0 	\\
 8 	& 4 	& $(8,0,0)$ 	& $(0,8,0)$ 	& $(0,0,8)$ 	& 256 	& 21 	& 762 	& 22 	& 940 	& 22 	& 940 	& 13 	& 66 	& 34 	& 1133 	\\
10	& 5 	& $(10,0,0)$ 	& $(0,10,0)$ 	& $(0,0,10)$ 	& 500 	& 27   	& 3066 	& 28	& 3436 	& 28	& 3436 	& 17 	& 122 	& 0 	& 0		\\
12 	& 6 	& $(12,0,0)$ 	& $(0,12,0)$ 	& $(0,0,12)$ 	& 864 	& 33   	& 12282	& 34	& 13036	& 34	& 13036	& 21 	& 258  	& 70 	&  	\\
\\
2 	& 1 	& $(- 2,2,2)$ 	& $(2,-2,2)$ 	& $(2,2,-2)$ 	& 16 	& 3 	& 6 	& 4 	& 16 	& 4 	& 16 	& 1 	& 2 	& 0 	& 0 	\\
4 	& 2 	& $(- 4,4,4)$ 	& $(4,-4,4)$ 	& $(4,4,-4)$ 	& 128 	& 9 	& 42 	& 10 	& 76 	& 10 	& 76 	& 5 	& 18 	& 34 	& 1377 	\\
6 	& 3 	& $(- 6,6,6)$ 	& $(6,-6,6)$ 	& $(6,6,-6)$ 	& 432 	& 15 	& 186 	& 16 	& 268 	& 16 	& 268 	& 9 	& 26 	& 12 	& 0 	\\
8 	& 4 	& $(- 8,8,8)$ 	& $(8,-8,8)$ 	& $(8,8,-8)$ 	& 1024 	& 21 	& 762 	& 22 	& 940 	& 22 	& 940 	& 13 	& 66 	& 82 	&  	\\
\end{tabular}
\end{ruledtabular}
\end{table*}

\bibliography{bibliography} 
\end{document}